\documentclass[a4paper,11pt]{article}
\pdfoutput=1
\usepackage{jcappub}

\def\laq{~\raise 0.4ex\hbox{$<$}\kern -0.8em\lower 0.62ex\hbox{$\sim$}~}
\def\gaq{~\raise 0.4ex\hbox{$>$}\kern -0.7em\lower 0.62ex\hbox{$\sim$}~}

\def\beq{\begin{equation}}
\def\eeq{\end{equation}}
\def\bea{\begin{eqnarray}}
\def\eea{\end{eqnarray}}

\def \pa {\partial}
\def \dd {\partial}
\def \ra {\rightarrow}
\def \ti {\widetilde}

\def \ga {\gamma}

\def \de {\delta}
\def \De {\Delta}

\def \Om {\Omega}

    \def\be{\begin{equation}}
    \def\ee{\end{equation}}
    \def\ba{\begin{eqnarray}}
    \def\ea{\end{eqnarray}}

\newcommand{\eq}{\begin{equation}}
\newcommand{\eqx}{\end{equation}}
\newcommand{\eqn}{\begin{eqnarray}}
\newcommand{\eqnx}{\end{eqnarray}}

\newcommand{\Ups}{\Upsilon}

\newcommand{\nv}{{\bf{n}}}

\newcommand{\etasB}{{\eta_s^{(0)}}}
\newcommand{\etasF}{{\eta_s^{(1)}}}
\newcommand{\etasS}{{\eta_s^{(2)}}}
\newcommand{\rsB}{{\deta_s^{(0)}}}
\newcommand{\rsF}{{\deta_s^{(1)}}}

\newcommand{\thetaF}{{\theta_s^{a (1)}}}

\newcommand{\HH}{\mathcal{H} }

\newcommand{\ndv}{{v_{||}}}
\newcommand{\vn}{{\bf n}}
\newcommand{\vk}{{\bf k}}

\newcommand{\deltaF}{{\delta^{(1)}}}
\newcommand{\deltaS}{{\delta^{(2)}}}

\newcommand{\INT}{{\int_{\eta_s}^{\eta_o}}}
\newcommand{\deltazF}{{\delta z^{(1)}}}
\newcommand{\deltazS}{{\delta z^{(2)} }}
\newcommand{\Gaunt}[6]{{\cal G}^{#1#2#3}_{#4#5#6}}

\newcommand{\psii}{{\psi^I}}
\newcommand{\psia}{{\psi^A}}

\newcommand{\dQint}{{\int_{\eta_s}^{\eta_o} d\eta' \partial_{\eta'} \psii \left( \eta' \right) }}

\newcommand{\Hcal}{\mathcal H}

\newcommand{\deta}{r}
\newcommand{\Qint}{{\int_{\eta_s}^{\eta_o} d\eta' \psii \! \left( \eta' \right) }}
\newcommand{\sigmaB}{{\left( 1 - \frac{1}{\HH_s \deta_s } \right)}}
\newcommand{\sigmaF}{{\left( 1 - \frac{\HH'_s}{\HH_s^2  } \right)}}

\newcommand{\Jint}{{ \INT d\eta' \frac{\eta' - \eta_s}{\eta_o -\eta'} \Delta_2 \psii \left( \eta' \right)}}
\newcommand{\thetaint}{{\INT d\eta' \gamma_0^{ab} \partial_b \int_{\eta'}^{\eta_o} d\eta'' \psii \left( \eta''\right)}}

\newcommand{\DERSis}{{\frac{1}{\HH_s^2}\left(\partial^2_r\psi^I_s\!+\!\partial^2_\eta\psi^I_s\!-\!\partial_{\eta}\partial_{r}\psi^I_s \right)}}
\newcommand{\DERSas}{{\frac{1}{\HH_s^2}\left(\partial^2_\eta\psi^A_s\!-\! \partial^2_r\psi^A_s\!-\!
\partial_{\eta}\partial_{r}\psi^A_s \right)}}

\title{Galaxy number counts to second order and their bispectrum}

\author[a]{Enea Di Dio }
\author[a]{, Ruth Durrer }
\author[a]{, Giovanni Marozzi }
\author[a]{, Francesco Montanari }

\affiliation[a]{
Universit\'e de Gen\`eve, D\'epartement de Physique Th\'eorique and CAP,
24 quai Ernest-Ansermet, CH-1211 Gen\`eve 4, Switzerland
}

\emailAdd{Enea.DiDio@unige.ch}
\emailAdd{Ruth.Durrer@unige.ch}
\emailAdd{Giovanni.Marozzi@unige.ch}
\emailAdd{Francesco.Montanari@unige.ch}

\abstract{
We determine the number counts to second order in cosmological perturbation theory in 
the Poisson gauge and allowing for anisotropic stress. The calculation is performed using an innovative approach based on the recently proposed "geodesic light-cone" gauge. This allows us to determine the number counts in a purely geometric way, without using Einstein's equation. The result is valid for general dark energy models and (most) modified gravity models.
We then evaluate numerically some relevant contributions to  the number counts bispectrum. In particular we consider the terms involving the density, redshift space distortion and lensing.  
}

\begin{document}
\maketitle

\section{Introduction}
\label{Sec1}
\setcounter{equation}{0}

Cosmology has entered the precision era. So far, mainly via observations of the cosmic microwave background (CMB) most recently performed by the Planck satellite~\cite{Ade:2013kta,Ade:2013zuv}. But also high precision observations of cosmic large scale
structure (LSS), i.e. the distribution of matter in the Universe, are being performed, see e.g.~\cite{Samushia:2013yga,Delubac:2014aqe}, and are under construction or planning, see~\cite{DES,euclid}\footnote{See also: \url{http://www.darkenergysurvey.org/science/} and \url{http://www.euclid-ec.org}} and references therein.

The interpretation of observations of LSS is more complicated than the CMB. First of all we observe galaxies and not the density field, this is the biasing problem, and secondly, fluctuations on small scales become large so that non-linearities become important. Nevertheless, if we can handle these problems, a detailed observation of LSS is 
very interesting because it contains much more information than the CMB. LSS is a three dimensional dataset, and if we can observe it from the scale of about $1h^{-1}$Mpc out to the Hubble scale, we have in principle $\simeq (3000)^3 =  2.7\times 10^{10}$ independent modes at our disposition. This has to be compared with the CMB with about $10^7$ modes (including $E$-polarization). 

Furthermore, non-linearities in the CMB are very small and CMB fluctuations are very Gaussian. Hence the power spectrum captures nearly all the information and (reduced) higher order correlations are very small. For large scale structure, at least on intermediate to small scales, non-linearities become important. The analysis of the highly non-linear regime can probably only be done via full numerical simulations, but in the intermediate, weakly non-linear regime higher order perturbation theory can be applied. As soon as non-linearities are present, higher order correlations develop i.e.  non-trivial three-point and reduced four-point functions are generated. These functions contain additional interesting information on gravitational clustering which can be very important for the Dark Energy problem or the contribution from massive neutrinos.

The present work inscribes in this direction. We first compute a basic LSS observable, namely the galaxy number counts to second order in perturbation theory and then determine the induced bispectrum. Our approach is novel in several aspects.

We  describe the number counts as a function of direction and observed redshift. 
This has the big advantage that it is directly related to observables and  that it is  model independent  contrary to the galaxy correlation function in real space or its power spectrum. The latter need the knowledge of a background Universe to convert measured redshifts and angles into distances and are therefore not well adapted, e.g., to estimate cosmological parameters.
This method was introduced in~\cite{Bonvin:2011bg} and implemented and tested numerically 
in~\cite{DiDio:2013bqa,DiDio:2013sea}. Galaxy number counts have the advantage of being directly measured and their correlation function $\xi(\theta,z,z')$ or the power spectrum in the form of $C_\ell(z,z')$ can be directly compared to observations.

The first order fluctuations of the number counts have been calculated in~\cite{Yoo:2009au,Yoo:2010ni,Bonvin:2011bg,Challinor:2011bk}. In this work we go to second order. Several second order calculations have already been performed.
Expressions for the redshift perturbations and cosmological distances were obtained in~\cite{BenDayan:2012wi,Fanizza:2013doa,Marozzi:2014kua}  and applications are discussed in~\cite{BenDayan:2012ct,BenDayan:2013gc,Ben-Dayan:2014swa}. Similar expressions are also obtained in~\cite{Umeh:2012pn,Umeh:2014ana}.
Results for the Newtonian density fluctuation were obtained in~\cite{Bernardeau:2001qr} and second order perturbation theory of  lensing is discussed in~\cite{Bernardeau:2011tc}.
Also expressions for the second order number counts have just appeared in~\cite{Bertacca:2014dra,Bertacca:2014wga,Yoo:2014sfa}. However, since these calculations are very involved, it is useful to have results from independent groups, obtained in different ways, which can then be compared. Here we employ a novel method to calculate the second order number counts. We use the geodesic light-cone gauge introduced in~\cite{Gasperini:2011us}. 
Actually, if we would know the density fluctuation exactly, our expression for the number counts would be fully non-perturbative. Then we perform a second order coordinate transformation to translate the result to the familiar Bardeen potentials, i.e. the  metric perturbations in longitudinal or Poisson gauge. We also allow for anisotropic stresses. We however neglect second order vector and tensor contributions which can be easily obtained from the first order results by inserting the second order vector and tensor perturbations induced by the Bardeen potentials, which have been calculated in the past~\cite{Acquaviva:2002ud,Ananda:2006af,Lu:2008ju}. We write the final result fully in terms of the Bardeen potentials, the peculiar velocities and the density fluctuations. Of course, the final expressions remain complicated, but they can be readily used and compared to other results in the literature.

Finally, we use our expressions to compute some of the most important new terms in the bispectrum, namely the contribution from redshift space distortion and the one from lensing, and we evaluate them numerically and compare them with the usual term from weakly non-linear Newtonian gravitational clustering.

The remainder of this paper is organised as follows: in Section~\ref{Sec2} we lay out the basic quantities which we want to compute. In Section~\ref{Sec3} we describe the geodesic light-cone gauge and its transformation to the Poisson or longitudinal gauge. In Section~\ref{Sec4} we compute the second order number counts first in geodesic light-cone gauge and then we transform it to Poisson gauge. 
We also present a simplified formula containing only the dominant terms.
In Section~\ref{Sec5} we derive and discuss some of the most relevant terms in the bispectrum for which we also plot numerical results using {\sc class}gal~\cite{DiDio:2013bqa}. In Section~\ref{Sec6} we conclude. Some useful  relations are collected in Appendix~\ref{AppA}.

\section{Galaxy Number Counts}
\label{Sec2}
\setcounter{equation}{0}

What we really observe in galaxy surveys is the number $N$ of galaxies in a redshift bin $dz$ and solid angle $d\Om$ as a function of the observed redshift $z$ and of the observation direction defined by the unit vector $-\nv$ (in this work $\nv$ denotes the direction of propagation of the  photon). 
Directly from these observable quantities we define the fluctuation of galaxy number counts as
\be \label{numcount}
\Delta \left(\nv ,z \right) \equiv \frac{N  \left(\nv ,z \right) - \langle N \rangle \left( z \right)  }{ \langle N \rangle \left( z \right) } \, ,
\ee
where $\langle \cdot \rangle$ denotes the average over directions. If we neglect bias, the observed number of galaxies can be expressed in terms of the density $\rho \left( \nv , z \right)$ and the volume $V \left( \nv , z \right)$  at fixed redshift according to
\be
N\left( \nv , z \right)=  \rho \left( \nv , z\right) V \left(\nv , z \right) \, .
\ee
To determine the galaxy number counts $\Delta \left(\nv ,z \right)$ up to second order in perturbation theory we need to expand the redshift density perturbation and the volume to second order.
The volume and the density have to be defined in function of the observed redshift and direction of observation, then their expansion to second order can be given  as
\bea
V\left( \vn, z \right)  &=& \bar V \left( z \right) + \delta V^{(1)} \left( \vn , z \right)+\delta V^{(2)} \left( \vn , z \right) =\bar V \left( z \right) \left( 1 + \frac{\delta V^{(1)}}{\bar V} + \frac{\delta V^{(2)}}{\bar V} \right)\,  \\
\rho \left( \vn, z \right)  &=& \bar \rho \left(\eta_s^{(0)}\right) \left(  1 + \delta^{(1)} + \delta^{(2)} \right) \,.
\label{GenDefrho2order}
\eea
Here, $z$ denotes the observed redshift, which is also perturbed, and we define the time coordinate of a fiducial background model of a source by
\be 
1+z_s=\frac{a(\eta_o)}{a(\eta_s^{(0)})}\,,
\ee
where $\eta_o$ denotes the conformal time of the observer. We also note that the background conformal distance is $r_s^{(0)}=\eta_o-\eta_s^{(0)}$.

Hence to some extent, the perturbations of the volume and the density above come from the perturbations of the redshift. In the next section we shall present the second order perturbations of the redshift, obtained in~\cite{Marozzi:2014kua}, and Section~\ref{Sec3}
is devoted to the calculation of the volume and the density fluctuations.

Once these perturbations are calculated, inserted in the definition~(\ref{numcount}), they yield to second order in perturbation theory the following result 
\be
\Delta \left( \nv ,z \right) = \left[ \deltaF + \frac{\delta V^{(1)}}{ \bar V} + \deltaF \frac{\delta V^{(1)}}{\bar  V} + \deltaS +  \frac{\delta V^{(2)}}{\bar  V}  - \langle \deltaF\frac{\delta V^{(1)}}{\bar  V} \rangle  - \langle \deltaS \rangle - \langle \frac{\delta V^{(2)}}{\bar  V}\rangle\right] 
\ee
Here we use the fact that the angular mean of first order perturbations vanishes.
We then identify the first order and the second order contributions to galaxy number counts, respectively,
\bea \label{numcount:first}
\Delta^{(1)} &=& \deltaF + \frac{\delta V^{(1)}}{\bar{V}}\, , \\
\Delta^{(2)} &=& \deltaF \frac{\delta V^{(1)}}{\bar{V}} + \deltaS + \frac{\delta V^{(2)}}{\bar{V}} - \langle \deltaF \frac{\delta V^{(1)}}{\bar{V}} \rangle - \langle \deltaS \rangle - \langle  \frac{\delta V^{(2)}}{\bar  V} \rangle\, .
\label{numcount:second}
\eea
The first order contribution~(\ref{numcount:first}) has been already computed in~\cite{Yoo:2009au,Yoo:2010ni,Bonvin:2011bg,Challinor:2011bk}. Here we are mainly interested in computing the second order term~(\ref{numcount:second}). For this purpose we  adopt the so-called geodesic light-cone (GLC) coordinates defined in~\cite{Gasperini:2011us}. In terms of GLC coordinates we determine an exact solution which we then expand in a more conventional gauge, namely the Poisson gauge,  to second order.

\section{From the Geodesic Light-Cone to the Poisson gauge}
\label{Sec3}
\setcounter{equation}{0}

Let us first introduce the geodesic light-cone (GLC) coordinates~\cite{Gasperini:2011us}.
They consist of a timelike coordinate $\tau$ (which can be  identified with the proper time of  synchronous gauge and, 
therefore, describes a geodesic observer that is static in this gauge~\cite{BenDayan:2012pp}),  of a null coordinate $w$ and of two angular coordinates $\tilde{\theta}^a$ ($a=1,2$). We denote them with a tilde to distinguish these exact screen space angles from the background angles in the Friedmann metric which we introduce below.

The line-element of the GLC metric takes the form  
\beq
\label{LCmetric}
ds^2 =\Upsilon^2 dw^2 - 2 \Upsilon  dw d\tau+\gamma_{ab}(d \tilde{\theta}^a-U^a dw)(d \tilde{\theta}^b-U^b dw)\, ,\quad a, b = 1,2\, .
 \eeq
It depends on six arbitrary functions ($\Upsilon$,  the 2-component vector $U^a$ and the symmetric $2\times2$ matrix $\gamma_{ab}$).
In matrix form:
\beq
\label{GLCmetric}
g_{\mu\nu} =
\left(
\begin{array}{ccc}
0 & -\Upsilon &  \vec{0} \\
-\Upsilon & \Upsilon^2 + U^2 & -U_b \\
\vec0^{\,T}  &-U_a^T  & \gamma_{ab} \\
\end{array}
\right)
\, ,\quad
g^{\mu\nu} =
\left(
\begin{array}{ccc}
-1 & -\Upsilon^{-1} & -U^b/\Upsilon \\
-\Upsilon^{-1} & 0 & \vec{0} \\
-(U^a)^T/ \Upsilon & \vec{0}^{\, T} & \gamma^{ab}
\end{array}
\right) \, ,
\eeq
where $\gamma_{ab}$ and its inverse  $\gamma^{ab}$ lower and  raise  the two-dimensional indices and $U^2=\ga_{ab}U^aU^b$. The 2-dimensional space described by the coordinates $(\tilde{\theta}^1,\tilde{\theta}^2)$ is the exact screen space with metric $\gamma_{ab}$.

In an unperturbed Friedmann background, our coordinates correspond to cosmic time $\tau=t$, background past light-cone 
$w=\eta+r$, where $\eta$ is conformal time so that $ad\eta=dt$, and the usual polar angles $\theta=\tilde{\theta}^1$ and $\phi=\tilde{\theta}^2$. The background metric components are $\Upsilon =a$, $U_a=0$ and $\gamma_{ab}d\tilde\theta^ad\tilde\theta^b= a^2r^2(d\theta^2 +\sin^2\!\theta d\phi^2)$. Here $a$ denotes the cosmic scale factor. 

Like in 
synchronous gauge, also the GLC gauge has some residual gauge freedom (for details see~\cite{Fanizza:2013doa}).
The condition $w=$ constant defines a null hypersurface ($\pa_\mu w \pa^\mu w=0$), corresponding to the past light-cone of a given observer, hereafter chosen to be 
the  geodesic observer which moves with  4-velocity
 $u_\mu = - \partial_{\mu} \tau$. The geodesic equation for $u^\mu$ is $\left( \pa^\nu \tau\right) \nabla_\nu \left( \pa_\mu \tau\right) = 0$. Let us also recall  that, in GLC gauge, the null geodesics connecting sources and observer are characterised simply by the tangent vector $k^{\mu} = - \omega g^{\mu \nu} \partial_{\nu} w = -  \omega g^{\mu w} = \omega  \Upsilon^{-1} \delta^{\mu}_{\tau}$ (where $ \omega$ is an arbitrary normalization constant), meaning that photons  travel at constant values of $w$ and $\tilde{\theta}^a$. 
This renders the calculation of  the redshift particularly simple in this gauge.

We now give exact, non-perturbative, expressions for  the redshift and the cosmological distances, like the luminosity and angular distances, in GLC gauge.  As in the previous section, subscripts ``o'' and ``s'', respectively,  denote quantities evaluated at the observer and the source space-time positions. We consider a light ray emitted by a geodesic source with 4-velocity $u^\mu_s = - (g^{\mu\nu}\partial_{\nu} \tau)_s$ lying on the past light-cone of  the geodesic observer defined by $w=w_o$ and on the spatial hypersurface $\tau= \tau_s$.
The light ray will be received by the geodesic observer at $\tau=\tau_o>\tau_s$. 
The exact non-perturbative expression of  the redshift $z_s$ associated with this light ray is then simply given by \cite{Gasperini:2011us}
\be
\label{redshift}
(1+z_s) = \frac{(k^{\mu} u_{\mu})_s }{(k^{\mu} u_{\mu})_o}  = \frac{(\partial^{\mu}w \pa_\mu \tau)_s }{(\partial^{\mu}w \pa_\mu \tau)_o}  = {\Ups(w_o, \tau_o, \ti \theta^a)\over \Ups(w_o, \tau_s, \ti \theta^a)} ~~.
\ee

On the other hand, in \cite{Fanizza:2013doa} an exact expression for the so-called Jacobi Map~\cite{SEF} is derived in GLC gauge and the following non-perturbative solution for the luminosity distance $d_L$ and the angular diameter distance $d_A$ is obtained:
\beq
\label{lumdist}
d_L^2 =  (1+z_s)^4 d_A^2 = 
(1+z_s)^4 \frac{4 \sqrt{\gamma_s}}{\left[\det \left(u_{\tau}^{-1} \pa_\tau{\gamma}^{ab}\right) \gamma^{3/2}\right]_{o}} \,, 
\eeq
where $\gamma$ denotes the determinant of the 2-dimensional matrix $\gamma_{ab}$.
 
We finally want to express our results in Poisson gauge.
Neglecting vector and tensor contributions, the Poisson gauge (PG) metric~\cite{PG,PG2} (sometimes also denoted 'Newtonian gauge' or 'longitudinal gauge')
takes the  form
\bea
ds_{PG}^2 &=& a^2(\eta) \left[ -(1+ 2 \Phi) d\eta^2  + (1- 2 \Psi)\delta_{ij}  dx^i dx^j \right]  \nonumber \\
&=&  a^2(\eta) \left[ -(1+ 2 \Phi) d\eta^2  + (1- 2 \Psi)( dr^2 + r^2 d^2 \Omega) \right] 
\label{PGmetricstandard}
\eea
where the (generalized) Bardeen potentials $\Phi$ and $\Psi$ are defined, up to second order, as follows:
\be
\Phi \equiv \phi + \frac{1}{2} \phi^{(2)} \, , \qquad \Psi \equiv \psi + \frac{1}{2} \psi^{(2)} ~~,
\ee
and we make no assumption on the anisotropic stress, so that $\Psi$ and $\Phi$ may differ also at first order. 

To express the redshift and the luminosity distance given in Eqs.(\ref{redshift}) and (\ref{lumdist}) in terms of standard PG variables we have to transform the GLC gauge quantities to quantities in PG. 
Following \cite{Marozzi:2014kua}, and employing suitable boundary conditions:
$i)$ the transformation is non singular around the observer position at $r=0$, and $ii)$  the two-dimensional spatial sections $r=$ const are locally parametrized at the observer position (for any time) by standard spherical coordinates $(\theta,\phi)\equiv(\theta^1,\theta^2) $, 
the coordinate transformation to second order  
is given by \cite{Marozzi:2014kua}
\bea
\tau &=& \tau^{(0)}+\tau^{(1)}+\tau^{(2)} \nonumber \\
 \text{with} & & \tau^{(0)} = \left( \int_{\eta_{in}}^\eta d\eta' a(\eta') \right) = t-t_{in} \quad,\quad \tau^{(1)} = a(\eta) P(\eta, r, \theta^a)\,, \nonumber \\
& &  \tau^{(2)}= \int_{\eta_{in}}^\eta d\eta' \frac{a(\eta')}{2} \left[ \phi^{(2)} - \phi^2 + ( \partial_r P )^2 + \gamma_0^{ab} ~ \partial_a P ~ \partial_b P \right] (\eta', r, \theta^a)~,
\label{tau2order} \\
w &=& w^{(0)}+w^{(1)}+w^{(2)} \nonumber \\
\text{with} & & w^{(0)}=\eta_+ \equiv \eta+r \quad,\quad w^{(1)}=Q(\eta_+, \eta_-, \theta^a) \,,
\nonumber \\
& & 
w^{(2)}= { \frac{1}{4} \int_{\eta_o}^{\eta_-} \!\!dx\left[ {\psi}^{(2)} + {\phi}^{(2)}+2(\psi^2-\phi^2) + 2 (\psi+\phi) \partial_+ Q + {\gamma}_0^{ab} ~ \partial_a Q ~ \partial_b Q \right] (\eta_+, x, \theta^a)}\,, \nonumber\\
\label{w2order} 
\eea
\bea
\!\tilde{\theta}^a \!&=&\! \tilde{\theta}^{a (0)}+\tilde{\theta}^{a (1)}+\tilde{\theta}^{a (2)} \nonumber \\
\text{with} 
& & \tilde{\theta}^{a (0)}=\theta^a 
\quad,\quad
\tilde{\theta}^{a (1)}={\frac12 \int_{\eta_o}^{\eta_-} dx~ \left[ {\gamma}_0^{ab} \partial_b Q \right] (\eta_+,x,\theta^a)} 
\nonumber \\
& &
\tilde{\theta}^{a (2)}= { \int_{\eta_o}^{\eta_-} dx~ 
\Big[ 
\frac{1}{2}
{\gamma}_0^{ac} \partial_c w^{(2)} 
+ {\psi} {\gamma}_0^{ac} \partial_c w^{(1)} 
+\frac{1}{2}{\gamma}_0^{dc} \partial_c w^{(1)} \partial_d   \tilde{\theta}^{a (1)}
+\frac{1}{2}(\psi+\phi)\partial_+\tilde{\theta}^{a (1)}}
\nonumber \\
& &
\hspace{2.7cm}+(\phi-\psi) \partial_-\tilde{\theta}^{a (1)}
-\partial_+ w^{(1)} \partial_-\tilde{\theta}^{a (1)} \Big](\eta_+,x,\theta^a) \,,
\label{thetatilde2orderShort}
\eea
where $(\ga_0^{ab}) = {\rm diag}(r^{-2},r^{-2} \sin^{-2}\theta)$, and $\eta_{in}$ represents an early enough time when the perturbations (or better their integrands) were negligible. We have also introduced the zeroth-order light-cone variables $\eta_\pm= \eta \pm r$, 
with corresponding partial derivatives:
\beq
\pa_\eta = \pa_+ + \pa_- \, , \qquad \pa_r = \pa_+ - \pa_- \, ,\qquad \pa_\pm= {\pa \over \pa \eta_\pm}={1\over 2} \left( \pa_\eta \pm \pa_r \right) \, .
\eeq
Furthermore, we define
\be
P(\eta, r, \theta^a) = \int_{\eta_{in}}^\eta d\eta' \frac{a(\eta')}{a(\eta)} \phi(\eta',r,\theta^a)
\, , \qquad Q(\eta_+, \eta_-, \theta^a) = \int_{\eta_o}^{\eta_-} dx~ \frac{1}{2}
\left(\psi+\phi\right)(\eta_+,x,\theta^a) \, .
\label{PQ}
\ee
Note that the integrals for $\tau$ go along the conformal time axis $\eta$ at fixed $r$, while the integrals for $w$ and $\tilde\theta^a$ are along the background light cone, $\eta_-$ at fixed $\eta_+$.

With this we can obtain the non-trivial entries of the GLC metric of Eq. (\ref{GLCmetric}) to second order in terms of the variables $(\eta,r,\theta^a)$:
\bea
\Upsilon^{-1} &=& \frac{1}{a(\eta)} \left[ 1 + \partial_+ Q - \partial_r P  +\frac{1}{2}(\psi-\phi)
+\partial_{\eta}w^{(2)} + \frac{2}{a}\partial_- \tau^{(2)} - \phi^{(2)} + 2\phi^2 
-\frac{1}{2}\phi(\phi+\psi)
\right. \nonumber \\
& & \left.
-\phi \partial_+ Q+\partial_r P  \left(\frac{1}{2}\phi-\frac{3}{2}\psi\right)- \partial_r P\partial_+ Q
- \gamma^{ab}_0 \partial_a P \partial_b Q\right] \,,
\label{Ups1}  
\eea
\bea
U^a &=& \partial_{\eta}\tilde{\theta}^{a (1)}-\frac{1}{a}\gamma_0^{ab}\partial_b \tau^{(1)}+\partial_{\eta}\tilde{\theta}^{a (2)}-
\frac{1}{a}\gamma^{ab}_0\partial_b \tau^{(2)} - \frac{1}{a} \partial_r \tau^{(1)} \partial_r \tilde{\theta}^{a(1)}  
\nonumber \\
& &  -\phi \partial_\eta \tilde{\theta}^{a (1)}-\frac{2}{a} \psi \gamma^{ab}_0 \partial_b \tau^{(1)} 
-\frac{1}{a}\gamma_0^{cd}\partial_c \tau^{(1)} \partial_d \tilde{\theta}^{a (1)} \nonumber \\
& &
+\left(\partial_+ Q - \partial_r P+\frac{1}{2}(\psi-\phi)\right) \left(-\partial_{\eta} \tilde{\theta}^{a (1)}+\frac{1}{a}\gamma^{ab}_0 \partial_b \tau^{(1)}\right)\,,
\label{Ua1}
\\ 
\gamma^{ab} &=& a^{-2}\left[ \gamma_0^{ab} \left(1 +  2 \psi\right) +2\gamma_0^{(a c} \partial_c \tilde{\theta}^{b) (1)}+ \gamma_0^{ab}
\left(\psi^{(2)} + 4 \psi^2 \right)-\partial_\eta \tilde{\theta}^{a (1)}\partial_\eta \tilde{\theta}^{b (1)}
\right. \nonumber \\
& & \left.
+\partial_r \tilde{\theta}^{a (1)}\partial_r \tilde{\theta}^{b (1)} +4 \psi \gamma_0^{(a c} \partial_c \tilde{\theta}^{b) (1)}+\gamma_0^{c d} \partial_c \tilde{\theta}^{a (1)}
 \partial_d \tilde{\theta}^{b (1)} 
 +2\gamma_0^{(a c} \partial_c \tilde{\theta}^{b) (2)} \right].
\label{gammaab}
\eea 
Here $X^{(a\cdots}Y^{\cdots b)\cdots}$ denotes symmetrization and $a$ and $b$.

As a first example of the  use of the coordinate transformation given above let us present the perturbed redshift up to second order in PG in the presence of anisotropic stress and in function of the observer's angular coordinates~\cite{Marozzi:2014kua}.
This is easily done starting from the non-perturbative solution~(\ref{redshift}) and using Eq.~(\ref{Ups1})
together with the fact that  $\tilde{\theta}^a$ actually are the standard angular coordinate at the observer position and are constant along the line of sight.

Therefore, any perturbative quantity can be written as a function of the observer's angular coordinates, i.e. the direction of  observation, by Taylor-expanding it around $\tilde{\theta}^a=\theta^a_o$.
To this purpose, and considering general quantity with an angle-independent background value, it is enough to invert Eq.~(\ref{thetatilde2orderShort}) to first order. Namely, we need the expansion
\be
\theta^a = {\theta}^{a (0)}+{\theta}^{a (1)}
= \theta^a_o - \int_{\eta}^{\eta_o} d\eta'~ \gamma_0^{ab} \partial_b \int_{\eta'}^{\eta_o} d\eta''~ \left(\psi+\phi\right)(\eta'', \eta_o-\eta'', \theta_o^a)\,.
\label{thetatilde1orderShort_Inverted}
\ee

To obtain familiar expressions in terms also of peculiar velocities, we recall that $\tau$ plays the role of a gauge-invariant velocity potential (see \cite{Fanizza:2013doa}), so that the spatial components of the perturbed velocity $v_\mu$ of the (geodesic) observer in polar coordinates in PG are given by
\bea
&&\hspace{5cm}(v_i)=(v_r+v_r^{(2)}, v_{\perp a}+v_{\perp a}^{(2)}) 
\nonumber \\
&&
\quad \text{with} \quad v_r=-\partial_r \tau^{(1)}\, , \quad v_r^{(2)}=-\partial_r \tau^{(2)} \,,\quad v_{\perp a}=-\partial_a \tau^{(1)} \,,\quad v_{\perp a}^{(2)}= -\partial_a \tau^{(2)}\,,
\label{GenVel}
\eea
where $ \tau^{(1)}$ and $ \tau^{(2)}$ are the first- and second-order part of the coordinate transformation $\tau= \tau (\eta, r, \theta^a)$ given in Eq.~(\ref{tau2order}). 
The unit vector $n^\mu$ along the direction connecting the source to the observer can be then expanded, 
in polar coordinates and to first order (which is enough for our purpose), 
\begin{equation}
n^\mu=\left(0, -\frac{1}{a}(1+\psi), 0, 0\right)\, , \qquad n_\mu=\left(0, -a(1-\psi), 0, 0\right).
\end{equation}  
Taking its scalar product with the spatial component of the perturbed velocity we obtain
\be 
{\bf v}\cdot {\bf n}=v_{||}+v_{||}^{(2)}\!=\! \partial_r P+ \psi \partial_r P +\frac{1}{2} \!\int_{\eta_{in}}^\eta \! \! d\eta' \frac{a(\eta')}{a(\eta)} \partial_r\! \left[ \phi^{(2)} - \phi^2 + ( \partial_r P )^2 + \gamma_0^{ab} ~ \partial_a P ~ \partial_b P \right] (\eta', r, \theta^a)\,.
\label{vpar}
\ee
For the velocity components which are perpendicular to the radial direction we have
\be
v_{\perp a} v_{\perp}^a = \gamma_0^{ab} \partial_a P \partial_b P \,.
\label{vper}
\ee
Let us also define the following useful variables
\be
\psi^I=\frac{\psi+\phi}{2}\, , \qquad \psi^A=\frac{\psi-\phi}{2}\,,
\label{VariableIsAs}
\ee
which define the isotropic and anisotropic part of the Bardeen potential; $\psi^I$ is also the potential of the scalar part of the Weyl tensor~\cite{book}.
In what follows we will use these variables to express our perturbed quantities.

Furthermore, to simplify the notation, hereafter we only indicate the conformal time $\eta$ as  the integration variable  along the line of sight, instead of all the arguments like $(\eta, \eta_o-\eta, \theta^a_o)$, of course the result also depends on the direction fixed by 
$(\theta^1_o,\theta^2_o)$. 
Finally, following~\cite{Marozzi:2014kua}, we obtain the observed redshift to second order and in function of the direction of observation
\be
1+z_s = \frac{a(\eta_o)}{a(\eta_s)} \left[1+ \delta z^{(1)} + \delta z^{(2)}  \right]
\ee
with
\be
\delta z^{(1)} = - v_{||s}-\psi^I_s+\psi^A_s-2 \int_{\eta_s}^{\eta_o} d \eta' \partial_{\eta'}\psi^I\left(\eta'\right)
\label{Redshift1}
\ee
\begin{eqnarray}
\delta z^{(2)}  &=& - v^{(2)}_{||s}-\frac{1}{2} \phi_s^{(2)}
-\frac{1}{2} \int_{\eta_s}^{\eta_0} d\eta' \partial_{\eta'}  \left[ \phi^{(2)} + \psi^{(2)}\right]\!\left( \eta'\right)
+\frac{1}{2}\left(v_{||s}\right)^2
+\frac{1}{2}\left(\psi^I_s\right)^2 \nonumber \\ & & 
+\left(- v_{||s}-\psi^I_s\right)\left(-\psi^I_s-2 \int_{\eta_s}^{\eta_o} d \eta' \partial_{\eta'}\psi^I\left(\eta'\right)\right)
+\frac{1}{2} v^a_{\perp s} v_{\perp a\,s} 
-2 a \, v^a_{\perp s} \partial_a \int_{\eta_s}^{\eta_o} d \eta' \psi^I\left(\eta'\right)
\nonumber \\ & &
+4\int_{\eta_s}^{\eta_0} d\eta' \left[\psi^I\left(\eta'\right) \partial_{\eta'}\psi^I\left(\eta'\right) 
+ \partial_{\eta'}\psi^I\left(\eta'\right) \int_{\eta'}^{\eta_o} d \eta'' \partial_{\eta''}\psi^I\left(\eta''\right)
\right.
\nonumber \\ & & \left.
+\psi^I\left(\eta'\right) \int_{\eta'}^{\eta_o} d \eta'' \partial^2_{\eta''}\psi^I\left(\eta''\right)
- \gamma_0^{ab} \partial_a \left( \int_{\eta'}^{\eta_o} d \eta'' \psi^I\left(\eta''\right) \right) 
\partial_b \left( \int_{\eta'}^{\eta_o} d \eta'' \partial_{\eta''}\psi^I\left(\eta''\right) \right) \right]
\nonumber \\  & & 
+ 2 \partial_a  \left(v_{||s}+\psi^I_s\right)  \int_{\eta_s}^{\eta_o}\!d \eta' \gamma_0^{ab} \partial_b \int_{\eta'}^{\eta_o}\!d \eta'' \psi^I\left(\eta''\right)
\nonumber \\
& & 
+4 \int_{\eta_s}^{\eta_o}\!\!d\!\eta'\partial_a \left(\partial_{\eta'}\psi^I\left(\eta'\right)\right)
\!\int_{\eta_s}^{\eta_o}\!\!\!d\!\eta''\gamma_0^{ab} \partial_b\!\int_{\eta''}^{\eta_o}\!d\!\eta'''\psi^I\left(\eta'''\right)
\nonumber 
\\ 
& &
-\psi^A_s   v_{||s}+\frac{3}{2} (\psi^A_s)^2
-3\psi^I_s\psi^A_s
-4 \int_{\eta_s}^{\eta_o} d \eta' \partial_{\eta'}(\psi^I\psi^A)\left(\eta'\right)
-2 \psi^A_s  \int_{\eta_s}^{\eta_0} d\eta' \partial_{\eta'} \psi^I\left( \eta'\right)
\nonumber \\ & &
- 2 \partial_a  \psi^A_s  \int_{\eta_s}^{\eta_o}\!d \eta' \gamma_0^{ab} \partial_b \int_{\eta'}^{\eta_o}\!d \eta'' \psi^I\left(\eta''\right)\,.
\label{Redshift2}
\end{eqnarray}
Let us stress that  the above integrals are along the line of sight, $f(\eta')=f(\eta',x^i(\eta'))$ and partial derivatives w.r.t.~$\eta'$ are only derivatives with respect to the first dependence. 

Above and hereafter we neglect the terms from the gravitational potential and the velocity at the observer position. They cannot be calculated within perturbation theory. Note, however, that in addition to the  monopole  and the dipole terms, they now also contribute to the quadrupole e.g.~via terms of the form $v_{\parallel o}^2 = (\vn\cdot{\bf v}_o)^2$. However, differences like $C_2(z_2)-C_2(z_1)$ are meaningful and can be obtained from our formula, see~\cite{DiDio:2013sea} for a discussion of $C_0(z_2)-C_0(z_1)$ for the power spectrum from linear perturbation theory. Furthermore, terms of the form $X_oY_s$ which are a product of a term $X_o$ at the observer position and a term $Y_s$ at the source position can contribute to all moments in the 4th order contributions to the power spectrum. Assuming that the correlation of $X_o$ and $Y_s$ can be neglected, for two sources at positions $s_1$ and $s_2$ they give correlators of the form $\langle X_o^2\rangle\langle Y_{s_1}Y_{s_2}\rangle$. We have checked that $X_o$ is always either the velocity or the gravitational potential at the observer position. The first can be removed by going to the CMB frame while the second is of the order $10^{-10}$ times the power spectrum and hence negligible.  Similar considerations apply for the (reduced) bispectrum.  

\section{Evaluation of the number counts to second order}
\label{Sec4}
\setcounter{equation}{0}

\subsection{Redshift density perturbation}

To second order the density perturbation at fixed PG coordinates can be written as 
\be
 \rho \left( \eta_s, \deta_s, \theta^a_s \right) = \bar \rho  \left( \eta_s\right) + \delta \rho^{(1)}  \left( \eta_s, \deta_s, \theta^a_s \right)  + \delta \rho^{(2)}  \left( \eta_s, \deta_s, \theta^a_s \right)
  = \bar\rho(\eta_s)\big(1 +\de^{(1)}_\rho + \de^{(2)}_\rho\big)\,.
\label{Dens2order}
\ee
We want to evaluate this at fixed observed redshift and with respect to the observer's angular coordinates $\theta_o^a$.
Therefore, we define a fiducial background model with coordinates 
$(\eta_s^{(0)}, \deta_s^{(0)}, \theta_o^a)$ for which the observed redshift and the past light-cone of our observer are given by
\be 
1+{z}_s=\frac{a(\eta_o)}{a(\eta_s^{(0)})}  \, , \qquad w=w_o=\eta_o=\eta_s^{(0)}+\deta_s^{(0)}  \,.
\ee
We then expand conformal time and radial PG coordinates around the coordinates of the fiducial model as $\eta_s=\eta_s^{(0)}+\eta_s^{(1)}
+\eta_s^{(2)}$ and $\deta_s=\deta_s^{(0)}+\deta_s^{(1)}+\deta_s^{(2)}$, and obtain the terms of these expansions by 
perturbatively solving the following system of 
equations:
\begin{eqnarray}
& & 1+{z}_s=\frac{a(\eta_o)}{a(\eta_s^{(0)})}=\frac{a(\eta_o)}{a(\eta_s)} \left[1+ \delta {z}^{(1)} + \delta {z}^{(2)} \right] \label{e:barz} \\
& & w =\eta_o= w^{(0)}+w^{(1)}+w^{(2)} \label{e:w}
\end{eqnarray}
where we  use Eqs.~(\ref{w2order}),~(\ref{Redshift1}) and~(\ref{Redshift2}) for the perturbations appearing in Eqs.~(\ref{e:barz}) and~(\ref{e:w}).
We then obtain the perturbed conformal time and radial coordinate needed so that our measured redshift corresponds to the background redshift of our fiducial model,
\be 
\eta_s^{(1)}=\frac{ \delta z^{(1)}}{\Hcal_s}
\label{etaS1}
\ee
\be
\eta_s^{(2)}=\frac{1}{\Hcal_s}\left[\delta z^{(2)}+{\delta {z}^{(1\rightarrow 2)}}-\frac{1}{2}\left(1+\frac{\Hcal_s'}{\Hcal_s^2}\right)(\delta z^{(1)})^2\right] \,.
\label{etaS2}
\ee
and
\be 
\deta_s^{(0)}=\eta_o-\eta_s^{(0)} \, ,\qquad \deta_s^{(1)}=-\eta_s^{(1)}+2 
 \int_{\eta_s^{(0)}}^{\eta_o} d \eta' \psi^I\left(\eta'\right)
\label{rS1}
\ee
\be
\deta_s^{(2)}=-\eta_s^{(2)}-w_s^{(2)}-w_s^{(1\rightarrow 2)}
\ee
where $\HH_s=(a'/a)(\eta_s^{(0)})$ is the comoving Hubble parameter of the fiducial background model, evaluated at the background conformal time $\eta_s^{(0)}$, and all the quantities above are evaluated at 
$(\eta_s^{(0)}, \deta_s^{(0)}, \theta_o^a)$ .
The quantities ${\delta  {z}^{(1\rightarrow 2)}}$ and $w_s^{(1\rightarrow 2)}$ stand for the second order contribution coming from Taylor expanding $\delta z^{(1)}$ and $w_s^{(1)}$ around the background fiducial model. They are given by 
\bea 
{\delta z^{(1\rightarrow 2)} } &=& \eta_s^{(1)}\left( \partial_\eta \psi^I_s+\partial_\eta \psi^A_s+\Hcal_s v_{||s}
+\partial_r v_{||s}\right)+\left[-2 \partial_\eta \psi^I_s-\partial_r \psi^I_s+\partial_r \psi^A_s
\right. \nonumber \\
 & & \left.
-\partial_r v_{||s}-2 \int_{\eta_s}^{\eta_o} d\eta'
\partial_{\eta'}^2 \psi^I(\eta')\right] \left(2 \int_{\eta_s}^{\eta_o} d\eta' \psi^I(\eta')\right)
\eea
\begin{eqnarray}
w_s^{(1\rightarrow 2)}&=&\left[-\psi^I_s
-2 \int_{\eta_s}^{\eta_o} d \eta' \partial_{\eta'}\psi^I\left(\eta'\right)\right]\left(2 \int_{\eta_s}^{\eta_o} d\eta' \psi^I(\eta')\right) +\psi_s^I \left[2 \eta_s^{(1)}-2 \int_{\eta_s}^{\eta_o} d\eta' \psi^I(\eta')\right] \nonumber \\
& &
+4 \int_{\eta_s}^{\eta_o}\!\!d\!\eta'\partial_a \psi^I\left(\eta'\right)
\!\int_{\eta_s}^{\eta_o}\!\!\!d\!\eta''\gamma_0^{ab} \partial_b\!\int_{\eta''}^{\eta_o}\!d\!\eta'''\psi^I\left(\eta'''\right)
\,,
\end{eqnarray}
where to second order we are allowed to drop the superscript $^{(0)}$ from $\eta_s^{(0)}$. 

Let us now expand the second order result~(\ref{Dens2order}) around the fiducial model, to obtain the second order density fluctuation at fixed observed redshift and direction of observation,
\bea
\rho \left({\bf n}, z_s \right) &=& \bar \rho + \partial_\eta \bar \rho \left( \etasF + \etasS \right) +\frac{1}{2} \partial_\eta^2\bar \rho 
\left(\etasF\right)^2 
\nonumber \\
&&
 + \delta\rho^{(1)} + \etasF \partial_\eta \delta\rho^{(1)} + \rsF \partial_r \delta\rho^{(1)} + \thetaF \partial_a \delta\rho^{(1)} + \delta\rho^{(2)}\, \\
  &=& \bar\rho\left( 1+\de^{(1)} + \de^{(2)}\right)\,.
\eea
Here  all the quantities have to be evaluated at $(\eta_s^{(0)}, \deta_s^{(0)}, \theta_o^a)$.
Finally, using Eqs.~(\ref{etaS1})-(\ref{rS1}) and the relation $\frac{d\bar\rho}{dz} = 3 \frac{\bar\rho}{1+z}$, 
we obtain the following perturbative result: at first order
\bea
\deltaF &=&\frac{1}{\bar \rho} \left( \etasF \partial_\eta \bar\rho + \delta\rho^{(1)} \right) 
\nonumber \\ 
&=&  -3 \, \delta z^{(1)}+\delta_\rho^{(1)}
\label{rhoRedAng1}
\eea
and at second order
\bea
\deltaS &=&   \frac{1}{\bar \rho} \left[ \etasS \partial_\eta \bar\rho + \frac{1}{2}\left( \etasF \right)^2 \partial^2_\eta \bar\rho + \etasF \partial_\eta \delta\rho^{(1)} + \rsF \partial_r \delta\rho^{(1)} + \thetaF \partial_a \delta\rho^{(1)} + \delta\rho^{(2)} \right] 
\nonumber \\
&=& -3 \deltazS + 6 \left( \deltazF \right)^2 
- 3 \left[ \frac{1}{\HH_s} \left( \partial_\eta \psi^I_s +  \partial_\eta \psi^A_s\right)  + v_{||s} + \frac{1}{\HH_s} \partial_r v_{||s}  \right] \deltazF 
\nonumber \\
&& 
+6 \left(  2 \partial_\eta \psi^I_s +\partial_r \psi^I_s -\partial_r \psi^A_s + \partial_r v_{||s}  + 2 \INT d\eta' \partial_{\eta'}^2 \psi^I \left( \eta' \right) \right)  \INT d\eta' \psi^I \left( \eta' \right) 
\nonumber \\
&& + \frac{1}{ \HH_s} \left[  \frac{1}{\bar \rho}\partial_\eta \left(\bar \rho \,\delta_\rho^{(1)}\right) - \partial_r \delta_\rho^{(1)} \right] \deltazF + 2\partial_r \delta_\rho^{(1)} \INT d\eta' \psi^I \left( \eta' \right)  \nonumber \\
&&
-2\partial_a  \delta_\rho^{(1)}
\INT d\eta' \gamma_0^{ab}\partial_b \int_{\eta'}^{\eta_o} d\eta'' \psi^I \left( \eta''  \right)
+\delta_\rho^{(2)}\,,
\label{rhoRedAng2}
\eea
where Eq.~(\ref{rhoRedAng1}) agrees with the first order result of~\cite{Bonvin:2011bg}.

\subsection{Volume perturbation}

In order to compute the contribution to the galaxy number counts due to the volume perturbation, we start considering the 3-dimensional volume element seen by a source with 4-velocity $u^\mu$
\be
dV = \sqrt{-g} \epsilon_{\mu \nu \alpha \beta} u^\mu dx^\nu dx^\alpha dx^\beta \, .
\ee
We then express this in terms of the observed quantities $\left(z, \theta_o, \phi_o \right)$
\be
dV = \sqrt{-g} \epsilon_{\mu \nu \alpha \beta} u^\mu  \frac{\partial x^\nu}{\partial z}\frac{\partial x^\alpha}{\partial \theta_s} \frac{\partial x^\beta}{\partial \phi_s} \left| \frac{\partial \left( \theta_s, \phi_s \right) }{\partial \left( \theta_o \phi_o \right)} \right| dz d\theta_o d \phi_o \equiv v \left( z , \theta_o, \phi_o \right) dz d\theta_o d\phi_o \, .
\ee
The volume perturbation is determined by
\be
\frac{\delta V}{\bar V} = \frac{v - \bar v}{\bar v} = \frac{\delta v}{\bar v}\,.
\ee

Let us  now express the  above quantities in GLC coordinates, $x^\mu = \left( \tau, w, \tilde \theta^a \right)$ with $a=1,2$, to show how the volume perturbation can be easily given fully  non-perturbatively in GLC gauge.
Considering that $\tilde \theta^a$ is constant along the geodesic of the light-cone $\tilde \theta^a=\theta_o^a$, defined through $w=\text{const}$, we have
\be
dV= - \sqrt{-g} u^w \frac{\partial \tau}{\partial z} dz d\theta_o d\phi_o\,.
\ee
Using also $u^w= -g^{w \tau}= \Upsilon^{-1}$ and $\sqrt{-g}=\sqrt{\left| \gamma \right|} \Upsilon$, we obtain
\be
dV = \sqrt{\left| \gamma \right|} \left( - \frac{d\tau}{dz} \right) dz d\theta_o d\phi_o\,, \qquad \mbox{ or } \qquad 
v =  \sqrt{\left| \gamma \right|} \left( - \frac{d\tau}{dz} \right) = \sqrt{\left| \gamma \right|} \frac{\Upsilon_s^2}{\Upsilon_o\dd_{\tau}\Upsilon_s} \,, 
\label{e:vol}
\ee
where we made use of Eq.~(\ref{redshift}) to calculate $d z/d\tau$. Note that Eq.~(\ref{e:vol}) is again a non-perturbative expression for the volume element at the source in terms of the observed redshift and the observation angles in GLC gauge.
 
To express the  above quantity in  Poisson gauge up to second order we have to calculate $\sqrt{\left| \gamma \right|}$ and $\frac{d\tau}{dz}$ in this gauge. 
We first note that $\sqrt{\left| \gamma \right|}$ can be obtained, inverting Eq.~(\ref{lumdist}), from the known result for the area distance given in~\cite{BenDayan:2012wi,Fanizza:2013doa,Marozzi:2014kua}.
In particular, neglecting the terms evaluated at the observer position, we have
\be 
\sqrt{\left| \gamma \right|}=\frac{a_o^2 \deta_s^{(0)\,2}}{(1+z_s)^2} \sin \theta_o \left[1+2 \bar{\delta}_S^{(1)}(z_s, \theta^a_o)+ \left(\bar{\delta}_S^{(1)}(z_s, \theta^a_o)\right)^2+2 \bar{\delta}_S^{(2)}(z_s, \theta^a_o) \right]\,,
\ee
where we have set
\beq
\frac{{d}_L(z_s, {\theta}_o^a)}{(1+z_s)a_o \deta_s^{(0)}}
= {{d}_L(z_s, {\theta}_o^a)\over d_L^{FLRW}(z_s)} =  1 + \bar{\delta}_S^{(1)}(z_s, {\theta}_o^a) + \bar{\delta}_S^{(2)}(z_s, {\theta}_o^a) ~~,
\eeq
and $\bar{\delta}_S^{(1)}(z_s, \theta^a_o)$ and $\bar{\delta}_S^{(2)}(z_s, \theta^a_o)$, the first and second order scalar perturbations of the luminosity distance,  are given in Sec.~III of~\cite{Marozzi:2014kua}.
Finally, from the results of Sec.~III of~\cite{Marozzi:2014kua} and after some algebraic manipulations,
we obtain the following useful perturbative result to first order
\begin{eqnarray}
\frac{\left(\sqrt{\left| \gamma \right|}\right)^{(1)} }{\left(\sqrt{\left| \gamma \right|}\right)^{(0)}} &=&
2 \left(1-\frac{1}{\HH_s r_s^{(0)}}\right) \delta z^{(1)} - 2 \left( \psi_s^I + \psi_s^A \right) + \frac{4}{\deta_s^{(0)}} \int_{\eta_s^{(0)}}^{\eta_o} d\eta' \psi^I\left(\eta'\right) 
\nonumber \\
&&
- \frac{2}{\deta_s^{(0)}} \int_{\eta_s^{(0)}}^{\eta_o} d\eta'  \frac{\eta'-\eta_s^{(0)}}{\eta_o-\eta'} \Delta_2 \psi^I\left(\eta'\right) \,.
\label{SGammaord1}
\end{eqnarray}
Here $\Delta_2$ denotes  the angular Laplacian.

To second order we have a more involved result
\be
\frac{\left(\sqrt{\left| \gamma \right|}\right)^{(2)} }{\left(\sqrt{\left| \gamma \right|}\right)^{(0)}} = 2\sigmaB \delta z^{(2)} + \left[ - \frac{1}{\HH_s \deta_s}  \sigmaF + \sigmaB^2 \right]\left( \delta   z^{(1)}\right)^2+\Xi_{IS}+\Xi_{AS}
\label{SGammaord2}
\ee
where
\bea
& & \Xi_{IS}=-\psi_s^{(2)}
-\frac{1}{2} \frac{1}{\deta_s}\int_{\eta_s}^{\eta_o}\!d \eta' \frac {\eta' - \eta_s}{\eta_o - \eta'} \Delta_2\left[
\psi^{(2)}+\phi^{(2)}\right]\!\left(\eta'\right)
+\frac{1}{\deta_s}\int_{\eta_s}^{\eta_o}\!d \eta'\left[
\psi^{(2)}+\phi^{(2)}\right]\!\left(\eta'\right)
\nonumber \\ 
& &
+2 \left(1-\frac{1}{ \Hcal_s \deta_s}\right) \left\{- \frac{1}{\HH_s}v_{||s}  \partial_r v_{||s}  - v_{||s} ^2 \right.
\nonumber \\
&& 
+ \left[ \psi_s^I -2 \dQint  - \frac{2}{\deta_s} \Qint 
+ \frac{2}{\deta_s} \Jint \right]v_{||s}   
 \nonumber \\
&& +  \left[-\psi_s^I - 2 \dQint - 2 \HH_s \Qint \right] \frac{1}{\HH_s} \partial_r v_{||s}
\nonumber \\
& & 
\left.
+\left[\partial_r\psi^I_s
+2 \partial_\eta\psi^I_s+2 \int_{\eta_s}^{\eta_o}\!d \eta' \partial^2_{\eta'}\psi^I\left(\eta'\right)\right]
\left(-2 \int_{\eta_s}^{\eta_o}\!d \eta' \psi^I\left(\eta'\right)\right)
\right. \nonumber \\ 
& & \left.
-\left(-\psi^I_s-2 \int_{\eta_s}^{\eta_o}\!d \eta' \partial_{\eta'}\psi^I\left(\eta'\right)\right)
\left[
\frac{2}{\deta_s}\int_{\eta_s}^{\eta_o}\!d \eta' \frac {\eta' - \eta_s}{\eta_o - \eta'} \Delta_2
\psi^I\left(\eta'\right)-\frac{2}{\deta_s}\int_{\eta_s}^{\eta_o}\!d \eta'\psi^I\left(\eta'\right)\right]\right\}
\nonumber 
\eea
\bea
& &
+ \left[ -\frac{2}{\deta_s} \Qint + \frac{2}{\HH_s \deta_s} \psi_s^I + \frac{1}{\HH_s^2 \deta_s} \partial_\eta \psi_s^I - \frac{1}{\HH_s} \partial_r \psi_s^I - \frac{1}{\HH_s  \deta_s^2} \INT d\eta' \Delta_2 \psii\left(\eta'\right) \right] 2 v_{||s} 
\nonumber \\
& &
{
-\frac{2}{ \Hcal_s} \partial_a v_{||s} \gamma_{0s}^{ab}  \int_{\eta_s}^{\eta_o}\!\!\!d\eta' \partial_b 
\psi^I(\eta')
}
+\frac{4}{\deta_s^2}\left(\int_{\eta_s}^{\eta_o}\!d \eta'\psi^I\left(\eta'\right)\right)^2
-\frac{4}{\deta_s} \psi^I_s   \int_{\eta_s}^{\eta_o}\!d \eta'\psi^I\left(\eta'\right)
\nonumber \\ 
& &
+8 \psi^I_s \int_{\eta_s}^{\eta_o}\!d \eta' \partial_{\eta'}\psi^I\left(\eta'\right)
+4 (\psi^I_s)^2+\frac{2}{\Hcal_s}\left(\partial_r \psi^I_s-\frac{1}{ \Hcal_s \deta_s}\partial_\eta \psi^I_s\right)\left(-\psi^I_s-2 \int_{\eta_s}^{\eta_o}\!d \eta' \partial_{\eta'}\psi^I\left(\eta'\right)\right)
\nonumber \\
& &
-4 \partial_r \psi^I_s\int_{\eta_s}^{\eta_o}\!d \eta'\psi^I\left(\eta'\right)
+\frac{4}{\deta_s}\int_{\eta_s}^{\eta_o}\!d \eta'\!\left[\psi^I\left(\eta'\right)\left(-\psi^I\left(\eta'\right) 
-2 \int_{\eta'}^{\eta_o}\!d\!\eta'' \partial_{\eta''}\psi^I\left(\eta''\right)\right) 
\right.
\nonumber \\
&&
\left.
+\gamma_0^{ab}\partial_a\!\left(\int_{\eta'}^{\eta_o}\!\!\!d\!\eta''\!\psi^I\left(\eta''\right)\right)\!\partial_b\!\left(\int_{\eta'}^{\eta_o}\!\!\!d\!\eta''\!\psi^I\left(\eta''\right)\right)\right] 
\nonumber 
\\ 
& &
+\left(\psi^I_s-\frac{2}{\deta_s}\int_{\eta_s}^{\eta_o}\!d \eta'\psi^I\left(\eta'\right)\right)
\frac{4}{\deta_s}\int_{\eta_s}^{\eta_o}\!d \eta' \frac {\eta' - \eta_s}{\eta_o - \eta'} \Delta_2\psi^I\left(\eta'\right)
+2 \left(\frac{1}{\deta_s}\int_{\eta_s}^{\eta_o}\!d \eta' \frac {\eta' - \eta_s}{\eta_o - \eta'} \Delta_2\psi^I\left(\eta'\right)\right)^2
\nonumber 
\\ 
& &
+\left[\frac{1}{ \Hcal_s \deta_s}\left(-\psi^I_s-2 \int_{\eta_s}^{\eta_o}\!d \eta' \partial_{\eta'}\psi^I\left(\eta'\right)\right)
-\frac{1}{\deta_s}\int_{\eta_s}^{\eta_o}\!d \eta'\psi^I\left(\eta'\right)\right]\frac{2}{\deta_s}\int_{\eta_s}^{\eta_o}\!d \eta'  \Delta_2\psi^I\left(\eta'\right)
\nonumber 
\\
& & 
-2 \left[\int_{\eta_s}^{\eta_o}\!\!\!d\eta' \frac{1}{(\eta_o-\eta')^2} \psi^I(\eta')+2
 \int_{\eta_s}^{\eta_o}\!\!\!d\eta'  \frac{1}{(\eta_o-\eta')^2} \int_{\eta'}^{\eta_o}\!\!\!d\eta''\partial_{\eta''}\psi^I(\eta'')\right]
  \int_{\eta_s}^{\eta_o}\!\!\!d\eta' \Delta_2 \psi^I(\eta')
\nonumber
\\
& &
+\!4\partial_a\!\psi^I_s
\!\int_{\eta_s}^{\eta_o}\!\!\!d\!\eta'\gamma_0^{ab}\partial_b\!\int_{\eta'}^{\eta_o}\!d\!\eta''\psi^I\left(\eta''\right)
-\frac{8}{r_s}\left[\int_{\eta_s}^{\eta_o}\!d \eta' \partial_a \psi^I\left(\eta'\right)
\!\int_{\eta_s}^{\eta_o}\!\!\!d\!\eta''\gamma_0^{ab}\partial_b\!\int_{\eta''}^{\eta_o}\!d\!\eta'''\psi^I\left(\eta'''\right)\right]
\nonumber
\\
& &
{
+2 \partial_a \left( \int_{\eta_s}^{\eta_o}\!d\!\eta'\psi^I\left(\eta'\right)\right)}
\left[4 \int_{\eta_s}^{\eta_o}\!\!\!d\eta' \frac{1}{(\eta_o-\eta')}\gamma_0^{ab}
\int_{\eta'}^{\eta_o}\!\!\!d\eta'' \partial_b \psi^I(\eta'') - 3 \int_{\eta_s}^{\eta_o}\!\!\!d\eta' \gamma_0^{ab}\partial_b\psi^I(\eta')
\right.
\nonumber 
\\ 
& &
\left. 
-6 \int_{\eta_s}^{\eta_o}\!\!\!d\eta' \gamma_0^{ab}
\int_{\eta'}^{\eta_o}\!\!\!d\eta'' \partial_b\partial_{\eta''}\psi^I(\eta'')
\right]
\nonumber 
\\ 
& &
+2 \partial_a\left(\int_{\eta_s}^{\eta_o}\!\!\!d\!\eta'\gamma_0^{bc}\partial_c\!\int_{\eta'}^{\eta_o}\!d\!\eta''\psi^I\left(\eta''\right)\right)
\partial_b\left(\int_{\eta_s}^{\eta_o}\!\!\!d\!\eta'\gamma_0^{ad}\partial_d\!\int_{\eta'}^{\eta_o}\!d\!\eta''\psi^I\left(\eta''\right)\right)
\nonumber 
\\
& &
-4\!\left(\int_{\eta_s}^{\eta_o}\!\!\!d \eta'\psi^I\left(\eta'\right)\right)\!\int_{\eta_s}^{\eta_o}\!\!\!d \eta'\left[-\frac{1}{(\eta_o-\eta')^3}
\!\int_{\eta'}^{\eta_o}\!\!\!d\!\eta''\Delta_2\psi^I\left(\eta''\right) \right.
\nonumber
\\
&&
\left.
+\frac{1}{(\eta_o \!-\!\eta')^2}\!\left(\!\frac{1}{2} \Delta_2\psi^I\!\left(\eta'\right)\!+\!\int_{\eta'}^{\eta_o}\!\!\!d\!\eta''\partial_{\eta''}\!\left(\Delta_2\psi^I\!\left(\eta''\right)\right)
\right)
\right]\!
+\!\frac{2}{\left(\sin{\theta_o}\right)^2}\!\left[\frac{1}{\deta_s}\!\int_{\eta_s}^{\eta_o}\!\!\!d \eta' \frac{\eta'\!-\!\eta_s}{\eta_o\!-\!\eta'}
\partial_{{\theta_o}}\psi^I\!\left(\eta'\right)\!\right]^2 
\nonumber \\ 
& &
+2 \partial_a \left\{\frac{1}{ \Hcal_s} \left[-\psi^I_s-2\int_{\eta_s}^{\eta_o}\!\!\!d\!\eta'\!\partial_{\eta'}\psi^I\left(\eta'\right)\right]
- \int_{\eta_s}^{\eta_o}\!\!\!d\eta'\psi^I(\eta')\right\}\gamma_{0s}^{ab}  \int_{\eta_s}^{\eta_o}\!\!\!d\eta' \partial_b \psi^I(\eta')
\nonumber \\ 
& &
+\frac{4}{\deta_s} \int_{\eta_s}^{\eta_o}\!\!\!d\eta'  \frac{\eta'-\eta_s}{\eta_o-\eta'} \partial_b \left[ \Delta_2 \left(\psi^I(\eta')\right)\right]
\int_{\eta_s}^{\eta_o} d \eta'' 
\gamma_0^{ab} \partial_a \int_{\eta''}^{\eta_o}\!d \eta'''  \psi^I(\eta''')
\nonumber
\\
&&-\frac{2}{\deta_s} \int_{\eta_s}^{\eta_o}\!\!\!d\!\eta' \frac{\eta'-\eta_s}{\eta_o - \eta'} \Delta_2 \left[
\psi^I\left(\eta'\right) \left(-\psi^I\left(\eta'\right)-2\int_{\eta'}^{\eta_o}\!\!\!d\!\eta''\!\partial_{\eta''}\psi^I\left(\eta''\right)\right)
\right.
\nonumber 
\\
& & \left.
+\gamma_0^{ab}\partial_a\left( \int_{\eta'}^{\eta_o}\!\!\!d \eta'' \psi^I\left(\eta''\right) \right) 
\partial_b \left( \int_{\eta'}^{\eta_o} d \eta'' \psi^I\left(\eta''\right) \right) \right] \nonumber
\nonumber 
\\
&&
-2 \int_{\eta_s}^{\eta_o} d \eta'\left\{ -2 \psi^I\left(\eta'\right)\frac{1}{\eta_o-\eta'}\int_{\eta'}^{\eta_o} d \eta''  \frac {\eta'' - \eta'}{\eta_o - \eta''}\Delta_2 \partial_{\eta''} \psi^I\left(\eta''\right) \right. 
 \nonumber \\
&&+2 \gamma_0^{ab}\partial_b \left(\int_{\eta'}^{\eta_o} d \eta'' \psi^I\left(\eta''\right)\right) \frac{1}{\eta_o-\eta'}\!\int_{\eta'}^{\eta_o}\!d \eta''  \frac {\eta''-\eta'}{\eta_o-\eta''}\partial_a \Delta_2 \psi^I\left(\eta''\right)
 \nonumber 
\eea
\bea
& & \left.
-\left(-2\psi^I\left(\eta'\right)-2\int_{\eta'}^{\eta_o}\!\!\!d\!\eta''\!\partial_{\eta''}\psi^I\left(\eta''\right)\right)
\frac{1}{(\eta_o-\eta')^2}\int_{\eta'}^{\eta_o}\!\!\!d\!\eta''\!\Delta_2 \psi^I\left(\eta''\right)\right.
\nonumber
\\ 
&& \left.
-2\partial_a \psi^I\left(\eta'\right)\int_{\eta'}^{\eta_o}\!\!\!d\!\eta'' \gamma_0^{ab}\partial_b \int_{\eta''}^{\eta_o}\!d \eta'''\partial_{\eta'''}\psi^I\left(\eta'''\right) \right.
\nonumber 
\\ 
& & \left.
+2 \partial_a\!\!\left[\gamma_0^{db}\partial_b\!\!\int_{\eta'}^{\eta_o}\!\!\!d\!\eta''\!\psi^I\left(\eta''\right)\right]\int_{\eta'}^{\eta_o}\!\!\!d\!\eta''\! \partial_d
\left[\gamma_0^{ac}\partial_c\!\!\int_{\eta''}^{\eta_o}\!\!\!d\!\eta'''\!\psi^I\left(\eta'''\right)\right] \right.
\nonumber
\\
&& 
\left.
+2\gamma_0^{ab}\partial_a\!\!\left(\psi^I\left(\eta'\right)+\int_{\eta'}^{\eta_o}\!\!\!d\!\eta''\!\partial_{\eta''}\psi^I\left(\eta''\right)\right)
\partial_b\!\!\int_{\eta'}^{\eta_o}\!\!\!d\eta''\!\psi^I\left(\eta''\right)
\right\}
\label{XiIS}
\eea
and
\bea
\Xi_{AS} = & & 2 \sigmaB \left\{ 2 \psi_s^A v_{||s} +\frac{\psi_s^A}{\HH_s}  \partial_r v_{||s} 
- \psi_s^A \left[ -\frac{2}{\deta_s} \Qint 
\right. \right.
\nonumber \\
&&
\left. \left.
- 2 \dQint 
+ \frac{2}{\deta_s} \Jint \right] 
- 2 \left( \psi_s^A\right)^2 
+ 2 \psi_s^I \psi_s^A 
 \right\}
 \nonumber \\
&&
+ 2 v_{||s}  \psi_s^A
+\left(\frac{1}{ \Hcal_s r_s}  \partial_\eta \psi^A_s-\partial_r \psi^A_s\right) \frac{2}{ \Hcal_s} v_{||s}
-\frac{4}{\Hcal_s r_s} \partial_r \psi^A_s \int_{\eta_s}^{\eta_o}\!d \eta'\psi^I\left(\eta'\right)
\nonumber \\
& &
+ 2 \psi_s^A \left[ - 2 \psi_s^I - \frac{2}{\deta_s} \Qint + \frac{2}{\deta_s} \Jint \right] 
\nonumber \\
& &
-\left(\frac{1}{ \Hcal_s r_s}\partial_\eta  \psi^A_s-\partial_r  \psi^A_s\right)\frac{2}{ \Hcal_s} \left[-(\psi^I_s-
\psi_s^A)-2\int_{\eta_s}^{\eta_o}\!\!\!d\!\eta'\!\partial_{\eta'}\psi^I\left(\eta'\right)\right]
\nonumber \\
& &
{-\left(\frac{1}{ \Hcal_s r_s}\partial_\eta  \psi^I_s-\partial_r  \psi^I_s\right)\frac{2}{ \Hcal_s} \psi^A_s
}
{ +\frac{2}{ \Hcal_s r_s^2} \psi^A_s \int_{\eta_s}^{\eta_o}\!d \eta' \Delta_2 \psi^I\left(\eta'\right)}
\nonumber \\
& &
+ 4 \partial_a \psi_s^A \thetaint 
+\frac{2}{\Hcal_s}\partial_a  \psi^A_s
\gamma_{0s}^{ab} \partial_b \int_{\eta_s}^{\eta_o} d \eta' \psi^I\left(\eta'\right)
\nonumber \\
&&
+\frac{8}{\deta_s} \INT d\eta' \left( \psii \psia \right) \left(\eta'\right) 
- \frac{4}{\deta_s} \INT d\eta' \frac{\eta' - \eta_s}{\eta_o - \eta' } \Delta_2 \left( \psi^I \psi^A \right)\left( \eta'\right) \,.
\eea

This second term $\Xi_{AS}$  vanishes if no anisotropic stresses are present. Hereafter, we use the subscripts $_{IS}$ to denote terms in which the first order square contributions come entirely from $\psi_I=(\psi+\phi)/2$ and $_{AS}$ for terms which come from the anisotropic stress, $\propto \psi_A =(\psi-\phi)/2$.

On the other hand, to compute $\frac{d\tau}{dz}$ to second order in perturbation theory in Poisson gauge we simply derive the coordinate transformation given in Eq.(\ref{tau2order}) with respect to the redshift after expanding it around the fiducial background $(\eta_s^{(0)}, \deta_s^{(0)}, \theta_o^a)$.
Proceeding step by step we first write
\be
\frac{d\tau}{dz} = \frac{d\etasB}{dz} \frac{d\tau}{d\etasB} = - \frac{1}{\HH_s} \frac{a\left( \etasB \right)}{a\left( \eta_o \right)} \frac{d\tau}{d\etasB}\,.
\ee
The second order expansion of $\tau$ around the fiducial model is given by
\be 
\tau  (\eta_s, \deta_s, \theta^a_s ) = \tau  
+  (\etasF+\etasS) \partial_\eta \tau
+ \rsF\partial_r \tau +  \thetaF\partial_a \tau+ \frac{( \etasF)^2}{2} \partial^2_\eta \tau \,,
\ee
where now all the terms are evaluated at $(\eta_s^{(0)}, \deta_s^{(0)}, \theta_o^a)$.
Then, using Eq.(\ref{tau2order}), and defining the perturbative expansion of $\tau  (\eta_s, \deta_s, \theta^a_s) = \tau_s^{(0)}+ \tau_s^{(1)}+ \tau_s^{(2)}$, we obtain the following results
\be
\tau_s^{(0)}= \int_{\eta_{in}}^\etasB d\eta' a (\eta')
\ee
\be
\tau_s^{(1)} = a (\eta_s^{(0)}) \eta_s^{(1)} + \int_{\eta_{in}}^{\eta_s^{(0)}} d\eta' a(\eta') (\psi^I-\psi^A)(\eta',  \rsB, \theta^a_o)
\ee
\bea
\!\!\!\!\!\!\!\!\!\!\!\!\tau_s^{(2)} &=& a (\eta_s^{(0)}) \eta_s^{(2)} +\frac{1}{2} \HH_s a (\eta_s^{(0)})  ( \etasF)^2
+ a (\eta_s^{(0)}) (\psi_s^I-\psi_s^A) \etasF + a (\eta_s^{(0)}) v_{||s} \deta_s^{(1)}  
\nonumber \\
&&
+ \int_{\eta_{in}}^\etasB \! d\eta' \frac{a(\eta')}{2} \left[ \phi^{(2)} - (\psi^I-\psi^A)^2 + (\ndv)^2 +v_{\perp a} v_{\perp}^a  \right] (\eta' , \rsB, \theta^a_o) 
\nonumber \\
& & 
+ \thetaF \partial_a \int_{\eta_{in}}^{\eta_s^{(0)}} d\eta' a(\eta') (\psi^I-\psi^A)(\eta',  \rsB, \theta^a_o)\,.
\eea

We now derive the expressions above with respect to $\eta_s^{(0)}$. After a long but straightforward
calculation (see Appendix~\ref{AppA}  for some useful relations), we obtain the following results
\be
\frac{1}{a(\eta_s^{(0)})} \frac{d \tau_s^{(0)} }{d\etasB} =1
\ee
\bea
\frac{1}{a(\eta_s^{(0)})} \frac{d \tau_s^{(1)} }{d\etasB} &=& \sigmaF \delta z^{(1)} + \frac{1}{\HH_s} \left[ \partial_\eta \psi_s^I+ \partial_ r v_{||s} \right] + \psi_s^I + \frac{1}{\HH_s} \partial_\eta \psi_s^A - \psi_s^A
\label{DerTau1}
\eea
\bea
\frac{1}{a(\eta_s^{(0)})} \frac{d \tau_s^{(2)} }{d\etasB} &=&   \sigmaF \delta z^{(2)} + \left[ - \frac{1}{2} \frac{\HH'_s}{\HH_s^2} + \frac{3}{2} \left( \frac{\HH'_s}{\HH_s^2} \right)^2 - \frac{1}{2} \frac{\HH''_s}{\HH_s^3} \right] \left( \delta z^{(1)} \right)^2 
\nonumber \\
&&
+ \left[\DERSis - \frac{1}{\HH_s} \partial_{r} \psi^I_s +\frac{2}{\HH_s} \left(1-\frac{3}{2}\frac{\HH'_s}{\HH_s^2}\right) \partial_\eta \psi^I_s
\right. 
\nonumber
\\
& & 
\left.
-\frac{\HH'_s}{\HH_s^2} v_{||s}
-\frac{3}{\HH_s}\frac{\HH'_s}{\HH_s^2}  \partial_r v_{||s} -\frac{1}{\HH_s^2} \partial^2_r v_{||s}+ \left( 1 - \frac{\HH'_s}{\HH_s^2} \right) \psi^I_s
\right]  \delta z^{(1)} 
\nonumber 
\\
& &
+\left[ \DERSas+\frac{1}{\HH_s} \partial_{r} \psi^A_s -\frac{3}{\HH_s}\frac{\HH'_s}{\HH_s^2} \partial_\eta \psi^A_s \right.
\nonumber \\
&& \left.
-\sigmaF \psi^A_s \right] \delta z^{(1)} 
+ \Pi_{IS} + \Pi_{AS} \,,
\label{DerTau2}
\eea
where
\bea
& &\Pi_{IS}=\frac{1}{2} \phi_s^{(2)} +\frac{1}{2 \HH_s} \partial_\eta \psi_s^{(2)}+ \frac{1}{\HH_s} \partial_r v_{||s}^{(2)}+
\left[2 \sigmaF \INT d\eta' \partial^2_{\eta'} \psii \left( \eta' \right) +2  \sigmaF\partial_\eta \psi^I_s \right.
\nonumber \\
&& \left.
-\frac{\HH'_s}{\HH_s^2  } \partial_r \psi^I_s - \frac{1}{\HH_s} \partial_{\eta} \partial{r} \psi^I_s
+\sigmaF \partial_r v_{||s} 
- \frac{1}{\HH_s} \partial^2_r v_{||s} \right] \left(-2 \Qint \right) 
\nonumber \\
&&
+ \left[\frac{1}{\HH_s} \partial_r v_{||s}+v_{||s}+\frac{1}{\HH_s} \partial_{\eta} \psi^I_s\right]\left(-2 \dQint\right)
-\frac{1}{2} (\psi_s^I)^2+\frac{2}{\HH_s}\psi_s^I \left(\partial_r v_{||s}+\partial_\eta \psi^I_s\right) 
\nonumber \\
&&
-\frac{1}{2} (v_{||s})^2-\psi_s^I v_{||s}+\frac{2}{\HH_s^2} \partial_\eta \psi^I_s \partial_r v_{||s}+
\frac{1}{\HH_s^2} \left(\partial_\eta \psi^I_s\right)^2+
\frac{1}{\HH_s^2} \left(\partial_r v_{||s}\right)^2
\nonumber
\\
& & 
-\frac{1}{2} \left( 1 - \frac{4}{\HH_s \deta_s } \right) v_{\perp a\,s} v_{\perp\,s}^a +\frac{2}{\HH_s} a v_{\perp\,s}^a \partial_a  v_{||s} - \frac{4}{\HH_s \deta_s }  a v_{\perp\,s}^a \partial_a \Qint
+\frac{1}{\HH_s}a v_{\perp\,s}^a \partial_a \psi^I_s 
\nonumber
\\
& & 
-\frac{4}{\HH_s} \gamma_{0 s}^{ab}\partial_a v_{||s}
\partial_b \Qint
-\frac{1}{\HH_s} v_{||s} \partial_r v_{||s}-\frac{1}{\HH_s} v_{||s} \partial_\eta \psi_s^I
\nonumber \\
&&
+2 \partial_a \left[ -  \psi_s^I
-\frac{1}{\HH_s} \left(\partial_\eta \psi_s^I+\partial_r v_{||s}\right)
\right] \int_{\eta_s}^{\eta_o}d \eta' \gamma_0^{ab}\partial_b \int_{\eta'}^{\eta_o}d \eta'' \psi^I \left(\eta''\right)
\eea
and 
\bea
& & \Pi_{AS}= \left[\frac{2}{\HH_s} \partial_{\eta}\partial{r} \psi_s^A -2 \frac{\HH'_s}{\HH_s^2  }\partial_{r} \psi_s^A
\right] \Qint-\frac{2}{\HH_s} \partial_{\eta} \psi_s^A \dQint 
\nonumber \\
&&
+\left[\frac{2}{\HH_s} \left(\psi_s^A+\psi_s^I\right)
-\frac{1}{\HH_s} v_{||s} 
+\frac{2}{\HH_s^2}\left(\partial_\eta \psi_s^I+ \partial_r v_{||s}\right)\right] \partial_{\eta} \psi_s^A 
+\frac{1}{\HH_s^2}\left( \partial_{\eta} \psi_s^A \right)^2+\frac{2}{\HH_s}\psi_s^A \partial_{\eta} \psi_s^I
\nonumber \\
&&
+\psi_s^A \psi_s^I-\frac{1}{2}\left(\psi_s^A\right)^2+\psi_s^A v_{||s}
+2 \partial_a\left[ \psi_s^A
-\frac{1}{\HH_s} \partial_\eta \psi_s^A\right]
\int_{\eta_s}^{\eta_o}\!d\eta'\! \gamma_0^{ab}\partial_b\! \int_{\eta'}^{\eta_o}d \eta'' \psi^I \left(\eta''\right) 
\nonumber
\\
& & 
+\frac{1}{\HH_s} a v_{\perp\,s}^a \partial_a \psi^A_s\, .
\eea

Finally, inserting the results given in Eqs.~(\ref{SGammaord1}), (\ref{SGammaord2}), (\ref{DerTau1}) and (\ref{DerTau2}) in Eq.~(\ref{e:vol}), we arrive at the following expressions for the first and second order volume perturbation
\bea
\frac{\delta V^{(1)}}{\bar V}&=& \left(3-\frac{2}{\HH_s r_s^{(0)}}-\frac{\HH'_s}{\HH_s^2}\right) \delta z^{(1)} -\psi_s^I+\frac{4}{r_s^{(0)}} \int_{\eta_s^{(0)}}^{\eta_o} d\eta' \psi^I\left(\eta'\right) 
\nonumber
\\
& & 
-\frac{2}{r_s^{(0)}}
\int_{\eta_s^{(0)}}^{\eta_o} d\eta' \frac{\eta'-\eta_s^{(0)}}{\eta_o-\eta'}\Delta_2 \psi^I\left(\eta'\right)
+\frac{1}{\HH_s} \left(\partial_\eta \psi^I_s+ \partial_r v_{||s} \right)-3\psi_s^A+\frac{1}{\HH_s}\partial_\eta \psi^A_s \, , \qquad
\label{VolPer1}
\eea

\bea
\frac{\delta V^{(2)}}{\bar V} &=&  \left(3-\frac{2}{\HH_s \deta_s}-\frac{\HH'_s}{\HH_s^2}\right) \delta z^{(2)}+
\left[ - \frac{1}{2} \frac{\HH'_s}{\HH_s^2} + \frac{3}{2} \left( \frac{\HH'_s}{\HH_s^2} \right)^2 - \frac{1}{2} \frac{\HH''_s}{\HH_s^3} - \frac{1}{\HH_s \deta_s} \sigmaF \right.
\nonumber
\\
& & 
\left.
+\left(1-\frac{1}{\HH_s \deta_s} \right)^2 
+2 \left(1-\frac{1}{\HH_s \deta_s} \right)\sigmaF
\right] \left( \delta z^{(1)} \right)^2 +\Lambda_{IS}+\Lambda_{AS}\,,
\label{VolPer2}
\eea
where
\bea
& & \!\!\!\!\!\Lambda_{IS} = -\psi_s^{(2)}+\frac{1}{2} \phi_s^{(2)}+\frac{1}{2 \HH_s} \partial_\eta  \psi_s^{(2)}
+\frac{1}{\HH_s} \partial_r v_{||s}^{(2)}
-\frac{1}{2} \frac{1}{\deta_s}\int_{\eta_s}^{\eta_o}\!d \eta' \frac {\eta' - \eta_s}{\eta_o - \eta'} \Delta_2\left[
\psi^{(2)}+\phi^{(2)}\right]\!\left(\eta'\right)
\nonumber \\ 
& &
+\frac{1}{\deta_s}\int_{\eta_s}^{\eta_o}\!d \eta'\left[
\psi^{(2)}+\phi^{(2)}\right]\!\left(\eta'\right)
+2 \left(1-\frac{1}{ \Hcal_s \deta_s}\right) \left\{
- \frac{2}{\HH_s} v_{||s}  \partial_r v_{||s}  - (v_{||s} )^2 -v_{\perp a\,s} v_{\perp\,s}^a \right.
\nonumber
\\
& & 
+\left[ -\frac{1}{\HH_s} \partial_\eta \psi_s^I
 -2 \dQint  - \frac{2}{\deta_s} \Qint 
+ \frac{2}{\deta_s} \Jint \right] 
v_{||s} 
\nonumber
\\
& & 
+ \left[-2 \psi_s^I - 4 \dQint - 2 \HH_s \Qint \right] \frac{1}{\HH_s} \partial_r v_{||s}
\nonumber
\\
& & 
+a v_{\perp\,s}^a \partial_a \Qint
+\left[\partial_r\psi^I_s
+2 \partial_\eta\psi^I_s
+2 \int_{\eta_s}^{\eta_o}\!d \eta' \partial^2_{\eta'}\psi^I\left(\eta'\right)\right]
\left(-2 \int_{\eta_s}^{\eta_o}\!d \eta' \psi^I\left(\eta'\right)\right)
\nonumber \\
&&
-\left(-\psi^I_s-2 \int_{\eta_s}^{\eta_o}\!d \eta' \partial_{\eta'}\psi^I\left(\eta'\right)\right)
\left[
\frac{2}{\deta_s}\int_{\eta_s}^{\eta_o}\!d \eta' \frac {\eta' - \eta_s}{\eta_o - \eta'} \Delta_2
\psi^I\left(\eta'\right) \right.
\nonumber \\
&&
\left. \left.
-\frac{2}{\deta_s}\int_{\eta_s}^{\eta_o}\!d \eta'\psi^I\left(\eta'\right)
-\frac{1}{\HH_s} \partial_\eta \psi_s^I-\psi_s^I
\right]\right\}
+\frac{3}{2} v_{\perp a\,s} v_{\perp\,s}^a+\frac{2}{\HH_s} a  v_{\perp\,s}^a \partial_a v_{||s}+\left(-\frac{1}{2}
+\frac{\HH'_s}{\HH_s^2  }\right)  ( v_{||s})^2
\nonumber 
\\
&&
+\left(-1+3 \frac{\HH'_s}{\HH_s^2  }\right)\frac{1}{\HH_s}v_{||s}\partial_r v_{||s}
+\frac{1}{\HH_s^2}\left[v_{||s}\partial^2_r v_{||s}+\left(\partial_r v_{||s}\right)^2\right]
+\left[ 
-\DERSis \right.
\nonumber
\\
& & 
 -\frac{1}{\HH_s}\partial_r \psi_s^I
-\frac{3}{\HH_s} \left(1- \frac{\HH'_s}{\HH_s^2  }-\frac{2}{3}
\frac{1}{\HH_s \deta_s}\right) \partial_\eta \psi_s^I 
-\frac{4}{\deta_s}\left(2- \frac{\HH'_s}{\HH_s^2  }\right) \Qint 
\nonumber
\\
& & 
+\sigmaF \frac{2}{\deta_s} \Jint
-2 \sigmaF \dQint
\nonumber
\\
& & 
\left.
+ \frac{4}{\HH_s \deta_s} \psi_s^I
- \frac{2}{\HH_s \deta_s^2} \INT d\eta' \Delta_2 \psii\left(\eta'\right)\right]v_{||s}
+\left[ -2\left(1- \frac{\HH'_s}{\HH_s^2  }-
\frac{2}{\HH_s \deta_s}\right) \partial_r v_{||s}
\right.
\nonumber
\\
& & 
\left.+\frac{2}{\HH_s}\partial^2_r v_{||s}\right] \Qint
+\left[ - \frac{2}{\HH_s}\left(1- 3 \frac{\HH'_s}{\HH_s^2  }\right) \partial_r v_{||s}+\frac{2}{\HH_s^2}\partial^2_r v_{||s}\right] \dQint
\nonumber
\\
& & 
-\frac{2}{\HH_s}\partial_r v_{||s}\frac{1}{\deta_s}\Jint
+ \frac{2}{\HH_s^2} \partial_\eta \psi_s^I \partial_r  v_{||s}+\frac{1}{\HH_s}\left(\frac{1}{\HH_s}\partial^2_r  v_{||s}
+3 \frac{\HH'_s}{\HH_s^2  } \partial_r  v_{||s}\right) \psi_s^I
\nonumber
\\
& & 
 -\frac{2}{\HH_s \deta_s }  a v_{\perp\,s}^a \partial_a \Qint
+\frac{1}{\HH_s}a v_{\perp\,s}^a \partial_a \psi^I_s -\frac{6}{\HH_s} \gamma_{0s}^{ab}\partial_a v_{||s}
\partial_b \Qint
\nonumber
\\
& & 
+\frac{4}{\deta_s^2}\left(\int_{\eta_s}^{\eta_o}\!d \eta'\psi^I\left(\eta'\right)\right)^2
+\left\{\left[2 \sigmaF \psi_s^I +4 \sigmaF \dQint 
\right.\right.
\nonumber
\\
& & 
\left.\left.
- \frac{2}{\HH_s} \partial_\eta \psi_s^I\right]\frac{1}{\deta_s}
+2 \sigmaF \INT d\eta' \partial^2_{\eta'} \psii \left( \eta' \right)+2 \sigmaF \partial_\eta \psi_s^I +
\left( 2 - \frac{\HH'_s}{\HH_s^2  } \right)\partial_r \psi_s^I 
\right.
\nonumber \\
& & 
\left.
\!- \frac{1}{\HH_s} \partial_{\eta}\partial_{r} \psi_s^I \right\}\!\left(\!-\!2\!\Qint\right)
\!+\!\left[ \frac{3}{\HH_s}\left(1\!-\!\frac{\HH'_s}{\HH_s^2} \!-\! \frac{2}{3}\frac{1}{\HH_s \deta_s}\right)\partial_\eta \psi_s^I\!+\!\frac{1}{\HH_s} \partial_{r} \psi_s^I\!-\!\left( 5 \!-\! \frac{\HH'_s}{\HH_s^2  } \right)\psi_s^I
\right.
\nonumber
\\
& & 
\left.
+\DERSis
\right]\left(-2 \dQint\right)
+\left[\left(4-2\frac{\HH'_s}{\HH_s^2} \right)\psi_s^I
\right.
\nonumber
\\
& & 
+4 \left(1-\frac{\HH'_s}{\HH_s^2} \right)\dQint
-\frac{8}{\deta_s}\Qint
\nonumber
\\
& & 
\left.
-\frac{2}{\HH_s}\partial_\eta \psi_s^I\right]
\frac{1}{\deta_s}\Jint
+2\left(\frac{1}{\deta_s}\Jint\right)^2
\nonumber
\eea
\bea
& & 
+\left[\frac{1}{ \Hcal_s \Delta \eta}\left(-\psi^I_s-2 \int_{\eta_s}^{\eta_o}\!d \eta' \partial_{\eta'}\psi^I\left(\eta'\right)\right)
-\frac{1}{\Delta \eta}\int_{\eta_s}^{\eta_o}\!d \eta'\psi^I\left(\eta'\right)\right]\frac{2}{r_s}\int_{\eta_s}^{\eta_o}\!d \eta' \Delta_2\psi^I\left(\eta'\right)
\nonumber \\ 
& &
-2 \left[ \int_{\eta_s}^{\eta_o}\!\!\!d\eta' \frac{1}{(\eta_o-\eta')^2} \psi^I(\eta')+2
 \int_{\eta_s}^{\eta_o}\!\!\!d\eta'  \frac{1}{(\eta_o-\eta')^2} \int_{\eta'}^{\eta_o}\!\!\!d\eta''\partial_{\eta''}\psi^I(\eta'')\right]
  \int_{\eta_s}^{\eta_o}\!\!\!d\eta' \Delta_2 \psi^I(\eta')
\nonumber
\\
& & 
+\left( \frac{5}{2} - \frac{\HH'_s}{\HH_s^2  } \right)\left(\psi_s^I\right)^2
+\frac{1}{\HH_s^2}\left(\partial_\eta \psi_s^I\right)^2
+\left[-\DERSis 
\right.
\nonumber
\\
& & 
\left.
+\frac{1}{\HH_s} \left(-2+3 \frac{\HH'_s}{\HH_s^2} + \frac{2}{\HH_s \deta_s}\right)\partial_\eta \psi_s^I
-\frac{1}{\HH_s}\partial_r \psi_s^I
\right]\psi_s^I 
\nonumber 
\\
& & 
-2 \partial_a \psi_s^I \int_{\eta_s}^{\eta_o}d \eta' \gamma_0^{ab}\partial_b \int_{\eta'}^{\eta_o}d \eta'' \psi^I \left(\eta''\right)
\nonumber
\\
& & 
-\frac{2}{\HH_s}\partial_a \left[\partial_\eta \psi_s^I+\partial_r v_{||s}\right] \int_{\eta_s}^{\eta_o}d \eta' \gamma_0^{ab}\partial_b \int_{\eta'}^{\eta_o}d \eta'' \psi^I \left(\eta''\right)
 -2  a v_{\perp\,s}^a \partial_a \Qint
\nonumber
\\
& & 
+\frac{4}{\deta_s}\int_{\eta_s}^{\eta_o}\!d \eta'\!\left[\psi^I\left(\eta'\right)\left(-\psi^I\left(\eta'\right) 
-2 \int_{\eta'}^{\eta_o}\!d\!\eta'' \partial_{\eta''}\psi^I\left(\eta''\right)\right)
\right.
\nonumber \\
&&
\left.
+\gamma_0^{ab}\partial_a\!\left(\int_{\eta'}^{\eta_o}\!\!\!d\!\eta''\!\psi^I\left(\eta''\right)\right)\!\partial_b\!\left(\int_{\eta'}^{\eta_o}\!\!\!d\!\eta''\!\psi^I\left(\eta''\right)\right)\right] 
+\!4\partial_a\!\psi^I_s
\!\int_{\eta_s}^{\eta_o}\!\!\!d\!\eta'\gamma_0^{ab}\partial_b\!\int_{\eta'}^{\eta_o}\!d\!\eta''\psi^I\left(\eta''\right)
\nonumber 
\\
&&
{
-\frac{8}{r_s}\left[\int_{\eta_s}^{\eta_o}\!d \eta' \partial_a \psi^I\left(\eta'\right)
\!\int_{\eta_s}^{\eta_o}\!\!\!d\!\eta''\gamma_0^{ab}\partial_b\!\int_{\eta''}^{\eta_o}\!d\!\eta'''\psi^I\left(\eta'''\right)\right]
}
\nonumber
\\
& &
{
+2 \partial_a \left(\!\int_{\eta_s}^{\eta_o}\!d\!\eta'\psi^I\left(\eta'\right)\right)}
\left[4 \int_{\eta_s}^{\eta_o}\!\!\!d\eta' \frac{1}{(\eta_o-\eta')}\gamma_0^{ab}
\int_{\eta'}^{\eta_o}\!\!\!d\eta'' \partial_b \psi^I(\eta'') - 3 \int_{\eta_s}^{\eta_o}\!\!\!d\eta' \gamma_0^{ab}\partial_b\psi^I(\eta')
\right.
\nonumber
\\
& &
\left.
{
-6 \int_{\eta_s}^{\eta_o}\!\!\!d\eta' \gamma_0^{ab}
\int_{\eta'}^{\eta_o}\!\!\!d\eta'' \partial_b\partial_{\eta''}\psi^I(\eta'')
}
\right]
\nonumber 
\\ 
& &
+2 \partial_a\left(\int_{\eta_s}^{\eta_o}\!\!\!d\!\eta'\gamma_0^{bc}\partial_c\!\int_{\eta'}^{\eta_o}\!d\!\eta''\psi^I\left(\eta''\right)\right)
\partial_b\left(\int_{\eta_s}^{\eta_o}\!\!\!d\!\eta'\gamma_0^{ad}\partial_d\!\int_{\eta'}^{\eta_o}\!d\!\eta''\psi^I\left(\eta''\right)\right)
\nonumber \\
& &
-4\!\left(\int_{\eta_s}^{\eta_o}\!\!\!d \eta'\psi^I\left(\eta'\right)\right)\!\int_{\eta_s}^{\eta_o}\!\!\!d \eta'\left[-\frac{1}{(\eta_o-\eta')^3}
\!\int_{\eta'}^{\eta_o}\!\!\!d\!\eta''\Delta_2\psi^I\left(\eta''\right)
+\frac{1}{(\eta_o-\eta')^2}\!\left(\frac{1}{2} \Delta_2\psi^I\left(\eta'\right)
\right. \right.
\nonumber \\
&&
\left. \left.
+\int_{\eta'}^{\eta_o}\!\!\!d\!\eta''\partial_{\eta''}\left(\Delta_2\psi^I\left(\eta''\right)\right)
\right)
\right]
+\frac{2}{\left(\sin{\theta_o}\right)^2}\left[\frac{1}{\deta_s}\int_{\eta_s}^{\eta_o}\!\!\!d \eta' \frac{\eta'-\eta_s}{\eta_o-\eta'}
\partial_{{\theta_o}}\psi^I\left(\eta'\right)\right]^2
\nonumber \\ 
& &
+2 \partial_a \left\{\frac{1}{ \Hcal_s} \left[-\psi^I_s-2\int_{\eta_s}^{\eta_o}\!\!\!d\!\eta'\!\partial_{\eta'}\psi^I\left(\eta'\right)\right]
- \int_{\eta_s}^{\eta_o}\!\!\!d\eta'\psi^I(\eta')\right\}\gamma_{0s}^{ab}  \int_{\eta_s}^{\eta_o}\!\!\!d\eta' \partial_b \psi^I(\eta')
\nonumber \\ 
& &
+\frac{4}{\deta_s} \int_{\eta_s}^{\eta_o}\!\!\!d\eta'  \frac{\eta'-\eta_s}{\eta_o-\eta'} \partial_b \left[ \Delta_2\left(\psi^I(\eta')\right)\right]
\int_{\eta_s}^{\eta_o} d \eta'' 
\gamma_0^{ab} \partial_a \int_{\eta''}^{\eta_o}\!d \eta'''  \psi^I(\eta''')
\nonumber 
\\
&&
-\frac{2}{\deta_s} \int_{\eta_s}^{\eta_o}\!\!\!d\!\eta' \frac{\eta'-\eta_s}{\eta_o - \eta'} \Delta_2 \left[
\psi^I\left(\eta'\right) \left(-\psi^I\left(\eta'\right)-2\int_{\eta'}^{\eta_o}\!\!\!d\!\eta''\!\partial_{\eta''}\psi^I\left(\eta''\right)\right)
\right.
\nonumber \\ 
& & \left.
+\gamma_0^{ab}\partial_a\left( \int_{\eta'}^{\eta_o}\!\!\!d \eta'' \psi^I\left(\eta''\right) \right) 
\partial_b \left( \int_{\eta'}^{\eta_o} d \eta'' \psi^I\left(\eta''\right) \right) \right]
\nonumber \\
&&
-2 \int_{\eta_s}^{\eta_o} d \eta'\left\{ -2 \psi^I\left(\eta'\right)\frac{1}{\eta_o-\eta'}\int_{\eta'}^{\eta_o} d \eta''  \frac {\eta'' - \eta'}{\eta_o - \eta''}\Delta_2 \partial_{\eta''} \psi^I\left(\eta''\right)
\right. \nonumber 
\\
& & \left. 
+2 \gamma_0^{ab}\partial_b \left(\int_{\eta'}^{\eta_o} d \eta'' \psi^I\left(\eta''\right)\right) \frac{1}{\eta_o-\eta'}\!\int_{\eta'}^{\eta_o}\!d \eta''  \frac {\eta''-\eta'}{\eta_o-\eta''}\partial_a \Delta_2 \psi^I\left(\eta''\right)
\right. \nonumber 
\eea
\bea
& & \left.
-\left(-2\psi^I\left(\eta'\right)-2\int_{\eta'}^{\eta_o}\!\!\!d\!\eta''\!\partial_{\eta''}\psi^I\left(\eta''\right)\right)
\frac{1}{(\eta_o-\eta')^2}\int_{\eta'}^{\eta_o}\!\!\!d\!\eta''\!\Delta_2 \psi^I\left(\eta''\right)
\right.
\nonumber 
\\
& & 
-2\partial_a \psi^I\left(\eta'\right)\int_{\eta'}^{\eta_o}\!\!\!d\!\eta'' \gamma_0^{ab}\partial_b \int_{\eta''}^{\eta_o}\!d \eta'''\partial_{\eta'''}\psi^I\left(\eta'''\right)
\nonumber \\
&&
\left.
+2 \partial_a\!\!\left[\gamma_0^{db}\partial_b\!\!\int_{\eta'}^{\eta_o}\!\!\!d\!\eta''\!\psi^I\left(\eta''\right)\right]\int_{\eta'}^{\eta_o}\!\!\!d\!\eta''\! \partial_d
\left[\gamma_0^{ac}\partial_c\!\!\int_{\eta''}^{\eta_o}\!\!\!d\!\eta'''\!\psi^I\left(\eta'''\right)\right]
\right.
\nonumber 
\\
& & 
\left.
+2\gamma_0^{ab}\partial_a\!\!\left(\psi^I\left(\eta'\right)
+\int_{\eta'}^{\eta_o}\!\!\!d\!\eta''\!\partial_{\eta''}\psi^I\left(\eta''\right)\right)
\partial_b\!\!\int_{\eta'}^{\eta_o}\!\!\!d\eta''\!\psi^I\left(\eta''\right)
\right\}
\eea
and 
\bea
& &\Lambda_{AS} = 2 \sigmaB \left\{ 3 \psi_s^A v_{||s} - \frac{v_{||s}}{\HH_s}  \partial_\eta \psi_s^A +2 \frac{\psi_s^A}{\HH_s}  \partial_r v_{||s}
\right.
\nonumber \\
&&
 - \psi_s^A \left[ -\frac{2}{\deta_s} \Qint 
- 4 \dQint + \frac{2}{\deta_s} \Jint \right]
\nonumber \\
&&
  + \frac{1}{\HH_s} \partial_\eta \psi_s^A \left[ - \left( \psi_s^I - \psi_s^A \right) - 2 \dQint \right]  
 \left. + \frac{1}{\HH_s} \psi_s^A \partial_\eta \psi_s^I 
- 3 \left( \psi_s^A\right)^2 + 4 \psi_s^I \psi_s^A 
 \right\} 
 \nonumber \\
 &&
+\left[-\DERSas\!
+\!\left(\! -1 \!+\!3 \frac{\HH'_s}{\HH_s^2}\! 
+\frac{2}{\HH_s \deta_s}
\right)\frac{1}{\HH_s}\partial_\eta \psi_s^A
\!-\!\frac{3}{\HH_s}\partial_r \psi_s^A\!
\right.
\nonumber 
\\
& & \left.
+\!\left( 6 \!-\!4 \frac{\HH'_s}{\HH_s^2  } \right)\psi_s^A\right]v_{||s}
+\left[-\frac{1}{\HH_s^2}\partial^2_r v_{||s}-\left(2 +3 \frac{\HH'_s}{\HH_s^2  } \right)
\frac{1}{\HH_s}\partial_r v_{||s}\right]\psi_s^A
+\frac{2}{\HH_s^2}\partial_\eta \psi_s^A \partial_r v_{||s}
\nonumber 
\\
& &
+\frac{1}{\HH_s}a v_{\perp s}^a\partial_a \psi_s^A
+ \left[ -\frac{2}{\HH_s}\partial_\eta \psi_s^A+\left(4 +2 \frac{\HH'_s}{\HH_s^2  } \right)\psi_s^A
\right]  \frac{1}{\deta_s} \Jint 
\nonumber 
\\
&&
{+\frac{2}{ \Hcal_s \deta_s^2} \psi^A_s  \int_{\eta_s}^{\eta_o}\!d \eta' \Delta_2 \psi^I\left(\eta'\right)}
+\left\{\left[-\frac{2}{\HH_s}\partial_\eta \psi_s^A+2\left(1 +\frac{\HH'_s}{\HH_s^2  } \right)\psi_s^A
\right] \frac{1}{\deta_s}
\right.
\nonumber 
\\
&&
\left.
-\frac{1}{\HH_s}\partial_{\eta}\partial_{r} \psi_s^A+\left(\frac{\HH'_s}{\HH_s^2}+\frac{2}{\HH_s \deta_s}\right) \partial_{r} \psi_s^A
\right\}\left(-2 \Qint\right) 
\nonumber 
\\
&&
+\left[\DERSas
+\left(1 -3 \frac{\HH'_s}{\HH_s^2}-\frac{2}{\HH_s \deta_s} \right)\frac{1}{\HH_s}\partial_{\eta} \psi_s^A
+\frac{3}{\HH_s}\partial_{r} \psi_s^A
\right.
\nonumber 
\\
& &
\left.
-3 \left(1 -\frac{\HH'_s}{\HH_s^2  } \right)\psi_s^A\right]\left(-2 \dQint\right)-\left(\frac{3}{2} -3 
\frac{\HH'_s}{\HH_s^2  } \right)\left(\psi_s^A\right)^2
+\frac{1}{\HH_s^2  }\left(\partial_\eta \psi_s^A\right)^2 
\nonumber \\
&&
+\!
\frac{2}{\HH_s^2  }\partial_\eta \psi_s^A\partial_\eta \psi_s^I
\!+\!\left[\DERSas+\DERSis \right.
\nonumber \\
&&
\left.
-\left(3\frac{\HH'_s}{\HH_s^2}+\frac{2}{\HH_s \deta_s}\right)
\frac{1}{\HH_s  }\partial_\eta \psi_s^A
+\left(2-3\frac{\HH'_s}{\HH_s^2}-\frac{2}{\HH_s \deta_s}\right)
\frac{1}{\HH_s}\partial_\eta \psi_s^I
-\left(1+\frac{\HH'_s}{\HH_s^2  }\right)\psi_s^I
+\frac{3}{\HH_s  }\partial_r \psi_s^A
\right.
\nonumber 
\\
&&
\left.
+\frac{1}{\HH_s  }\partial_r \psi_s^I
\right] \psi_s^A 
+\left[-\DERSas
+\left(3\frac{\HH'_s}{\HH_s^2}+\frac{2}{\HH_s \deta_s}\right)
\frac{1}{\HH_s}\partial_\eta \psi_s^A
\right.
\nonumber 
\\
& &
\left.-\frac{3}{\HH_s  }\partial_r \psi_s^A\right]\psi_s^I
+ 6\, \partial_a \psi_s^A \! \thetaint  \!
\nonumber 
\\
& &
{
+\frac{2}{\Hcal_s}\partial_a  \psi^A_s
\gamma_{0s}^{ab} \partial_b \int_{\eta_s}^{\eta_o} d \eta' \psi^I\left(\eta'\right)
}
-\! \frac{2}{\HH_s}\partial_a \left(\partial_\eta \psi_s^A\right) \int_{\eta_s}^{\eta_o}\!\!d \eta' \gamma_0^{ab}\partial_b \int_{\eta'}^{\eta_o}\!\!d \eta'' \psi^I \left(\eta''\right)
\nonumber \\
&& 
+\frac{8}{\deta_s} \INT d\eta' \left( \psii \psia \right) \left(\eta'\right) - \frac{4}{\deta_s} \INT d\eta' \frac{\eta' - \eta_s}{\eta_o - \eta' } \Delta_2 \left( \psi^I \psi^A \right)\left( \eta'\right) \,.
\label{LamAS}
\eea

\subsection{Number Counts}

We have now all we need to evaluate the galaxy number counts at observed redshift and in the direction of observation.
Using the relations given in Eqs.~(\ref{numcount:first}) and~(\ref{numcount:second}) and inserting the results obtained for the density, Eqs.~(\ref{rhoRedAng1}) and~(\ref{rhoRedAng2}), and for the volume fluctuation, Eqs.~(\ref{VolPer1})-(\ref{LamAS}), at fixed observed redshift and observer direction, we obtain to first order
\bea
\Delta^{(1)} &=& \left( \frac{2}{\HH_s\deta_s^{(0)}} +\frac{\HH'_s}{\HH_s^2} \right) \left( v_{||s} + \psi^I_s-\psi^A_s + 2 \int_{\eta_s^{(0)}}^{\eta_o} d\eta' \partial_{\eta'} \psi^I \left( \eta' \right)  \right) 
 - \psi^I_s + \frac{4}{\deta_s^{(0)}}  \int_{\eta_s^{(0)}}^{\eta_o} d\eta'  \psi^I \left( \eta' \right) 
\nonumber \\
&& 
- \frac{2}{\deta_s^{(0)}}\int_{\eta_s^{(0)}}^{\eta_o} d\eta' \frac{\eta' - \etasB}{\eta_0 -\eta'} \Delta_2 
\psi^I \left( \eta' \right)
+ \frac{1}{\HH_s} \left( \partial_\eta \psi^I_s + \partial_r  v_{||s} \right)-3 \psi_s^A + \frac{1}{\HH_s} \partial_\eta \psi^A_s
+\delta_\rho^{(1)} \nonumber
\,, \\
\eea
which agrees with~\cite{Bonvin:2011bg}, where it was derived directly in Poisson gauge.

To second order we obtain of course a much more involved expression:
\be\label{e:Delta2}
\Delta^{(2)} = \Sigma - \langle \Sigma \rangle
\ee
where 
\be
\Sigma=\Sigma_{IS}+\Sigma_{AS}
\ee
and 
\bea  \label{e:SigmaIS}
& &\Sigma_{IS} =  
\left( -\frac{2}{\HH_s\deta_s} -\frac{\HH'_s}{\HH_s^2} \right)
\left\{ - v^{(2)}_{||s}-\frac{1}{2} \phi_s^{(2)}
-\frac{1}{2} \int_{\eta_s}^{\eta_0} d\eta' \partial_{\eta'}  \left[ \phi^{(2)}\left( \eta'\right) + \psi^{(2)}\left( \eta'\right)\right]
+\frac{1}{2}\left(v_{||s}\right)^2
\right.
\nonumber \\ & & 
\left.
+\frac{1}{2}\left(\psi^I_s\right)^2
+\left(- v_{||s}-\psi^I_s\right)\left(-\psi^I_s-2 \int_{\eta_s}^{\eta_o} d \eta' \partial_{\eta'}\psi^I\left(\eta'\right)\right)
+\frac{1}{2} v^a_{\perp s} v_{\perp a\,s} 
\right.
\nonumber \\ & &
\left.
-2 a \, v^a_{\perp s} \partial_a \int_{\eta_s}^{\eta_o} d \eta' \psi^I\left(\eta'\right)
+4\int_{\eta_s}^{\eta_0} d\eta' \left[\psi^I\left(\eta'\right) \partial_{\eta'}\psi^I\left(\eta'\right) 
+ \partial_{\eta'}\psi^I\left(\eta'\right) \int_{\eta'}^{\eta_o} d \eta'' \partial_{\eta''}\psi^I\left(\eta''\right)
\right.
\right.
\nonumber \\ & & 
\left.
\left.
+\psi^I\left(\eta'\right) \int_{\eta'}^{\eta_o} d \eta'' \partial^2_{\eta''}\psi^I\left(\eta''\right)
- \gamma_0^{ab} \partial_a \left( \int_{\eta'}^{\eta_o} d \eta'' \psi^I\left(\eta''\right) \right) 
\partial_b \left( \int_{\eta'}^{\eta_o} d \eta'' \partial_{\eta''}\psi^I\left(\eta''\right) \right) \right]
\right.
\nonumber \\  & & 
+ 2 \partial_a  \left(v_{||s}+\psi^I_s\right)  \int_{\eta_s}^{\eta_o}\!d \eta' \gamma_0^{ab} \partial_b \int_{\eta'}^{\eta_o}\!d \eta'' \psi^I\left(\eta''\right)
\nonumber 
\\
& & 
\left.+4 \int_{\eta_s}^{\eta_o}\!\!d\!\eta'\partial_a \left(\partial_{\eta'}\psi^I\left(\eta'\right)\right)
\!\int_{\eta_s}^{\eta_o}\!\!\!d\!\eta''\gamma_0^{ab} \partial_b\!\int_{\eta''}^{\eta_o}\!d\!\eta'''\psi^I\left(\eta'''\right)
\right\}
\nonumber \\  & &
+ \left[\frac{1}{2} \frac{\HH'_s}{\HH_s^2}+\frac{3}{2} \left( \frac{\HH'_s}{\HH_s^2}\right)^2 - \frac{1}{2}\frac{\HH''_s}{\HH_s^3} +\frac{1}{\HH_s\deta_s} \left(1+ 3 \frac{\HH'_s}{\HH_s^2} +\frac{1}{\HH_s \deta_s}\right) \right] 
\left[ \left(v_{||s}\right)^2 + (\psi^I_s)^2 + 2 \psi^I_s v_{||s} 
\right.
\nonumber 
\\  
& & 
\left.
+ 4 \left( v_{||s} + \psi^I_s \right) \INT 
d\eta' \partial_{\eta'} \psi^I \left( \eta' \right) + 4 \left(\INT d\eta' \partial_{\eta'} \psi^I \left( \eta' \right) \right)^2\right]-\psi_s^{(2)}+\frac{1}{2} \phi_s^{(2)}+\frac{1}{2 \HH_s} \partial_\eta  \psi_s^{(2)} \nonumber 
\\
& &
+\frac{1}{\HH_s} \partial_r v_{||s}^{(2)}
-\frac{1}{2} \frac{1}{\deta_s}\int_{\eta_s}^{\eta_o}\!d \eta' \frac {\eta' - \eta_s}{\eta_o - \eta'} \Delta_2\left[
\psi^{(2)}+\phi^{(2)}\right]\left(\eta'\right)
+\frac{1}{\deta_s}\int_{\eta_s}^{\eta_o}\!d \eta'\left[
\psi^{(2)}+\phi^{(2)}\right]\left(\eta'\right)
\nonumber  \\ 
& &
+2 \left(1-\frac{1}{ \Hcal_s \deta_s}\right) \left\{
- \frac{2}{\HH_s} v_{||s}  \partial_r v_{||s}  - (v_{||s} )^2 -v_{\perp a\,s} v_{\perp\,s}^a
+\left[ -\frac{1}{\HH_s} \partial_\eta \psi_s^I  -2 \dQint 
\right.\right.
\nonumber
\\
& & 
 \left.
 - \frac{2}{\deta_s} \Qint 
+ \frac{2}{\deta_s} \Jint \right] 
v_{||s} 
+ \left[-2 \psi_s^I - 4 \dQint 
\right.
\nonumber \\
&&
\left.\left.
- 2 \HH_s \Qint \right] \frac{1}{\HH_s} \partial_r v_{||s}
+a v_{\perp\,s}^a \partial_a \Qint
+\left[\partial_r\psi^I_s
+2 \partial_\eta\psi^I_s
\right.
\right.
\nonumber
\eea
\bea
& & 
\left.
+2 \int_{\eta_s}^{\eta_o}\!d \eta' \partial^2_{\eta'}\psi^I\left(\eta'\right)\right]
\left(-2 \int_{\eta_s}^{\eta_o}\!d \eta' \psi^I\left(\eta'\right)\right)
\nonumber \\
&&
\left. -\left(-\psi^I_s-2 \int_{\eta_s}^{\eta_o}\!d \eta' \partial_{\eta'}\psi^I\left(\eta'\right)\right)
\left[
\frac{2}{\deta_s}\int_{\eta_s}^{\eta_o}\!d \eta' \frac {\eta' - \eta_s}{\eta_o - \eta'} \Delta_2
\psi^I\left(\eta'\right)
-\frac{2}{\deta_s}\int_{\eta_s}^{\eta_o}\!d \eta'\psi^I\left(\eta'\right)
\right.
\right.
\nonumber
\\
& & 
\left.
\left.
-\frac{1}{\HH_s} \partial_\eta \psi_s^I-\psi_s^I
\right]\right\}
+\frac{3}{2} v_{\perp a\,s} v_{\perp\,s}^a+\frac{2}{\HH_s} a  v_{\perp\,s}^a \partial_a v_{||s}+\left(\frac{5}{2}
+\frac{\HH'_s}{\HH_s^2  }\right) ( v_{||s})^2
\nonumber
\\
& & 
+\left(5+3 \frac{\HH'_s}{\HH_s^2  }\right)\frac{1}{\HH_s}v_{||s}\partial_r v_{||s}
+\frac{1}{\HH_s^2}\left[v_{||s}\partial^2_r v_{||s}+\left(\partial_r v_{||s}\right)^2\right]
+\left[ 
-\DERSis
\right.
\nonumber
\\
& & 
\left.
-\frac{1}{\HH_s}\partial_r \psi_s^I
-\frac{3}{\HH_s} \left(-1- \frac{\HH'_s}{\HH_s^2  }-\frac{2}{3}
\frac{1}{\HH_s \deta_s}\right) \partial_\eta \psi_s^I 
-\frac{4}{\deta_s}\left(-1- \frac{\HH'_s}{\HH_s^2  }\right) \Qint 
\right.
\nonumber
\\
& & 
\left.
+\left(-2- \frac{\HH'_s}{\HH_s^2  }\right) 
 \frac{2}{\deta_s} \Jint
-2 \left(-2- \frac{\HH'_s}{\HH_s^2  }\right)  \dQint
\right.
\nonumber
\\
& & 
\left.
+ \frac{4}{\HH_s \deta_s} \psi_s^I
- \frac{2}{\HH_s \deta_s^2} \INT d\eta' \Delta_2 \psii\left(\eta'\right) \right]v_{||s}
+\left[ 2\left(2+ \frac{\HH'_s}{\HH_s^2 }
+\frac{2}{\HH_s \deta_s}\right) \partial_r v_{||s}
\right.
\nonumber
\\
& & 
\left.
+\frac{2}{\HH_s}\partial_r^2 v_{||s}\right] \Qint
+\left[\frac{2}{\HH_s}\left(5+ 3 \frac{\HH'_s}{\HH_s^2  }\right) \partial_r v_{||s}+\frac{2}{\HH_s^2}\partial^2_r v_{||s}\right] \dQint
\nonumber
\\
&& 
-\frac{2}{\HH_s}\partial_r v_{||s}\frac{1}{\deta_s}\Jint
+ \frac{2}{\HH_s^2} \partial_\eta \psi_s^I \partial_r  v_{||s}+\frac{1}{\HH_s}\left[\frac{1}{\HH_s}\partial^2_r  v_{||s}
\right.
\nonumber
\\
& &
\left. 
+\left(6+3 \frac{\HH'_s}{\HH_s^2  }\right) \partial_r  v_{||s}\right]\psi_s^I
 -\frac{2}{\HH_s \deta_s }  a v_{\perp\,s}^a \partial_a \Qint
+\frac{1}{\HH_s}a v_{\perp\,s}^a \partial_a \psi^I_s 
\nonumber
\\
& & 
-\frac{6}{\HH_s} \gamma_{0s}^{ab}\partial_a v_{||s}
\partial_b \Qint
+\frac{4}{\deta_s^2}\left(\int_{\eta_s}^{\eta_o}\!d \eta'\psi^I\left(\eta'\right)\right)^2
+\left\{\left[2 
\left(-2- \frac{\HH'_s}{\HH_s^2  }\right) 
 \psi_s^I 
 \right.\right.
\nonumber
\\
& & 
\left.
 +4 \left(-2- \frac{\HH'_s}{\HH_s^2  }\right)  \dQint 
- \frac{2}{\HH_s} \partial_\eta \psi_s^I\right]\frac{1}{\deta_s}
+2 \left(-2- \frac{\HH'_s}{\HH_s^2  }\right)  \INT d\eta' \partial^2_{\eta'} \psii \left( \eta' \right)
\nonumber
\\
& & 
\left.
+2 \left(-2- \frac{\HH'_s}{\HH_s^2  }\right)  \partial_\eta \psi_s^I +
\left( -1 - \frac{\HH'_s}{\HH_s^2  } \right)\partial_r \psi_s^I 
- \frac{1}{\HH_s} \partial_{\eta}\partial_{r} \psi_s^I \right\}\left(-2\Qint\right)
\nonumber
\\
& & 
+\left[ \frac{3}{\HH_s}\left(-1-\frac{\HH'_s}{\HH_s^2} - \frac{2}{3}\frac{1}{\HH_s \deta_s}\right)\partial_\eta \psi_s^I+\frac{1}{\HH_s} \partial_{r} \psi_s^I-\left(2 - \frac{\HH'_s}{\HH_s^2  } \right)\psi_s^I
\right.
\nonumber \\
&&
\left.
+\DERSis
\right]
\left(-2 \dQint\right)
+\left[\left(-2-2\frac{\HH'_s}{\HH_s^2}\right)\psi_s^I
\right.
\nonumber
\\
& & 
\left.
+4 \left(-2-\frac{\HH'_s}{\HH_s^2} \right)\dQint
-\frac{8}{\deta_s}\Qint
\right.
\nonumber 
\\
& & 
\left. -\frac{2}{\HH_s}\partial_\eta \psi_s^I\right]
\frac{1}{\deta_s}\Jint
+2\left(\frac{1}{\deta_s}\Jint\right)^2
\nonumber
\\
& & 
+\left[\frac{1}{ \Hcal_s \Delta \eta}\left(-\psi^I_s-2 \int_{\eta_s}^{\eta_o}\!d \eta' \partial_{\eta'}\psi^I\left(\eta'\right)\right)
-\frac{1}{\Delta \eta}\int_{\eta_s}^{\eta_o}\!d \eta'\psi^I\left(\eta'\right)\right]\frac{2}{\deta_s}\int_{\eta_s}^{\eta_o}\!d \eta' \Delta_2\psi^I\left(\eta'\right)
\nonumber \\ 
& &
-2 \left[ \int_{\eta_s}^{\eta_o}\!\!\!d\eta' \frac{1}{(\eta_o-\eta')^2} \psi^I(\eta')+2
 \int_{\eta_s}^{\eta_o}\!\!\!d\eta'  \frac{1}{(\eta_o-\eta')^2} \int_{\eta'}^{\eta_o}\!\!\!d\eta''\partial_{\eta''}\psi^I(\eta'')\right]
  \int_{\eta_s}^{\eta_o}\!\!\!d\eta' \Delta_2 \psi^I(\eta')
\nonumber
\\
& & 
+\left( -\frac{1}{2} - \frac{\HH'_s}{\HH_s^2  } \right)\left(\psi_s^I\right)^2
+\frac{1}{\HH_s^2}\left(\partial_\eta \psi_s^I\right)^2
+\left[-\DERSis
\right.
\nonumber
\\
& & 
\left.
+\frac{1}{\HH_s} \left(4+3 \frac{\HH'_s}{\HH_s^2} + \frac{2}{\HH_s \deta_s}\right)\partial_\eta \psi_s^I
-\frac{1}{\HH_s}\partial_r \psi_s^I
\right]\psi_s^I
\nonumber
\eea
\bea
& & 
+2 \partial_a \left[ -  \psi_s^I
-\frac{1}{\HH_s}\left(\partial_\eta \psi_s^I+\partial_r v_{||s}\right)
\right] \int_{\eta_s}^{\eta_o}d \eta' \gamma_0^{ab}\partial_b \int_{\eta'}^{\eta_o}d \eta'' \psi^I \left(\eta''\right)
\nonumber
\\
& & 
 -2  a v_{\perp\,s}^a \partial_a \Qint
+\frac{4}{\deta_s}\int_{\eta_s}^{\eta_o}\!d \eta'\!\left[\psi^I\left(\eta'\right)\left(-\psi^I\left(\eta'\right) 
-2 \int_{\eta'}^{\eta_o}\!d\!\eta'' \partial_{\eta''}\psi^I\left(\eta''\right)\right)
\right.
\nonumber 
\\
&&
\left.
+\gamma_0^{ab}\partial_a\!\left(\int_{\eta'}^{\eta_o}\!\!\!d\!\eta''\!\psi^I\left(\eta''\right)\right)\!\partial_b\!\left(\int_{\eta'}^{\eta_o}\!\!\!d\!\eta''\!\psi^I\left(\eta''\right)\right)\right] 
+\!4\partial_a\!\psi^I_s
\!\int_{\eta_s}^{\eta_o}\!\!\!d\!\eta'\gamma_0^{ab}\partial_b\!\int_{\eta'}^{\eta_o}\!d\!\eta''\psi^I\left(\eta''\right)
\nonumber 
\\ 
& &
{
-\frac{8}{\deta_s}\left[\int_{\eta_s}^{\eta_o}\!d \eta' \partial_a \psi^I\left(\eta'\right)
\!\int_{\eta_s}^{\eta_o}\!\!\!d\!\eta''\gamma_0^{ab}\partial_b\!\int_{\eta''}^{\eta_o}\!d\!\eta'''\psi^I\left(\eta'''\right)\right]
}
\nonumber
\\
& &
{
+2 \partial_a \left(\!\int_{\eta_s}^{\eta_o}\!d\!\eta'\psi^I\left(\eta'\right)\right)}
\left[4 \int_{\eta_s}^{\eta_o}\!\!\!d\eta' \frac{1}{(\eta_o-\eta')}\gamma_0^{ab}
\int_{\eta'}^{\eta_o}\!\!\!d\eta'' \partial_b \psi^I(\eta'') - 3 \int_{\eta_s}^{\eta_o}\!\!\!d\eta' \gamma_0^{ab}\partial_b\psi^I(\eta')
\right.
\nonumber
\\
& &
\left.
{
-6 \int_{\eta_s}^{\eta_o}\!\!\!d\eta' \gamma_0^{ab}
\int_{\eta'}^{\eta_o}\!\!\!d\eta'' \partial_b\partial_{\eta''}\psi^I(\eta'')
}
\right]
\nonumber 
\\
& &
+2 \partial_a\left(\int_{\eta_s}^{\eta_o}\!\!\!d\!\eta'\gamma_0^{bc}\partial_c\!\int_{\eta'}^{\eta_o}\!d\!\eta''\psi^I\left(\eta''\right)\right)
\partial_b\left(\int_{\eta_s}^{\eta_o}\!\!\!d\!\eta'\gamma_0^{ad}\partial_d\!\int_{\eta'}^{\eta_o}\!d\!\eta''\psi^I\left(\eta''\right)\right)
\nonumber 
\\
& &
-4\!\left(\int_{\eta_s}^{\eta_o}\!\!\!d \eta'\psi^I\left(\eta'\right)\right)\!\int_{\eta_s}^{\eta_o}\!\!\!d \eta'\left[-\frac{1}{(\eta_o-\eta')^3}
\!\int_{\eta'}^{\eta_o}\!\!\!d\!\eta''\Delta_2\psi^I\left(\eta''\right)
+\frac{1}{(\eta_o-\eta')^2}\!\left(\frac{1}{2} \Delta_2\psi^I\left(\eta'\right)
\right. \right.
\nonumber \\
&&
\left. \left.
+\int_{\eta'}^{\eta_o}\!\!\!d\!\eta''\partial_{\eta''}\left(\Delta_2\psi^I\left(\eta''\right)\right)
\right)
\right]+\frac{2}{\left(\sin{\theta_o}\right)^2}\left[\frac{1}{\deta_s}\int_{\eta_s}^{\eta_o}\!\!\!d \eta' \frac{\eta'-\eta_s}{\eta_o-\eta'}
\partial_{{\theta_o}}\psi^I\left(\eta'\right)\right]^2
\nonumber \\ 
& &
+2 \partial_a \left\{\frac{1}{ \Hcal_s} \left[-\psi^I_s-2\int_{\eta_s}^{\eta_o}\!\!\!d\!\eta'\!\partial_{\eta'}\psi^I\left(\eta'\right)\right]
- \int_{\eta_s}^{\eta_o}\!\!\!d\eta'\psi^I(\eta')\right\}\gamma_{0s}^{ab}  \int_{\eta_s}^{\eta_o}\!\!\!d\eta' \partial_b \psi^I(\eta')
\nonumber \\ 
& &
+\frac{4}{\deta_s} \int_{\eta_s}^{\eta_o}\!\!\!d\eta'  \frac{\eta'-\eta_s}{\eta_o-\eta'} \partial_b \left[ \Delta_2\left(\psi^I(\eta')\right)\right]
\int_{\eta_s}^{\eta_o} d \eta'' 
\gamma_0^{ab} \partial_a \int_{\eta''}^{\eta_o}\!d \eta'''  \psi^I(\eta''')
\nonumber 
\\ 
& &
-\frac{2}{\deta_s} \int_{\eta_s}^{\eta_o}\!\!\!d\!\eta' \frac{\eta'-\eta_s}{\eta_o - \eta'} \Delta_2 \left[
\psi^I\left(\eta'\right) \left(-\psi^I\left(\eta'\right)-2\int_{\eta'}^{\eta_o}\!\!\!d\!\eta''\!\partial_{\eta''}\psi^I\left(\eta''\right)\right)
\right.
\nonumber \\ 
& & \left.
+\gamma_0^{ab}\partial_a\left( \int_{\eta'}^{\eta_o}\!\!\!d \eta'' \psi^I\left(\eta''\right) \right) 
\partial_b \left( \int_{\eta'}^{\eta_o} d \eta'' \psi^I\left(\eta''\right) \right) \right]
\nonumber \\
&&
-2 \int_{\eta_s}^{\eta_o} d \eta'\left\{ -2 \psi^I\left(\eta'\right)\frac{1}{\eta_o-\eta'}\int_{\eta'}^{\eta_o} d \eta''  \frac {\eta'' - \eta'}{\eta_o - \eta''}\Delta_2 \partial_{\eta''} \psi^I\left(\eta''\right)
\right. 
\nonumber 
\\ 
& & 
\left. 
+2 \gamma_0^{ab}\partial_b \left(\int_{\eta'}^{\eta_o} d \eta'' \psi^I\left(\eta''\right)\right) \frac{1}{\eta_o-\eta'}\!\int_{\eta'}^{\eta_o}\!d \eta''  \frac {\eta''-\eta'}{\eta_o-\eta''}\partial_a \Delta_2 \psi^I\left(\eta''\right)
\right. \nonumber 
\\ 
& & 
\left.
-\left(-2\psi^I\left(\eta'\right)-2\int_{\eta'}^{\eta_o}\!\!\!d\!\eta''\!\partial_{\eta''}\psi^I\left(\eta''\right)\right)
\frac{1}{(\eta_o-\eta')^2}\int_{\eta'}^{\eta_o}\!\!\!d\!\eta''\!\Delta_2 \psi^I\left(\eta''\right)\right.
\nonumber \\
&&
-2\partial_a \psi^I\left(\eta'\right)\int_{\eta'}^{\eta_o}\!\!\!d\!\eta'' \gamma_0^{ab}\partial_b \int_{\eta''}^{\eta_o}\!d \eta'''\partial_{\eta'''}\psi^I\left(\eta'''\right) 
\nonumber 
\\ 
& & 
+2 \partial_a\!\!\left[\gamma_0^{db}\partial_b\!\!\int_{\eta'}^{\eta_o}\!\!\!d\!\eta''\!\psi^I\left(\eta''\right)\right]\int_{\eta'}^{\eta_o}\!\!\!d\!\eta''\! \partial_d
\left[\gamma_0^{ac}\partial_c\!\!\int_{\eta''}^{\eta_o}\!\!\!d\!\eta'''\!\psi^I\left(\eta'''\right)\right]
\nonumber \\
&&
\left.
+2\gamma_0^{ab}\partial_a\!\!\left(\psi^I\left(\eta'\right)+\int_{\eta'}^{\eta_o}\!\!\!d\!\eta''\!\partial_{\eta''}\psi^I\left(\eta''\right)\right)
\partial_b\!\!\int_{\eta'}^{\eta_o}\!\!\!d\eta''\!\psi^I\left(\eta''\right)
\right\}
\nonumber \\
&& 
+\left[ \left(\frac{2}{\HH_s\deta_s}+\frac{\HH'_s}{\HH_s^2} \right) \left(v_{||s}+\psi^I_s+2 \int_{\eta_s}^{\eta_o} d \eta' \partial_{\eta'}\psi^I\left(\eta'\right)\right)-3 v_{||s}
+\frac{1}{\HH_s}\partial_r v_{||s}-4 \psi_s^I
\right.
\nonumber 
\eea
\bea
&&
\left.
-6 \dQint +\frac{4}{\deta_s} \Qint
-\frac{2}{\deta_s}\Jint 
+\frac{1}{\HH_s}\partial_\eta \psi_s^I
\right]\delta_\rho^{(1)}
\nonumber \\
&&
+\left[  \frac{1}{\bar \rho}\partial_\eta \left(\bar \rho \,\delta_\rho^{(1)}\right) - \partial_r \delta_\rho^{(1)} \right] 
\frac{1}{\HH_s}  \left(  - v_{||s}-\psi^I_s-2 \int_{\eta_s}^{\eta_o} d \eta' \partial_{\eta'}\psi^I\left(\eta'\right)\right)
\nonumber 
\\ 
& &
+2\partial_r \delta_\rho^{(1)}\Qint - 2 \partial_a \delta_\rho^{(1)}
\INT d\eta' \gamma_0^{ab} \partial_b \int_{\eta'}^{\eta_o} d\eta'' \psii \left( \eta''\right)
+\delta_\rho^{(2)}\,. \label{e:SIS}
\eea
The contributions proportional to $\psi^{A}$ which vanish for $\Lambda$CDM are given by
\bea  
\label{e:SigmaAS}
& & \Sigma_{AS} =   \left( -\frac{2}{\HH_s\deta_s} -\frac{\HH'_s}{\HH_s^2} \right)\left\{
-\psi^A_s   v_{||s}+\frac{3}{2} (\psi^A_s)^2
-3\psi^I_s\psi^A_s
-4 \int_{\eta_s}^{\eta_o} d \eta' \partial_{\eta'}(\psi^I\psi^A)\left(\eta'\right)
\right.
\nonumber \\
&&
\left.
-2 \psi^A_s  \int_{\eta_s}^{\eta_0} d\eta' \partial_{\eta'} \psi^I\left( \eta'\right)
- 2 \partial_a  \psi^A_s  \int_{\eta_s}^{\eta_o}\!d \eta' \gamma_0^{ab} \partial_b \int_{\eta'}^{\eta_o}\!d \eta'' \psi^I\left(\eta''\right)
\right\}
+ \left[\frac{1}{2} \frac{\HH'_s}{\HH_s^2}+\frac{3}{2} \left( \frac{\HH'_s}{\HH_s^2}\right)^2
\right.
\nonumber \\
&&
\left. 
- \frac{1}{2}\frac{\HH''_s}{\HH_s^3} +\frac{1}{\HH_s\deta_s} \left(1+ 3 \frac{\HH'_s}{\HH_s^2} 
+\frac{1}{\HH_s \deta_s}\right) \right] \left[\left(\psi_s^A\right)^2+2 \psi_s^a \left(  - v_{||s}-\psi^I_s-2 \int_{\eta_s}^{\eta_o} d \eta' \partial_{\eta'}\psi^I\left(\eta'\right)\right)\right]
\nonumber 
\\ 
& &
+2 \sigmaB \left\{ 3 \psi_s^A v_{||s} -\frac{1}{\HH_s} v_{||s} \partial_\eta \psi_s^A +2 \frac{\psi_s^A}{\HH_s}  \partial_r v_{||s}
\right.
\nonumber \\
&&
- \psi_s^A \left[ -\frac{2}{\deta_s} \Qint 
- 4 \dQint + \frac{2}{\deta_s} \Jint \right]  
\nonumber \\
&&
\left.
+ \frac{1}{\HH_s} \partial_\eta \psi_s^A \left[ - \left( \psi_s^I - \psi_s^A \right) - 2 \Qint \right]  
   + \frac{1}{\HH_s} \psi_s^A \partial_\eta \psi_s^I 
- 3 \left( \psi_s^A\right)^2 + 4 \psi_s^I \psi_s^A 
 \right\}
 \nonumber 
 \\
 &&
+\left[-\DERSas
+\left(5 +3 \frac{\HH'_s}{\HH_s^2}+\frac{2}{\HH_s \deta_s} \right)\frac{1}{\HH_s}\partial_\eta \psi_s^A
-\frac{3}{\HH_s}\partial_r \psi_s^A
\right.
 \nonumber 
 \\
 &&
 \left.
-\left(6+4 \frac{\HH'_s}{\HH_s^2  } \right)\psi_s^A\right]v_{||s}
+\left[-\frac{1}{\HH_s^2}\partial^2_r v_{||s}-\left(8 +3 \frac{\HH'_s}{\HH_s^2  } \right)
\frac{1}{\HH_s}\partial_r v_{||s}\right]\psi_s^A
+\frac{2}{\HH_s^2}\partial_\eta \psi_s^A \partial_r v_{||s}
\nonumber 
\\
& &
+\frac{1}{\HH_s}a v_{\perp s}^a\partial_a \psi_s^A+ \left[ -\frac{2}{\HH_s}\partial_\eta \psi_s^A 
+\left(10 +2 \frac{\HH'_s}{\HH_s^2  } \right)\psi_s^A
\right]  \frac{1}{\deta_s} \Jint 
\nonumber 
\\
& &
{+\frac{2}{ \Hcal_s \deta_s^2} \psi^A_s  \int_{\eta_s}^{\eta_o}\!d \eta' \Delta_2 \psi^I\left(\eta'\right)}
+\left\{\left[-\frac{2}{\HH_s}\partial_\eta \psi_s^A+2\left(4 +\frac{\HH'_s}{\HH_s^2  } \right)\psi_s^A
\right] \frac{1}{\deta_s} 
\right.
\nonumber 
\\
& &
\left.
-\frac{1}{\HH_s}\partial_{\eta}\partial_{r} \psi_s^A
+\left(3+\frac{\HH'_s}{\HH_s^2}+ \frac{2}{\HH_s \deta_s} \right)\partial_{r} \psi_s^A
\right\}\left(-2 \Qint\right)
\nonumber 
\\
&&
+\left[\DERSas 
-\left(5 +3 \frac{\HH'_s}{\HH_s^2}+\frac{2}{\HH_s \deta_s} \right)\frac{1}{\HH_s}\partial_{\eta} \psi_s^A
+\frac{3}{\HH_s}\partial_{r} \psi_s^A
\right.
\nonumber 
\\
&&
+\left.
3 \left(2 +\frac{\HH'_s}{\HH_s^2  } \right)\psi_s^A\right]\left(-2 \dQint\right)
+\left(\frac{15}{2} +3 
\frac{\HH'_s}{\HH_s^2  } \right)\left(\psi_s^A\right)^2+\frac{1}{\HH_s^2  }\left(\partial_\eta \psi_s^A\right)^2
\nonumber 
\\
&&
+\frac{2}{\HH_s^2  }\partial_\eta \psi_s^A\partial_\eta \psi_s^I+\left[\DERSas 
+\DERSis
\right.
\nonumber 
\\
& &
\left.
+\left(2\!-\!3\frac{\HH'_s}{\HH_s^2}-\frac{2}{\HH_s \deta_s}\right)\frac{1}{\HH_s  }\partial_\eta \psi_s^A
\!+\!\left(4\!-\!3\frac{\HH'_s}{\HH_s^2}-\frac{2}{\HH_s \deta_s}\right)
\frac{1}{\HH_s}\partial_\eta \psi_s^I\!-\!\left(7\!+\!\frac{\HH'_s}{\HH_s^2  }\right)\psi_s^I
\right.
\nonumber 
\\
& &
\left.
+\frac{3}{\HH_s  }\partial_r \psi_s^A+\frac{1}{\HH_s  }\partial_r \psi_s^I
\right] \psi_s^A
+\left[-\DERSas
\right.
\nonumber 
\\
&&
\left.
+\left(6+3\frac{\HH'_s}{\HH_s^2}+\frac{2}{\HH_s \deta_s}\right)
\frac{1}{\HH_s}\partial_\eta \psi_s^A
-\frac{3}{\HH_s  }\partial_r \psi_s^A\right]\psi_s^I
\nonumber 
\eea
\bea
&&
+ 6\, \partial_a \psi_s^A \thetaint 
{
+\frac{2}{\Hcal_s}\partial_a  \psi^A_s
\gamma_{0s}^{ab} \partial_b \int_{\eta_s}^{\eta_o} d \eta' \psi^I\left(\eta'\right)
}
\nonumber 
\\
& & 
-\frac{2}{\HH_s}\partial_a \left(\partial_\eta \psi_s^A\right) \int_{\eta_s}^{\eta_o}d \eta' \gamma_0^{ab}\partial_b \int_{\eta'}^{\eta_o}d \eta'' \psi^I \left(\eta''\right)+\frac{8}{\deta_s} \INT d\eta' \left( \psii \psia \right) \left(\eta'\right) 
\nonumber \\
&& 
- \frac{4}{\deta_s} \INT d\eta' \frac{\eta' - \eta_s}{\eta_o - \eta' } \Delta_2 \left( \psi^I \psi^A \right)\left( \eta'\right)
+\left[ \left( -\frac{2}{\HH_s\deta_s} -\frac{\HH'_s}{\HH_s^2} \right)\psi_s^A+\frac{1}{\HH_s} \partial_\eta \psi_s^A
\right]\delta_\rho^{(1)}
\nonumber \\
&& 
+\left[  \frac{1}{\bar \rho}\partial_\eta \left(\bar \rho \,\delta_\rho^{(1)}\right) - \partial_r \delta_\rho^{(1)} \right] 
\frac{1}{\HH_s} \psi_s^A \,.
\eea

\subsection{Number Counts: leading terms}

Let us finally give a more concise expression for the second order number counts in Eqs.~(\ref{e:SigmaIS}) and~(\ref{e:SigmaAS}).
We will neglect all the subleading terms, keeping only the leading (potentially observable) terms in the large redshift and small angle limit, where the pure Doppler and potential terms can be neglected. This means we keep only the terms with four spatial derivatives of  metric perturbations (the velocity has one spatial derivative at first order, 
see Eq.~(\ref{vpar})) and terms with the density and two spatial derivatives. We also keep the density, the redshift space distortions and the lensing at second order. 
Restricting to these dominant terms we obtain, after some algebraic manipulation, the following leading order result for $\Sigma$:

\bea  
& & 
\!\!\!\!\!\! \Sigma^{Leading} \simeq
\frac{1}{\HH_s} \partial_r v_{||s}^{(2)}
-\frac{1}{2} \frac{1}{\deta_s}\int_{\eta_s}^{\eta_o}\!d \eta' \frac {\eta' - \eta_s}{\eta_o - \eta'} \Delta_2\left(
\psi^{(2)}\left(\eta'\right)+\phi^{(2)}\left(\eta'\right)\right)+\boldsymbol{\delta_\rho^{(2)}}
\nonumber
\\
& & 
+\frac{1}{\HH_s^2}\left[v_{||s}\partial^2_r v_{||s}+\left(\partial_r v_{||s}\right)^2\right]
-\frac{2}{\HH_s}\partial_r v_{||s}\frac{1}{\deta_s}\Jint
\nonumber
\\
& & 
-\frac{2}{\HH_s}\partial_a \left(\partial_r v_{||s}\right)  \int_{\eta_s}^{\eta_o}d \eta' \gamma_0^{ab}\partial_b \int_{\eta'}^{\eta_o}d \eta'' \psi^I \left(\eta''\right)
+2\left(\frac{1}{\deta_s}\Jint\right)^2
\nonumber \\ 
& &
+\frac{4}{\deta_s} \int_{\eta_s}^{\eta_o}\!\!\!d\eta'  \frac{\eta'-\eta_s}{\eta_o-\eta'} \left\{ \partial_b \left( \Delta_2 \psi^I(\eta')\right)
\int_{\eta_s}^{\eta_o} d \eta'' 
\gamma_0^{ab} \partial_a \int_{\eta''}^{\eta_o}\!d \eta'''  \psi^I(\eta''')
\right.
\nonumber 
\\
& &
\left. +\Delta_2\left[-\frac{1}{2}
\gamma_0^{ab}\partial_a\left( \int_{\eta'}^{\eta_o}\!\!\!d \eta'' \psi^I\left(\eta''\right) \right) 
\partial_b \left( \int_{\eta'}^{\eta_o} d \eta'' \psi^I\left(\eta''\right) \right) \right]\right\} 
\nonumber 
\\ 
& &
-4 \int_{\eta_s}^{\eta_o}\!\!\!d\eta'  \left\{\gamma_0^{ab}\partial_b \left(\int_{\eta'}^{\eta_o} d \eta'' \psi^I\left(\eta''\right)\right) \frac{1}{\eta_o-\eta'}\!\int_{\eta'}^{\eta_o}\!d \eta''  \frac {\eta''-\eta'}{\eta_o-\eta''}\partial_a \Delta_2 \psi^I\left(\eta''\right)
\right\}
\nonumber 
\\
&& 
+\left[
\boldsymbol{\frac{1}{\HH_s}\partial_r v_{||s}}\boldsymbol{-\frac{2}{\deta_s}\Jint }
\right]\boldsymbol{\delta_\rho^{(1)}}
+ \frac{1}{\HH_s} v_{||s} \partial_r \delta_\rho^{(1)} 
\nonumber 
\\ 
& &
- 2 \partial_a \delta_\rho^{(1)}
\INT d\eta' \gamma_0^{ab} \partial_b \int_{\eta'}^{\eta_o} d\eta'' \psii \left( \eta''\right)\,. 
\label{SigmaLeading}
\eea

All these terms come from density fluctuations, redshift space distortions and lensing-like terms. 
The $\Lambda$CDM expression corresponding to Eq.~(\ref{SigmaLeading}) can be easily obtained substituting $\psii$ with $\psi$ (in {this} limit $\psii=\psi$ and $\psia=0$).

The terms indicated in bold face are the once which we evaluate numerically in Section~\ref{Sec5} for $\Lambda$CDM.
The other terms, which are really new physical effects (like e.g.~second order lensing), can also contribute significantly as they have  four transverse derivatives 
We shall study them in detail in a forthcoming publication~\cite{DDMM}.

\section{The Bispectrum and its numerical evaluation}
\label{Sec5}
\setcounter{equation}{0}

In this section we compute two new contributions to the bispectrum of the number count fluctuations and evaluate them numerically for some configurations, namely the contribution from redshift space distortions and the one from lensing. Since we consider the truly observed galaxy number count, and not fluctuations on some unobservable spatial hypersurface, the 3-point function is in general a function of 3 directions\footnote{We remember that $\bf n$ is the photon direction, so from the source to the observer. Hence we observer {the} galaxy in direction $- \bf n$. But, for the sake of simplicity, we prefer to use $\bf n$ to indicate the galaxy position.} and 3 redshifts,
\be
B \left( \nv_1, \nv_2, \nv_3, z_1, z_2 ,z_3 \right) = \langle \Delta\left( \nv_1, z_1 \right) \Delta \left( \nv_2, z_2 \right) \Delta\left( \nv_3 ,z_3 \right) \rangle \, ,
\ee
The leading non-vanishing terms are in the form
\bea
\langle  \Delta^{(2)}\left( \nv_1, z_1 \right) \Delta^{(1)} \left( \nv_2, z_2 \right) \Delta^{(1)}\left( \nv_3 ,z_3 \right) \rangle &=& \langle  \Sigma\left( \nv_1, z_1 \right) \Delta^{(1)} \left( \nv_2, z_2 \right) \Delta^{(1)}\left( \nv_3 ,z_3 \right) \rangle \nonumber \\
&&\hspace{-6pt} - \langle \Sigma\left( \nv_1, z_1 \right) \rangle \langle   \Delta^{(1)} \left( \nv_2, z_2 \right) \Delta^{(1)}\left( \nv_3 ,z_3 \right) \rangle
\eea
plus the permutations with respect to the galaxy positions $(\nv_i, z_i )$.
Expanding the direction dependence of $\Delta$ in spherical harmonics,
$$
 \Delta\left( \nv, z \right) = \sum_{\ell,m}a_{\ell m}(z)Y_{\ell m}(\nv) \,,
$$
 we can write
\be
B \left( \nv_1, \nv_2, \nv_3, z_1, z_2 ,z_3 \right) = \sum_{\ell_1,\ell_2,\ell_3,m_1,m_2,m_3}B_{\ell_1\ell_2\ell_3}^{m_1m_2m_3}(z_1,z_2,z_3)Y_{\ell_1 m_1}(\nv_1) Y_{\ell_2 m_2}(\nv_2) Y_{\ell_3 m_3}(\nv_3) \,, 
\ee
where
\bea \label{bi_harm}
B_{\ell_1\ell_2\ell_3}^{m_1m_2m_3}(z_1,z_2,z_3) &=& \langle a_{\ell_1 m_1}(z_1)a_{\ell_2 m_2}(z_2)a_{\ell_3 m_3}(z_3)\rangle \nonumber \\
&=& \!\int \!d\Omega_1 d\Omega_2 d\Omega_3 B \left( \nv_1, \nv_2, \nv_3, z_1, z_2 ,z_3 \right) Y^*_{\ell_1 m_1} \!\left( \vn_1 \right) Y^*_{\ell_2 m_2 } \!\left( \vn_2 \right) Y^*_{\ell_3 m_3 }\! \left( \vn_3 \right) \, . \nonumber \\
\eea
Statistical isotropy requires that
\be\label{bi_harm_red}
B_{\ell_1\ell_2\ell_3}^{m_1m_2m_3}(z_1,z_2,z_3) =\Gaunt{m_1,}   {m_2,}   {m_3} {\ell_1,} {\ell_2,} {\ell_3} b_{\ell_1,\ell_2,\ell_3}(z_1,z_2,z_3)\,,
\ee
where $\Gaunt{m_1,}   {m_2,}   {m_3} {\ell_1,} {\ell_2,} {\ell_3} $ is the Gaunt integral given by
\bea 
\Gaunt{m_1,}   {m_2,}   {m_3} {\ell_1,} {\ell_2,} {\ell_3} &=& \int d \Omega \ Y_{\ell_1 m_1 } \left( \vn \right)Y_{\ell_2 m_2 } \left( \vn \right) Y_{\ell_3 m_3 } \left( \vn \right)  \\
&=&\left(\begin{array}{ccc} \ell_1 & \ell_2 & \ell_3 \\ 0 & 0 & 0\end{array}\right)\left(\begin{array}{ccc} \ell_1 & \ell_2 & \ell_3 \\ m_1 & m_2 & m_3\end{array}\right)
\sqrt{\frac{(2\ell_1+1)(2\ell_2+1)(2\ell_3+1)}{4\pi}} \, .
\eea
On the second line we have expressed the Gaunt integral in terms of the Wigner $3j$ symbols, 
see e.g.~\cite{Abram}.
This integral vanishes if $\ell_3\not\in \left[ |\ell_1-\ell_2|~,~\ell_1+\ell_2\right]$, if the sum $\ell_1 + \ell_2 + \ell_3$ is odd, or if $m_1+m_2+m_3\neq 0$. 
The quantity $b_{\ell_1,\ell_2,\ell_3}(z_1,z_2,z_3)$ is the reduced bispectrum of the number counts. For Gaussian initial conditions it vanishes in linear perturbation theory and its first contributions are terms of the form $\langle \Delta^{(2)}\left( \nv_1, z_1 \right) \Delta^{(1)} \left( \nv_2, z_2 \right) \Delta^{(1)}\left( \nv_3 ,z_3 \right) + \Delta^{(1)}\left( \nv_1, z_1 \right) \Delta^{(2)} \left( \nv_2, z_2 \right) \Delta^{(1)}\left( \nv_3 ,z_3 \right) + \Delta^{(1)}\left( \nv_1, z_1 \right) \Delta^{(1)} \left( \nv_2, z_2 \right) \Delta^{(2)}\left( \nv_3 ,z_3 \right) \rangle$. Here we want to determine 
the presumably largest contributions to this expression. In subsequent work~\cite{DDMM} we shall study it in more detail.

We write the perturbations in terms of Fourier modes and we express them in terms of the power spectrum of the primordial curvature perturbation $R_{in}\left( \vk \right)$. In this section, we assume simple adiabatic Gaussian initial perturbations from inflation which are given by the curvature power spectrum,
\be
\langle R_{in} \left( \vk \right) R_{in} \left( \vk' \right) \rangle = \left( 2 \pi \right)^3 \delta_D \left( \vk + \vk' \right) P_R\left(k \right)\,.
\ee
For a given variable $A$ we define the transfer function $T_A(\eta,k)$ by
\be
A \left( \eta, \vk \right) =T_A(\eta,k)R_{in}(\vk) \,.
\ee
We will use also the angular power spectra defined as
\be
c^{AB}_\ell \left( z_1, z_2 \right) = 4 \pi \int \frac{dk}{k} \mathcal{P}_R (k) \Delta^A_\ell (z_1,k)\Delta^B_\ell (z_2,k) = \frac{2}{\pi} \int dk k^2 P_R (k)  \Delta^A_\ell (z_1,k)\Delta^B_\ell (z_2,k) 
\ee
where $\mathcal{P}_R (k) = \frac{k^3}{2 \pi^2} P_R (k)$ is the dimensionless primordial power spectrum. $\Delta^A_\ell \left( z, k \right)$ denote the transfer functions in angular and redshift space for different sources. We will define the transfer functions  explicitly for the cases analyzed in this work. The simplest relation is the one for the density fluctuations $\de_\rho^{(1)}$ where
\be\label{trans_den}
\Delta^\de_\ell \left( z, k \right)=T_\de\left(\eta(z) , k \right)j_\ell\left(k(\eta_o-\eta(z))\right)\,, 
\ee
where $j_\ell(x)$ is the spherical Bessel function of order $\ell$. 

The largest terms in $\Delta^{(1)}$ are the density fluctuation $\delta_\rho^{(1)}$, the redshift space distortion $\dd_r v_{\parallel s}/\HH_s$,
 and the lensing term 
$$
-\frac{2}{\deta_s^{(0)}}\int_{\eta_s^{(0)}}^{\eta_o} d\eta' \frac{\eta' - \etasB}{\eta_0 -\eta'} \Delta_2 \psi^I \left( \eta' \right)\,.
$$
In the following subsections we calculate some of their contributions to the bispectrum and compare the numerical results for some combinations of  $\ell_i$ and $z_i$. 

\subsection{Density}
We start by computing the leading term in the bispectrum, namely 
\be \label{defdenbi}
B_D \left( \nv_1, \nv_2, \nv_3, z_1, z_2 ,z_3 \right) =  \langle \delta^{(2)}_\rho \left( \nv_1, z_1 \right) \delta^{(1)}_\rho \left( \nv_2, z_2 \right)\delta^{(1)}_\rho \left( \nv_3, z_3 \right) \rangle_c  + \text{permutations} \, ,
\ee
where the $\langle \rangle_c$ denotes the connected part, and
\be
\delta^{(2)}_\rho \left( \nv ,z \right) = \frac{1}{\left( 2 \pi \right)^6} \int d^3k d^3k_1 d^3k_2 \delta_D \left( \vk - \vk_1 - \vk_2 \right) f_2 \left( \vk_1 , \vk_2 \right) \delta^{(1)}_\rho \left( \vk_1 \right) \delta^{(1)}_\rho \left( \vk_2 \right) e^{i \vk \cdot \nv r}  \, ,
\ee
with
\be
f_2\left( \vk_1 , \vk_2 \right) = \frac{5}{7}+\frac{1}{2} \frac{\vk_1 \cdot \vk_2 }{k_1 k_2 } \left( \frac{k_1}{k_2} + \frac{k_2}{k_1} \right) + \frac{2}{7} \left( \frac{\vk_1 \cdot \vk_2 }{k_1 k_2 } \right)^2 \, .
\ee
This expression for $f_2$ is valid for $\Omega_M=1$ and Newtonian gravity~\cite{Bernardeau:2001qr}.
We aim to estimate the order of magnitude of the different effects, leaving a detailed analysis to a future project~\cite{DDMM}. So we adopt the approximation $f_2\left( \vk_1 , \vk_2 \right)  \sim 1$.
This is of course not a very good approximation for certain angles of for very asymmetric situations, e.g.~$k_1\gg k_2$, but in this first analysis we just want to gain insight to the order of magnitude of the different terms.  Approximating $f_2\simeq 1$ significantly simplifies the calculation and still gives the correct order of magnitude for many cases. We defer the  study of asymmetrical cases, like the squeezed limit, to future work~\cite{DDMM}.
Within this  approximation, the second-order density perturbation is then given by 
\be
\delta^{(2)}_\rho \left( \nv ,z \right) \simeq \frac{1}{\left( 2 \pi \right)^6}\int d^3k d^3k_1 \delta^{(1)}_\rho \left( \vk \right) \delta^{(1)}_\rho \left( \vk_1 - \vk \right) e^{i \vk_1 \cdot \nv \deta } \, .
\ee

Evaluating~(\ref{defdenbi}) we find
\bea
&&\langle \delta^{(2)}_\rho \left( \nv_1, z_1 \right) \delta^{(1)}_\rho \left( \nv_2, z_2 \right)\delta^{(1)}_\rho \left( \nv_3, z_3 \right) \rangle_c  \nonumber \\
&=&\frac{1}{\left( 2 \pi \right)^{12}} \int d^3 k d^3k_1 d^3k_2 d^3k_3 e^{i \left( \vk_1 \cdot \nv_1 \deta_1 + \vk_2 \cdot \nv_2 \deta_2 + \vk_3 \cdot \vn_3 \deta_3 \right)} 
\nonumber \\
&&\hspace{1.2cm} \times T_\delta\left( \eta_1, k \right) T_\delta\left( \eta_1, \left| \vk_1 - \vk \right|\right)T_\delta\left(\eta_2, k_2 \right)  T_\delta\left( \eta_3, k_3\right) \nonumber \\
&&\hspace{1.2cm} \times \left[ \langle R_{in} \left( \vk \right) R_{in} \left( \vk_1 \!-\! \vk \right) R_{in} \left( \vk_2 \right) R_{in} \left(\vk_3 \right) \rangle\! -\! \langle R_{in} \left( \vk \right) R_{in} \left( \vk_1 \!-\! \vk \right) \rangle\langle R_{in} \left( \vk_2 \right) R_{in} \left(\vk_3 \right) \rangle \right] \nonumber \\
 &=& \frac{2}{\left( 2 \pi \right)^6} \int d^3k_2 d^3k_3 e^{-i \left( \vk_2 \cdot \nv_1 \deta_1 + \vk_3 \cdot \nv_1 \deta_1 \right)} e^{i \left( \vk_2 \cdot \nv_2 \deta_2+ \vk_3 \cdot \nv_3 \deta_3 \right)} 
 \nonumber \\
&&\hspace{1.2cm} \times T_\delta\left( \eta_1, k_2 \right) T_\delta\left( \eta_1,k_3 \right)T_\delta\left(\eta_2, k_2 \right)  T_\delta\left( \eta_3, k_3\right) P_R \left( k_2 \right) P_R \left( k_3 \right) 
\eea
where we have used the Wick theorem to compute the term into square brackets,
\bea
[\cdots]&=&\left( 2 \pi \right)^6 P_R \left( k_2 \right) P_R \left( k_3 \right) \delta_D \left( \vk+ \vk_2 \right) \delta_D \left( \vk_1  -\vk +\vk_3 \right) \nonumber \\
&& \hspace{-3mm}
+ \left( 2 \pi \right)^6 P_R \left( k_2 \right) P_R \left( k_3 \right) \delta_D \left( \vk+ \vk_3 \right) \delta_D \left( \vk_1  -\vk +\vk_2 \right) 
\eea

We now expand the Fourier modes in spherical harmonics and Bessel functions
\be
e^{i \vk \cdot {\bf n} r } = 4 \pi \sum_{ \ell m } i^\ell j_\ell \left( k r \right) Y_{\ell m } \left( \nv \right) Y^*_{\ell m } \left( {\bf \hat  k} \right) \, .
\ee
Integrating over the angles $ {\bf \hat  k_2}$ and $ {\bf \hat  k_3}$ and applying the orthogonality of spherical harmonics we find the three-point function
\bea \label{bi_den}
B_D \left( \nv_1, \nv_2, \nv_3, z_1, z_2 ,z_3 \right) &=&\frac{4}{\pi^2}  \sum_{\ell, \ell', m , m'}   Y_{\ell m } \left( \vn_1 \right) Y_{\ell' m'} \left( \vn_1 \right) Y^*_{\ell m } \left( \nv_2 \right) Y^*_{\ell' m'} \left( \nv_3 \right)  F_{\ell \ell'} \left( z_1 , z_2 ,z_3 \right) \nonumber \\ &&+ \text{permutations} \, ,
\eea
with
\bea
 F_{\ell \ell'} \left( z_1, z_2 , z_3 \right) &=&2\int dk_2 dk_3 k_2^2 k_3^2 P_R \left( k_2 \right) P_R \left( k_3 \right)  T_\delta\left( \eta_1, k_2 \right) T_\delta\left( \eta_1,k_3 \right)T_\delta\left(\eta_2, k_2 \right)  T_\delta\left( \eta_3, k_3\right) \nonumber \\
 && \qquad  \times j_\ell \left( k_2 \deta_1 \right) j_{\ell'} \left( k_3 \deta_1\right) j_\ell \left( k_2 \deta_2 \right) j_{\ell'} \left( k_3 \deta_3 \right)  \, . \label{e:Fll}
\eea
Then, with the expansion~(\ref{bi_den}) in spherical harmonics we obtain the reduced bispectrum defined in Eqs.~(\ref{bi_harm}) and~(\ref{bi_harm_red}),
\be
b^{\delta\delta\delta}_{\ell_1 \ell_2 \ell_3}=\frac{4}{\pi^2}  F_{\ell_2 \ell_3} \left( z_1 ,z_2 ,z_3 \right) + \text{permutations} \, .
\ee
The double integral in~(\ref{e:Fll}) is simply a product of two single integrals (due to the approximation $f_2(\vk_1,\vk_2)\simeq 1$ which we have adopted), so that we can simplify the result to
\bea
b^{\delta\delta\delta}_{\ell_1 \ell_2 \ell_3}&=& 2 \left[\frac{2}{\pi}\int dk_2  k_2^2  P_R \left( k_2 \right)   T_\delta\left( \eta_1, k_2 \right) T_\delta\left(\eta_2, k_2 \right)
  j_{\ell_2} \left( k_2 \deta_1 \right)  j_{\ell_2} \left( k_2 \deta_2 \right) \right] \nonumber \\
 &&\times  \left[\frac{2}{\pi} \int dk_3  k_3^2  P_R \left( k_3 \right)   T_\delta\left( \eta_1, k_3 \right) T_\delta\left(\eta_3, k_3 \right) 
  j_{\ell_3} \left( k_3 \deta_1 \right)  j_{\ell_3} \left( k_3 \deta_3 \right) \right] + \text{permutations} \nonumber \\
 &=& 2 c_{\ell_2}^{\delta\delta} \left( z_1,z_2 \right) c_{\ell_3}^{\delta\delta} \left( z_1,z_3 \right) + 2 c_{\ell_1}^{\delta\delta} \left( z_1,z_2 \right) c_{\ell_3}^{\delta\delta} \left( z_2,z_3 \right) +2 c_{\ell_1}^{\delta\delta} \left( z_1,z_3 \right) c_{\ell_2}^{\delta\delta} \left( z_2,z_3 \right)\, .
\eea

Here $c_\ell^{\de\de}(z_i,z_j)$ is the contribution  to the number count angular power spectrum from density fluctuations. It can be calculated with the publicly available code CLASSgal described in~\cite{DiDio:2013bqa}.
We set the bias b between matter and galaxies equal to one, but it is easy to add a linear and scale-independent bias as outlined in CLASSgal.

\subsection{Redshift space distortions}

After the density term, computed in the previous section, the second leading contribution is the quadratic term containing one density and one redshift space distortion,  $\propto \de^{(1)}_\rho\,\dd_rv_{\parallel}/\HH$, see the second boldface term in Eq.~(\ref{SigmaLeading}). This term gives the following contribution to the reduced 3-point function:
\be
B_v \left( \nv_1, \nv_2, \nv_3, z_1, z_2 ,z_3 \right) = 
\langle \Delta^{(2RSD)} \left( \nv_1, z_1 \right) \delta^{(1)}_\rho \left( \nv_2 , z_2 \right) \delta^{(1)}_\rho\left( \nv_3 , z_3 \right) \rangle_c + \text{permutations}
\ee
where
\bea
\Delta^{(2RSD)} \left( \nv, z \right) &=& \frac{1}{\HH_s}  \left[ \partial_r\ndv \left( \nv, z \right) \right] \delta^{(1)}_\rho \left( \nv , z \right)  \nonumber \\
&=& \frac{1}{\left( 2 \pi \right)^{6}\HH_s} \int d^3k d^3k' \ndv \left( \vk,\eta_s \right) \delta^{(1)}_\rho \left( \vk',\eta_s \right) \left( \partial_r e^{i \vk \cdot \nv r } \right)e^{i \vk' \cdot \nv r } \nonumber \\
&=& \frac{1}{\left( 2 \pi \right)^{6}\HH_s} \int d^3k d^3k' k V \left( \vk,\eta_s \right) \delta^{(1)}_\rho \left( \vk',\eta_s \right) \left( \partial^2_{(kr)} e^{i \vk \cdot \nv r } \right)e^{i \vk' \cdot \nv r }
\eea
with ${\bf v} \equiv i {\bf \hat k} V$, the longitudinal gauge velocity with velocity potential $V$ so that
\be
\ndv e^{i {\bf \hat k} \cdot \nv kr} = V \partial_{(kr)} e^{i {\bf \hat k} \cdot \nv kr} \, .
\ee

With the same approach as in previous section we arrive at
\bea
\hspace{-1cm}&&\frac{1}{\left( 2 \pi \right)^{12} \HH \left( z_1 \right)} \int d^3k d^3k_1 d^3k_2 d^3k_3 k e^{i \left( \vk_1 \cdot \nv_1 \deta_1 +\vk_2 \cdot \nv_2 \deta_2 +\vk_3 \cdot \nv_3 \deta_3 \right)}\left( \partial^2_{(kr_1)} e^{i \vk \cdot \nv_1 \deta_1 } \right) 
\nonumber \\
&&\hspace{2cm} \times
T_V \left( \eta_1 , k \right) T_\delta \left( \eta_1 , k_1 \right) T_\delta \left( \eta_2 , k_2 \right) T_\delta \left( \eta_3 , k_3 \right) \nonumber \\
&&\hspace{2cm} \times \left[ \langle R_{in} \left( \vk \right) R_{in} \left( \vk_1  \right) R_{in} \left( \vk_2 \right) R_{in} \left(\vk_3 \right) \rangle - \langle R_{in} \left( \vk\right) R_{in} \left( \vk_1 \right) \rangle \langle R_{in} \left( \vk_2 \right) R_{in} \left( \vk_3 \right) \rangle \right]  \nonumber 
\eea
\bea
&=&\!\! \frac{1}{\left( 2 \pi \right)^6}\!\! \int \!\!d^3k_2 d^3k_3 \frac{k_2}{\HH \left(\! z_1 \!\right)}e^{i \left( \vk_2 \cdot \nv_2 \deta_2 +  \vk_3 \cdot \nv_3 \deta_3 \right)} e^{ -i \vk_3 \cdot \nv_1 \deta_1}  
\partial^2_{( k_2 \deta_1)}  
e^{-i \vk_2 \cdot \nv_1 \deta_1 }
\nonumber \\
&& \hspace{1cm} \times
 T_V \!\left( \eta_1 , k_2 \!\right) T_\delta \!\left( \eta_1 , k_3\! \right)
 T_\delta \!\left( \eta_2 , k_2\! \right) T_\delta\! \left( \eta_3 , k_3 \!\right) P_R\! \left( k_2 \right) P_R \!\left( k_3 \right) \nonumber \\
&&\!\! +  \frac{1}{\left( 2 \pi \right)^6}\!\! \int \!\!d^3k_2 d^3k_3 \frac{k_3}{\HH \left(\! z_1 \!\right)}e^{i \left( \vk_2 \cdot \nv_2 \deta_2 +  \vk_3 \cdot \nv_3 \deta_3 \right)} e^{- i \vk_2 \cdot \nv_1 \deta_1}  
\partial^2_{( k_3 \deta_1)}
e^{-i \vk_3 \cdot \nv_1 \deta_1 }
\nonumber \\
&& \hspace{1cm} \times
T_V \!\left( \eta_1 , k_3 \!\right) T_\delta \!\left( \eta_1 , k_2\! \right)
 T_\delta \!\left( \eta_2 , k_2\! \right) T_\delta\! \left( \eta_3 , k_3 \!\right) P_R\! \left( k_2 \right) P_R \!\left( k_3 \right) \, .
\eea
Expanding the exponentials in spherical harmonics and Bessel functions we obtain
\bea
B_v \left( \nv_1, \nv_2, \nv_3, z_1, z_2 ,z_3 \right) &=&\frac{4}{\pi^2}  \sum_{\ell, \ell', m , m'}   Y_{\ell m } \left( \vn_1 \right) Y_{\ell' m'} \left( \vn_1 \right) Y^*_{\ell m } \left( \nv_2 \right) Y^*_{\ell' m'} \left( \nv_3 \right) Z_{\ell \ell'} \left( z_1 , z_2 ,z_3 \right) \nonumber \\
&&+ \text{permutations} \, ,
\eea
where
\bea
 Z_{\ell \ell'} \left( z_1 ,z_2 ,z_3 \right) \!&=&\! \frac{1}{\HH \left( z_1 \right) }\!\int \!\!d k_2 d k_3 k_2^3 k_3^2 P_R\! \left( k_2 \right) P_R \!\left( k_3 \right) T_V \!\left( \eta_1 , k_2 \right) T_\delta \!\left( \eta_1 , k_3 \right) T_\delta \!\left( \eta_2 , k_2 \right) T_\delta \!\left( \eta_3 , k_3 \right)\nonumber \\
 && \hspace{5cm} \times j''_\ell \left( k_2 \deta_1 \right) j_{\ell'} \left( k_3 \deta_1\right) j_\ell \left( k_2 \deta_2 \right) j_{\ell'} \left( k_3 \deta_3 \right) \nonumber \\
 &&\hspace{-6pt}+ \!\frac{1}{\HH \left( z_1 \right) }\!\int \!\!d k_2 d k_3 k_3^3 k_2^2 P_R\! \left( k_2 \right) P_R\! \left( k_3 \right) T_V \!\left( \eta_1 , k_3 \right) T_\delta\! \left( \eta_1 , k_2 \right) T_\delta\! \left( \eta_2 , k_2 \right) T_\delta\! \left( \eta_3 , k_3 \right)\nonumber \\
 && \hspace{5cm} \times j''_{\ell'} \left( k_3 \deta_1 \right) j_{\ell} \left( k_2 \deta_1\right) j_\ell \left( k_2 \deta_2 \right) j_{\ell'} \left( k_3 \deta_3 \right) \, .
\eea
Finally, the spherical harmonics expansion leads to the reduced bispectrum
\be
b^{v\delta\delta}_{\ell_1 \ell_2 \ell_3} = \frac{4}{\pi^2} Z_{\ell_2 \ell_3 } \left( z_1 , z_2 , z_3 \right) + \text{permutations} \, .
\ee
Again, we can express it in terms of the product of angular power spectra
\bea
b^{v\delta\delta}_{\ell_1 \ell_2 \ell_3} &=& c_{\ell_2}^{v\delta} ( z_1, z_2 ) c_{\ell_3}^{\delta\delta} ( z_1,z_3 ) +  c_{\ell_2}^{\delta\delta} (z_1,z_2) c_{\ell_3}^{v\delta} (z_1, z_3) \nonumber \\
&&\hspace{-6pt}+  c_{\ell_1}^{v\delta} ( z_2, z_1 ) c_{\ell_3}^{\delta\delta} ( z_2,z_3 ) +  c_{\ell_1}^{\delta\delta} (z_1,z_2) c_{\ell_3}^{v\delta} (z_2, z_3) \nonumber \\
&& \hspace{-6pt}+  c_{\ell_1}^{v\delta} ( z_3, z_1 ) c_{\ell_2}^{\delta\delta} ( z_2,z_3 ) +  c_{\ell_1}^{\delta\delta} (z_1,z_3) c_{\ell_2}^{v\delta} (z_3, z_2) \, , 
\eea
 Here the convention is that the first $z$ in the argument refers to the first superscript and we have used~(\ref{trans_den}) and
\be
\Delta^{RSD}_\ell \left( z,k \right) = \frac{k}{\HH(z)} T_V \left( \eta , k \right) j_\ell'' \left( k r \right) \, .
\ee

\subsection{Lensing}

Finally, we also compute the contribution in the third order term which is given by the quadratic term combining density and lensing, see third boldface term in Eq.~(\ref{SigmaLeading}).
\be
B_L \left( \nv_1, \nv_2, \nv_3, z_1, z_2 ,z_3 \right) = \langle \Delta^{(2L)} \left( \nv_1, z_1 \right)\delta_\rho^{(1)} \left( \nv_2, z_2 \right) \delta_\rho^{(1)} \left( \nv_3 ,z_3 \right) \rangle_c + \text{permutations} \, ,
\ee
where
{\be 
\Delta^{(2L)} \left( \nv , z \right) =\Delta^{(L)} \left( \nv , z \right) 
\delta^{(1)}_\rho \left( \nv ,z \right) 
= \left[ - \int_0^r d\deta' \frac{\deta- \deta'}{\deta\deta'} \Delta_2 \left( \psi + \phi \right) \left( \nv \deta' , \eta' \right) \right] \delta^{(1)}_\rho \left( \nv ,z \right) \, . 
\ee }
Repeating the steps for the previous terms and using that 
\be\label{e:trans-lens}
\Delta_\ell^L \left( z, k \right) = \ell \left( \ell +1 \right) \int_0^{\deta} d\deta' \frac{\deta- \deta'}{\deta\deta'} T_{\psi + \phi} \left( \eta, k \right) j_\ell \left( k \deta' \right) \, .
\ee
leads to
\bea \label{BL}
B_L \left( \nv_1, \nv_2, \nv_3, z_1, z_2 ,z_3 \right) &=& \frac{4}{\pi^2} \sum_{\ell, \ell', m , m'}  Y_{\ell m } \left( \vn_1 \right) Y_{\ell' m'} \left( \vn_1 \right) Y^*_{\ell m } \left( \nv_2 \right) Y^*_{\ell' m'} \left( \nv_3 \right)  L_{\ell \ell'} \left( z_1 , z_2 ,z_3 \right) \nonumber \\
&&+ \text{permutations} \, ,
\eea
where 
\bea
 L_{\ell \ell'} \left( z_1, z_2 , z_3 \right) &=& \ell \left( \ell +1 \right)  \int dk_2 dk_3 k_2^2 k_3^2 P_R \left( k_2 \right) P_R \left( k_3 \right) \nonumber \\ 
 && \hspace{1.7cm}\int _0^{\deta_1} d\deta'  \frac{\deta_1- \deta'}{\deta_1\deta'} T_{\psi+ \phi}
 \left( \eta', k_2 \right) T_\delta \left( \eta_1 , k_3 \right) T_\delta \left( \eta_2 , k_2 \right) T_\delta \left( \eta_3 , k_3 \right) \nonumber \\
 && \hspace{4cm} \times j_\ell \left( k_2 \deta' \right) j_{\ell'} \left( k_3 \deta_1\right) j_\ell \left( k_2 \deta_2 \right) j_{\ell'} \left( k_3 \deta_3 \right) \nonumber \\
 &&\hspace{-6pt}+ \ell' \left( \ell' +1 \right) \int dk_2 dk_3 k_2^2 k_3^2 P_R \left( k_2 \right) P_R \left( k_3 \right) 
 \nonumber \\ 
 && \hspace{1.7cm}
 \int _0^{\deta_1} d\deta'  \frac{\deta_1- \deta'}{\deta_1 \deta'} T_{\psi+ \phi}
 \left( \eta', k_3 \right) T_\delta \left( \eta_1 , k_2 \right) T_\delta \left( \eta_2 , k_2 \right) T_\delta \left( \eta_3 , k_3 \right) \nonumber \\
 && \hspace{4cm} \times j_{\ell'} \left( k_3 \deta' \right) j_{\ell} \left( k_2 \deta_1\right) j_\ell \left( k_2 \deta_2 \right) j_{\ell'} \left( k_3 \deta_3 \right)\,.
\eea
With this  we obtain the following expression for the reduced bispectrum containing one lensing term:
\be
b^{L \delta\delta}_{\ell_1 \ell_2 \ell_3 } =  \frac{4}{\pi^2}  L_{\ell_2 \ell_3 }  \left( z_1 ,z_2, z_3 \right) + \text{permutations}\, .
\ee
This can be rewritten as
\bea
b^{L \delta \delta}_{\ell_1 \ell_2 \ell_3} \!&=&\! \left[ \frac{2}{\pi}  \ell_2 \left( \ell_2 \!+\!1 \right)\!  \int\!\! dk_2  k_2^2  P_R \!\left( k_2 \right)  \!\int _0^{\deta_1}\!\! d\deta'  \frac{\deta_1\!-\! \deta'}{\deta_1\deta'} T_{\psi+ \phi}\!
 \left( \eta', k_2 \right) T_\delta \!\left( \eta_2 , k_2 \right) j_{\ell_2} \!\left( k_2 \deta' \right) j_{\ell_2}\! \left( k_2 \deta_2 \right)  \right] \nonumber \\
 && \times   \left[{\frac{2}{\pi}} \int dk_3 k_3^2 P_R \left( k_3 \right)  T_\delta \left( \eta_1 , k_3 \right) T_\delta \left( \eta_3 , k_3 \right) j_{\ell_3} \left( k_3 \deta_1 \right) j_{\ell_3} \left( k_3 \deta_3 \right) \right] \nonumber \\
 &+&\!\left[ \frac{2}{\pi}  \ell_3 \left( \ell_3 \!+\!1 \right)\!  \int\!\! dk_3  k_3^2  P_R \!\left( k_3 \right) \! \int _0^{\deta_1}\!\! d\deta'  \frac{\deta_1\!-\! \deta'}{\deta_1\deta'} T_{\psi+ \phi}\!
 \left( \eta', k_3 \right) T_\delta\! \left( \eta_3 , k_3 \right)  j_{\ell_3} \!\left( k_3 \deta' \right) j_{\ell_3} \!\left( k_3 \deta_3 \right)\right]   \nonumber \\
  && \times   \left[ {\frac{2}{\pi}} \int dk_2 k_2^2 P_R \left( k_2 \right)  T_\delta \left( \eta_1 , k_2 \right) T_\delta \left( \eta_2 , k_2 \right) j_{\ell_2} \left( k_2 \deta_1 \right) j_{\ell_2 }\left( k_2 \deta_2 \right)\right] + \text{permutations}\nonumber \\
&=& c^{L\delta}_{\ell_2} ( z_1, z_2 ) c_{\ell_3}^{\delta\delta} (z_1,z_3) +  c^{L\delta}_{\ell_3} ( z_1, z_3 ) c_{\ell_2}^{\delta\delta} (z_1,z_2) \nonumber \\
&&\hspace{-6pt}+ c^{L\delta}_{\ell_1} ( z_2, z_1 ) c_{\ell_3}^{\delta\delta} (z_2,z_3) +  c^{L\delta}_{\ell_3} ( z_2, z_3 ) c_{\ell_1}^{\delta\delta} (z_1,z_2) \nonumber \\
&&\hspace{-6pt}+ c^{L\delta}_{\ell_1} ( z_3, z_1 ) c_{\ell_2}^{\delta\delta} (z_2,z_3) +  c^{L\delta}_{\ell_2} ( z_3, z_2 ) c_{\ell_1}^{\delta\delta} (z_1,z_3)\, ,
\eea
where we have used~(\ref{trans_den}) and (\ref{e:trans-lens}).

Terms of a similar amplitude  as the ones calculated here would be
$$\langle\Delta^{(1RSD)} \left( \nv_1, z_1 \right) \deltaF \left( \nv_2 , z_2 \right) \deltaS\left( \nv_3 , z_3 \right) \rangle_c + \text{permutations}$$
and
$$\langle\Delta^{(1L)} \left( \nv_1, z_1 \right) \deltaF \left( \nv_2 , z_2 \right) \deltaS\left( \nv_3 , z_3 \right) \rangle_c + \text{permutations}\, ,$$
but for brevity, and since we expect them to yield similar results to the terms already evaluated here, we do not discuss them in this work. However, they have to be added when one wants a good estimate for the total amplitude of the bispectrum.

\subsection{Numerical Results}

We now evaluate and compare the amplitude of different configurations for the density, redshift space distortion and lensing terms.
We use a $\Lambda$CDM model with cosmological parameters consistent with Planck~\cite{Ade:2013kta,Ade:2013zuv} and compute angular power spectra with {\sc CLASS}gal \cite{DiDio:2013bqa}.

\begin{figure}[!htbp]
\begin{center}
\includegraphics[width=0.48\textwidth]{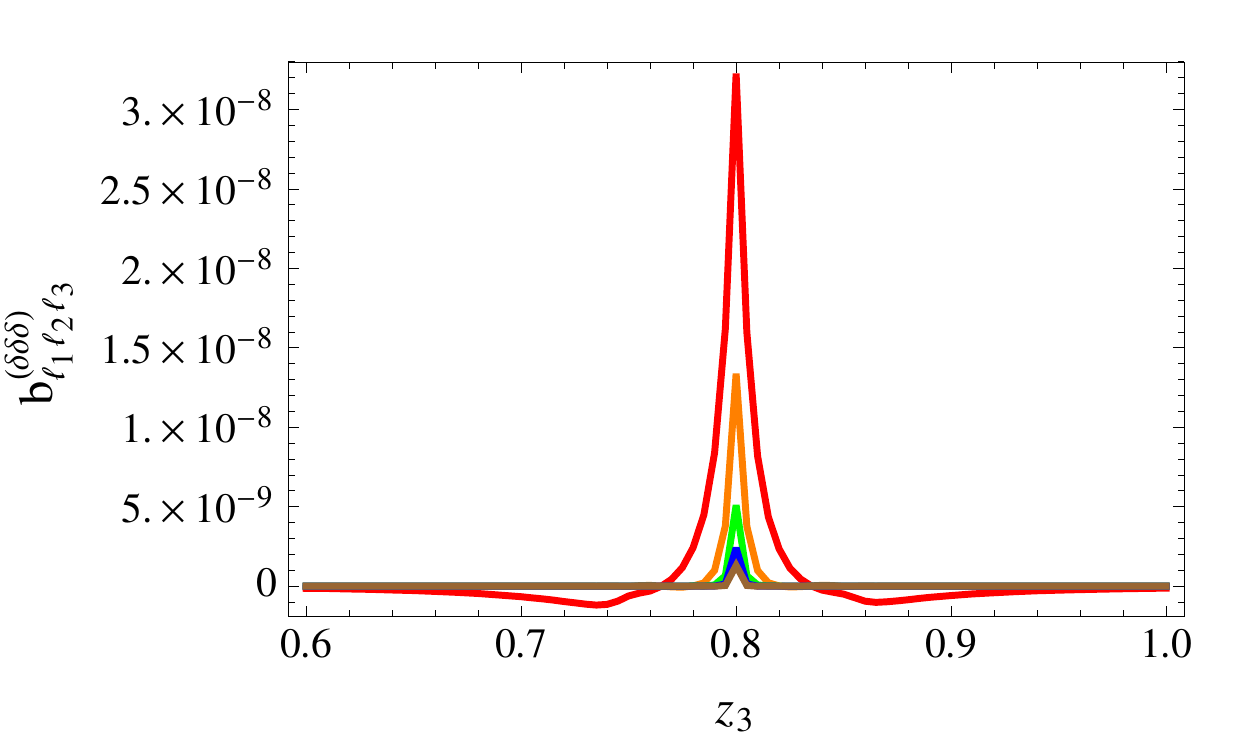}
\includegraphics[width=0.48\textwidth]{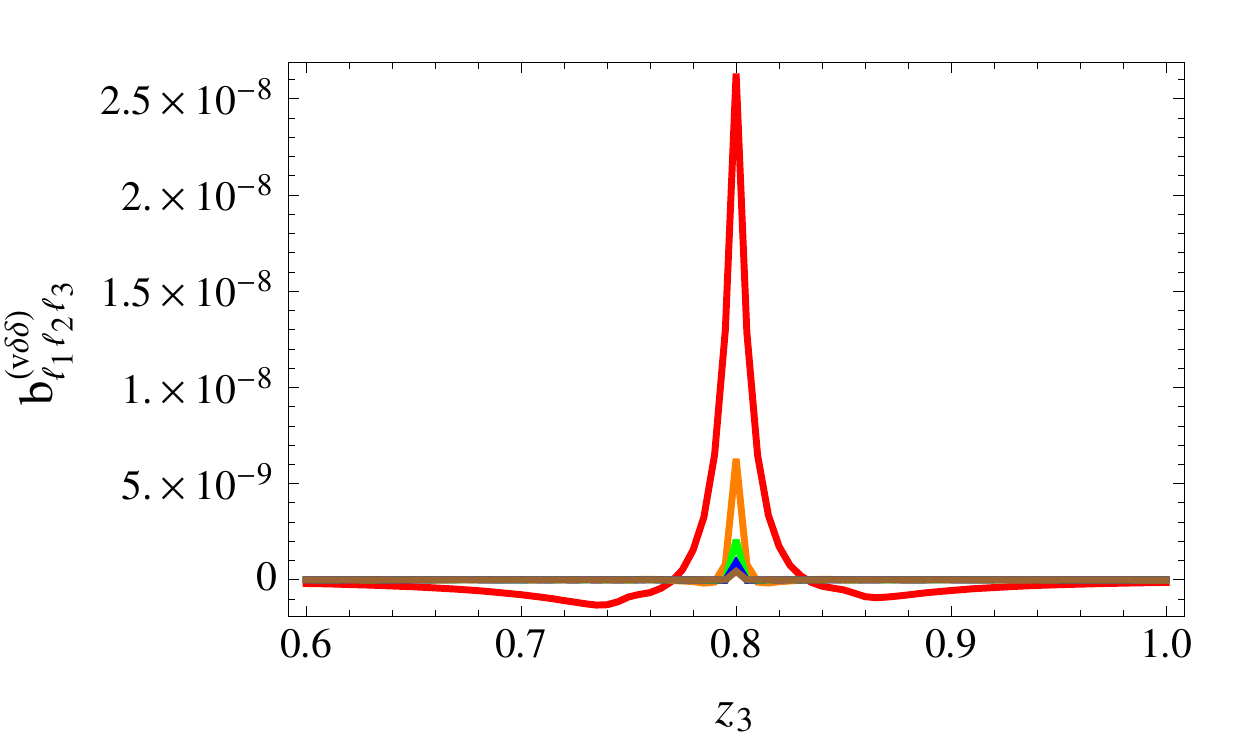}
\includegraphics[width=0.5\textwidth]{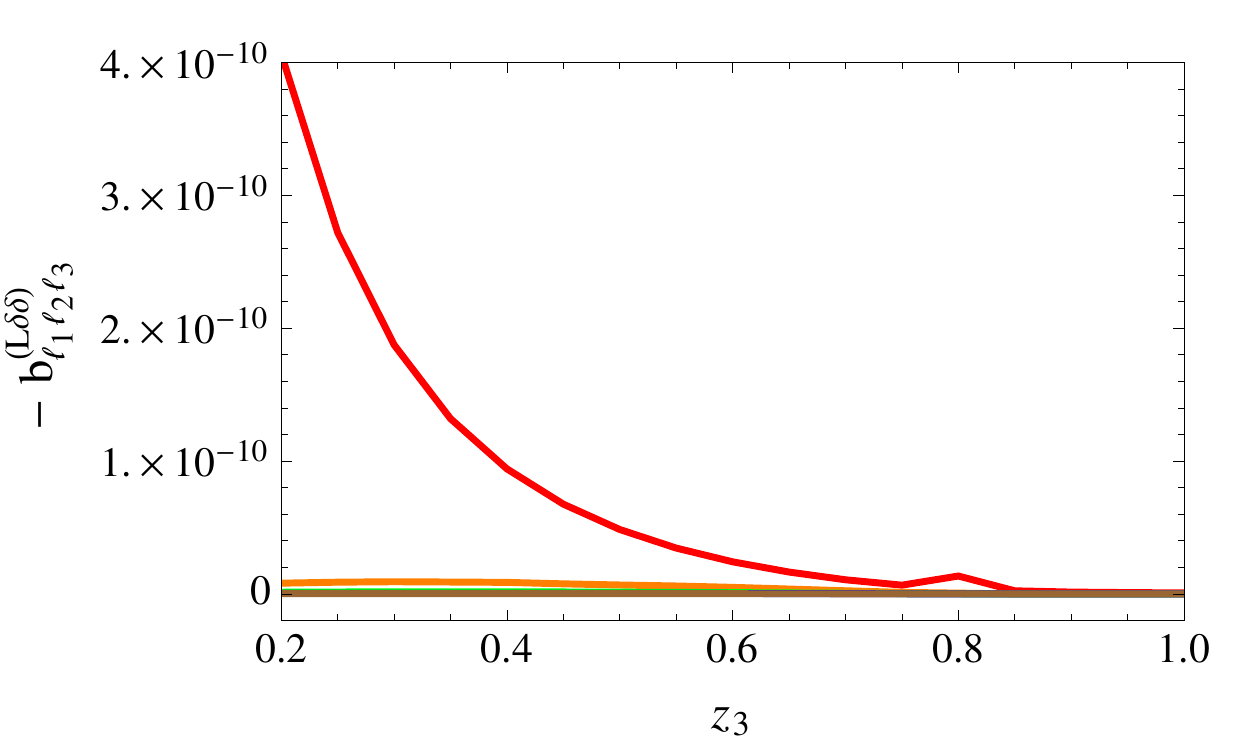}
\caption{We plot the contribution $b^{\delta \delta \delta}_{\ell_1 \ell_2 \ell_3}$ (upper left panel), $b^{v \delta \delta}_{\ell_1 \ell_2 \ell_3}$ (upper right panel), $-b^{L \delta \delta}_{\ell_1 \ell_2 \ell_3}$ (bottom panel) to the bispectrum for $z_1=z_2=0.8$ as a function of $z_3$ for different values of $\ell_1=\ell_2=\ell_3/2$ ($3$ red, $103$ orange, $203$ green, $303$ blue and $403$ brown). Note that the lensing term is typically two orders of magnitude smaller than the density and redshift space distortion terms.  
}
\label{fig:b_z3}
\end{center}
\end{figure}

In Fig.~\ref{fig:b_z3} we plot the contribution to the bispectrum given by the density term $b^{\delta \delta \delta}_{\ell_1 \ell_2 \ell_3}$, redshift space distortions $b^{v \delta \delta}_{\ell_1 \ell_2 \ell_3}$, and lensing $b^{L \delta \delta}_{\ell_1 \ell_2 \ell_3}$ by fixing two redshifts $z_1=z_2=0.8$ while varying the third one around this value.
Results are shown for different values of the multipole scale $\ell$ in the configuration $\ell=\ell_1=\ell_2=\ell_3/2$ (note that in particular with this choice $\ell_1+\ell_2+\ell_3=$ even, as required by the Gaunt integral).
Density and redshift space distortions show a peak when $z_3$ is equal to the other two redshifts.
The density peak is symmetric, and also redshift space distortions give a nearly symmetric contribution (deviations are not visible by eye).
In both cases, correlations vanish on large scales (large redshift differences) and can also lead to a negative bispectrum.
At higher multipoles, the terms oscillate more and pass through zero for the first time already at smaller redshift difference $\Delta z$. In Fig.~\ref{fig:b_z3} only the $\ell=3$ term is sufficiently well resolved to trust its oscillations.

On the other hand, lensing (which involves an integral along the line of sight) shows an asymmetric contribution with respect to $z_1=z_2$.
Fig.~\ref{fig:b_z3} suggests that high values of the lensing term in the bispectrum are obtained for configurations where a third galaxy is placed at a redshift $z_3$ much lower than the other two redshifts.
This comes from the fact that $-c^{L\de}_\ell(z_1,z_3)$ decreases with the redshift $z_3$, simply since the density fluctuation  decreases.
Note however a small peak around $z_3=0.8=z_1=z_2$, due to redshift auto-correlations of density-density  contribution to $b^{L \delta \delta}_{\ell_1 \ell_2 \ell_3}$.
Lensing gives a negative contribution, since its sign is given by the anti-correlation with density perturbations as already found, e.g., in~\cite{Bonvin:2011bg,DiDio:2013sea}.
Finally, from Fig.~\ref{fig:b_z3} it is already clear that redshift space distortions contributes to the bispectrum as much as density, while in this configuration lensing is subleading.

\begin{figure}[!htbp]
\begin{center}
\includegraphics[width=0.45\textwidth]{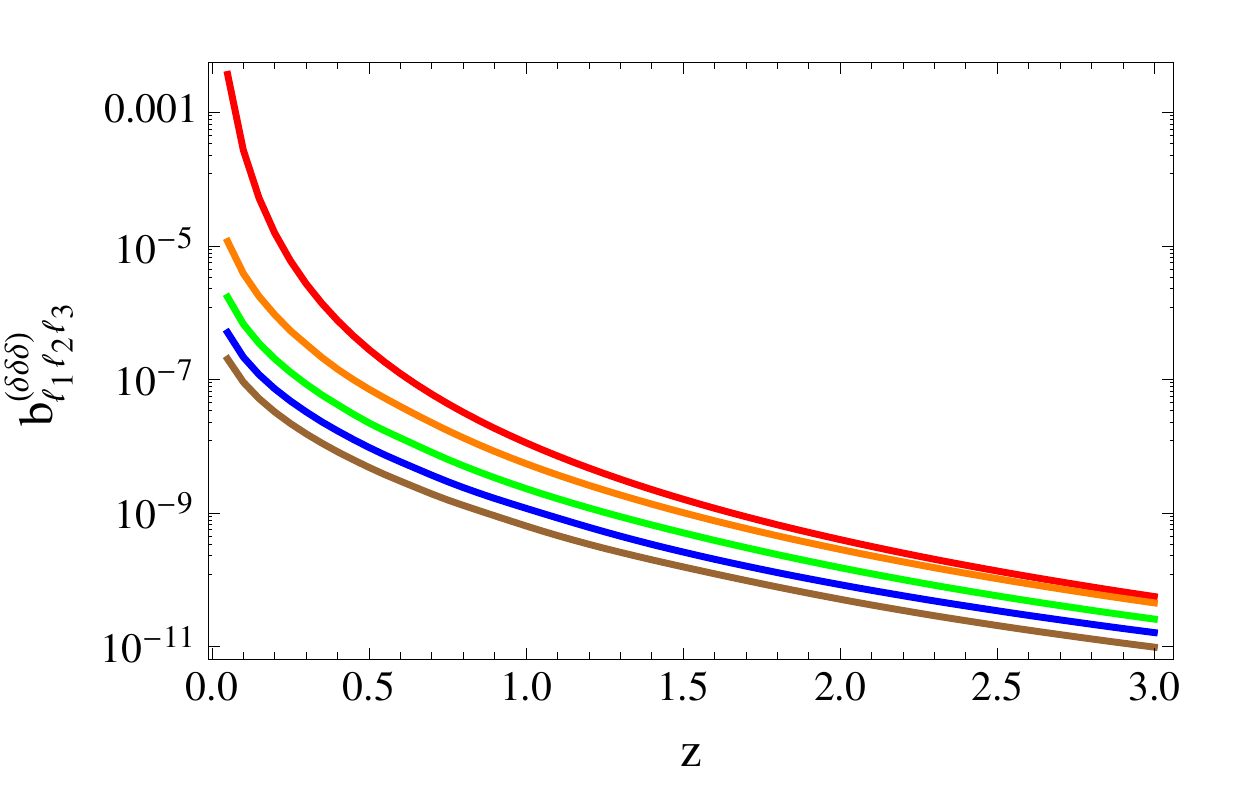}
\includegraphics[width=0.45\textwidth]{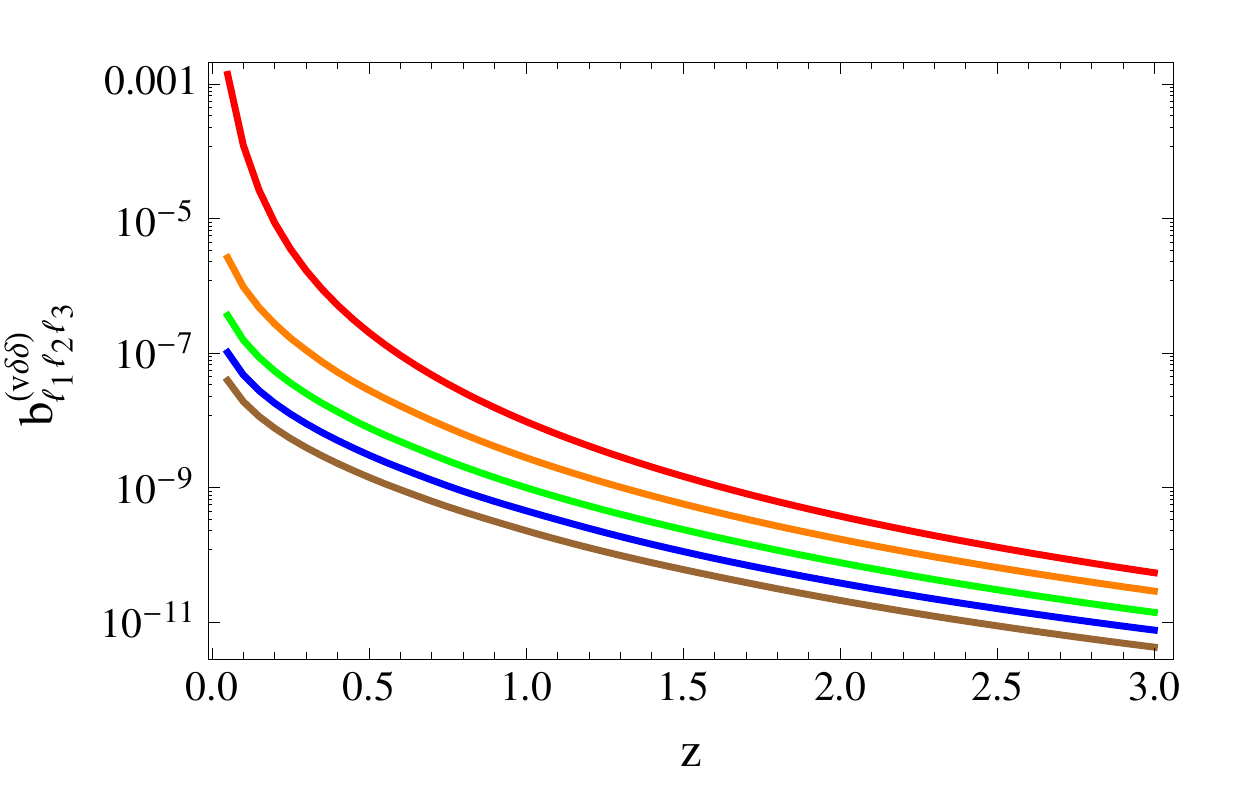}
\includegraphics[width=0.45\textwidth]{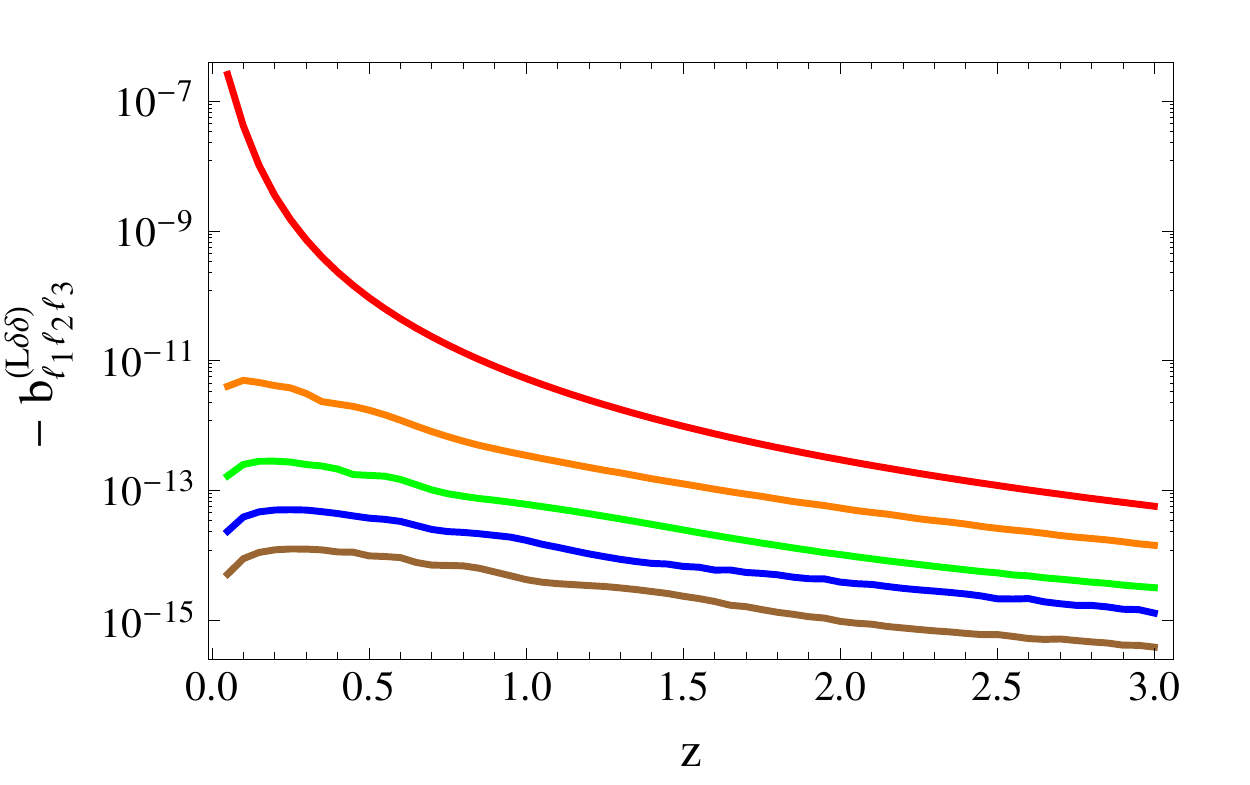}
\caption{
We plot the contributions of $b^{\delta \delta \delta}_{\ell_1 \ell_2 \ell_3}$ (up left), $b^{v \delta \delta}_{\ell_1 \ell_2 \ell_3}$ (up right), $-b^{L \delta \delta}_{\ell_1 \ell_2 \ell_3}$ (bottom) to the bispectrum for $z_1=z_2=z_3=z$ as a function of $z$ for different $\ell=\ell_1=\ell_2=\ell_3/2$ ($3$ red, $103$ orange, $203$ green, $303$ blue, $403$ brown).
}
\label{fig:b_zequal}
\end{center}
\end{figure}

Fig.~\ref{fig:b_zequal} shows the different contributions to the bispectrum when considering three galaxies at the same redshift $z$ while varying the multipoles $\ell=\ell_1=\ell_2=\ell_3/2$ scale.
All contributions decrease with $z$.
This is expected since we only consider deviation from a zero bispectrum {coming from non-linearities}, which increase with time. There is a small turnover visible in the lensing bispectrum for $\ell\ge 100$ which comes from the suppression of lensing at small redshift. For $\ell=3$ this is not visible as the signal is dominated by very large scale structures. Not that this is the pure $b^{L \delta \delta}_{\ell \ell \ell}$, not multiplied with any factors of $\ell$ which enhances the high $\ell$ signal.

\begin{figure}[!htbp]
\begin{center}
\includegraphics[width=0.45\textwidth]{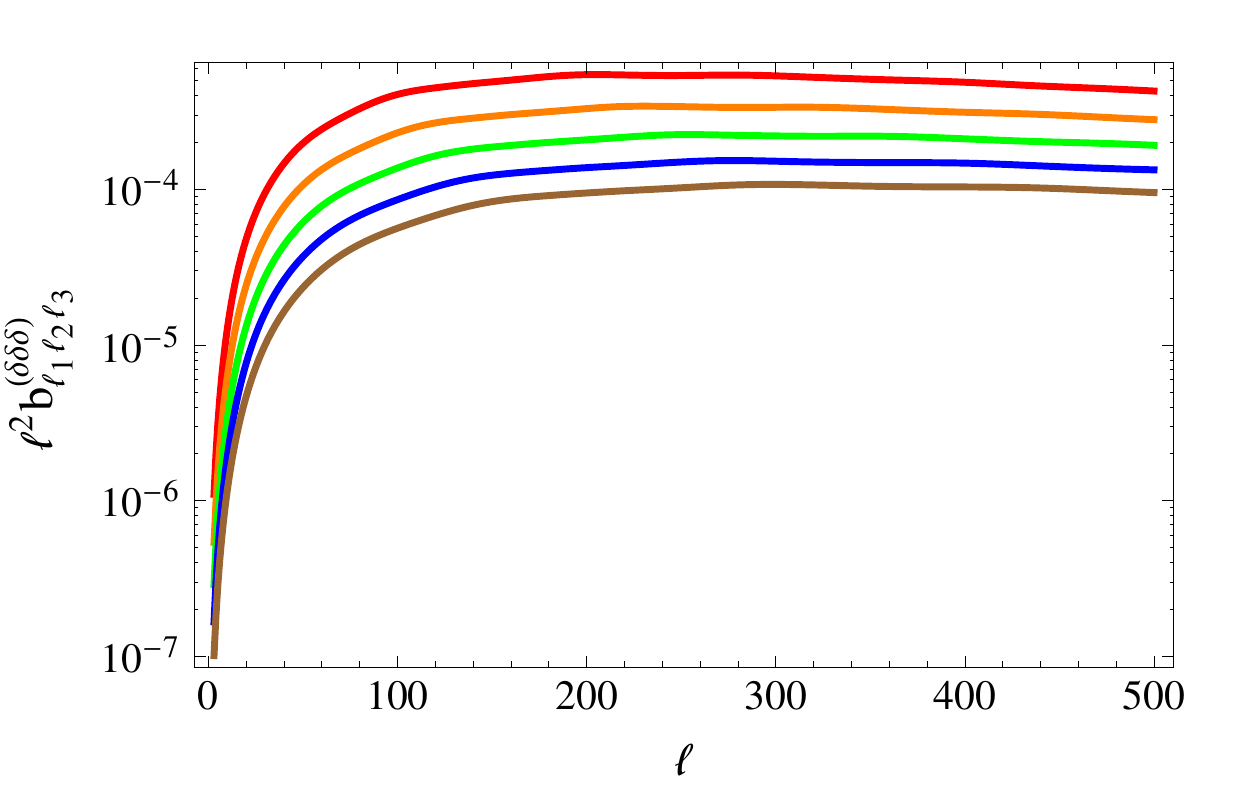}
\includegraphics[width=0.45\textwidth]{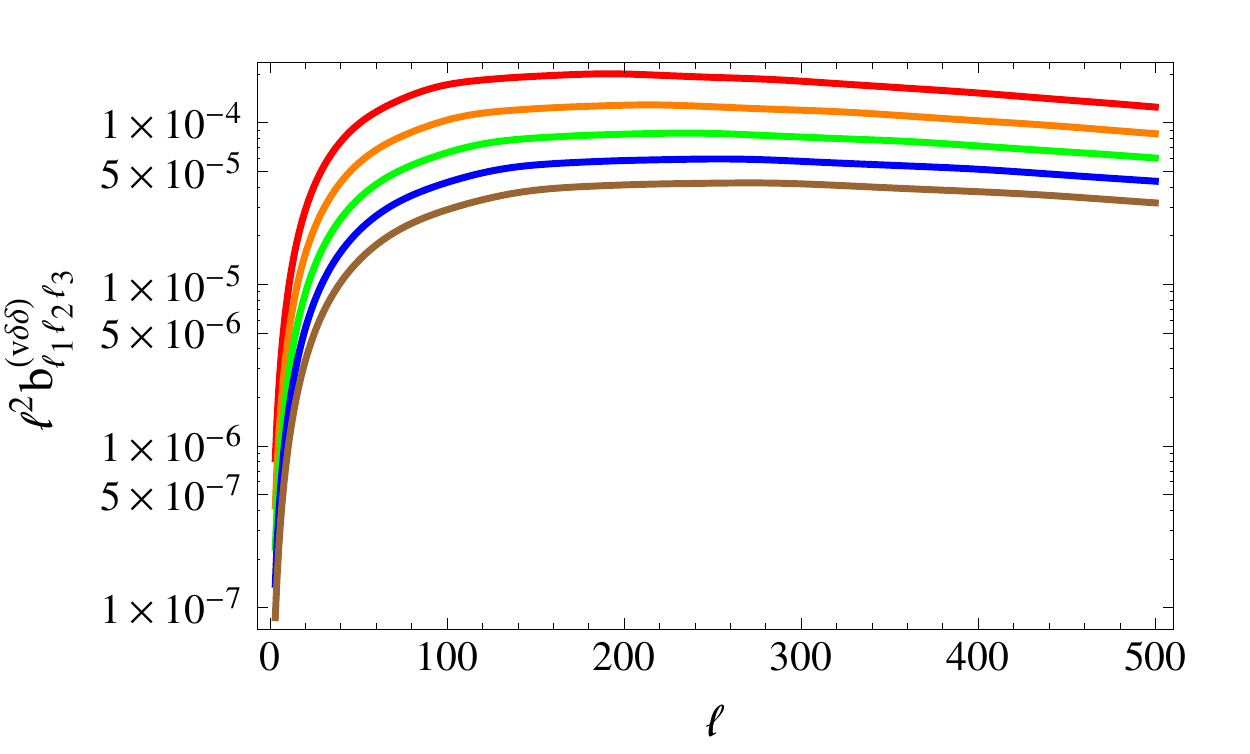}
\includegraphics[width=0.45\textwidth]{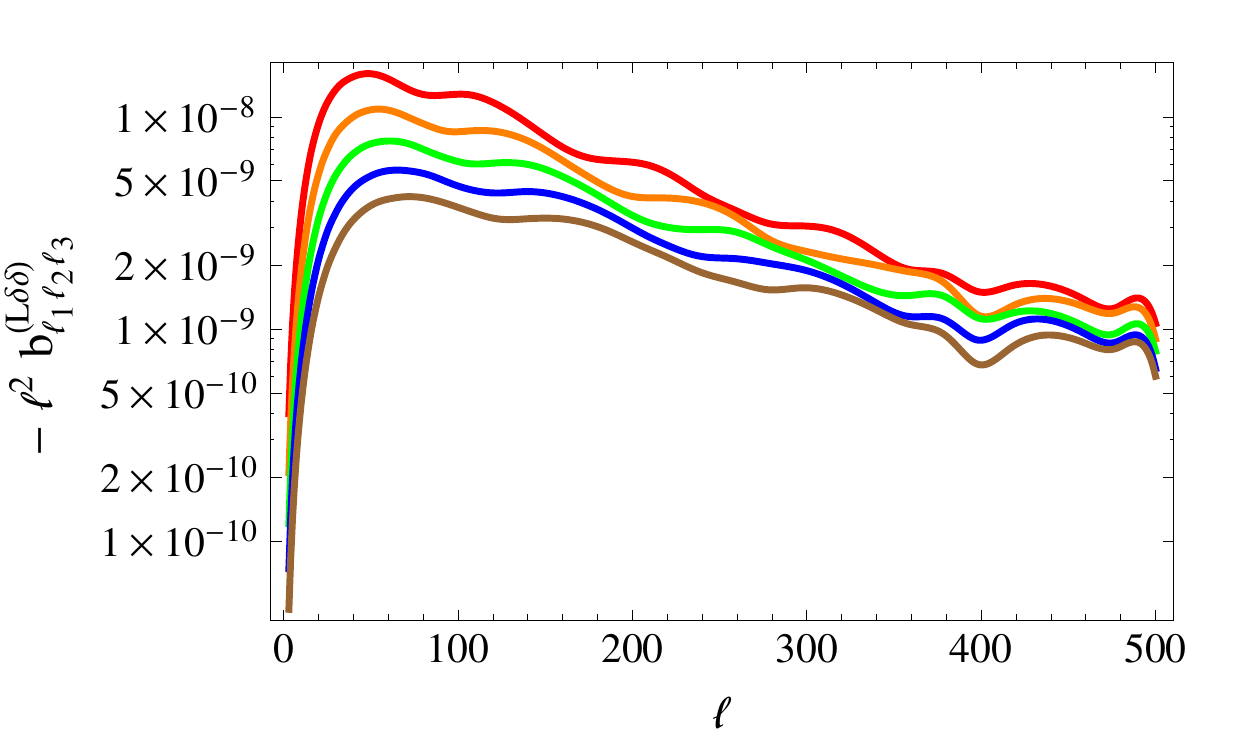}
\caption{
We plot the contributions of $\ell^2b^{\delta \delta \delta}_{\ell_1 \ell_2 \ell_3}$ (up left), $\ell^2b^{v \delta \delta}_{\ell_1 \ell_2 \ell_3}$ (up right), $-\ell^2b^{L \delta \delta}_{\ell_1 \ell_2 \ell_3}$ (bottom) to the the bispectrum for different $z_1=z_2=z_3=z$ ($z=0.6$ red, $z=0.7$ orange, $z=0.8$ green, $z=0.9$ blue, $z=1$ brown) as a function of $\ell_1=\ell_2=\ell_3/2=\ell$. Contrary to the density and redshift space distortion which are scale invariant, the lensing contribution is decreasing with increasing $\ell$;  $b^{L \delta \delta}_{\ell_1 \ell_2 \ell_3}$ roughly behaves like $\ell^{-3}$.
}
\label{fig:b_of_ell}
\end{center}
\end{figure}

In Fig.~\ref{fig:b_of_ell} we plot the contributions to the bispectrum again by fixing $z_1=z_2=z_3$, but now as a function of the multipole scale $\ell_1=\ell_2=\ell_3/2=\ell$. For the density and redshift space distortions the $\ell$-dependence of the bispectrum is similar to the one of the power spectrum. A rapid increase up to $\ell \simeq 100$ followed by a nearly scale invariant bispectrum, $\ell^2 b_{\ell\ell\ell}(z,z,z)\simeq$ constant. The lensing contribution however decays for $\ell\gtrsim50$.
The oscillations in the lensing contribution (bottom panel) are numerical. They can be removed by enhancing the redshift resolution. Since this is numerically quite costly we have refrained from doing it in this order of magnitude discussion.

Any real experiment can only count galaxies within a certain redshift bin. Hence to compare our theoretical prediction with real observations we need to integrate the theoretical reduced bispectrum $b_{\ell_1 \ell_2 \ell_3} (z_1,z_2,z_3)$ over three window functions describing the redshift bins. Therefore we define the observed reduced bispectrum as
\begin{equation}
b^W_{\ell_1 \ell_2 \ell_3} (z_1,z_2,z_3) = \int dz'_1 dz'_2 dz'_3 W(z_1,z'_1) W(z_2,z'_2) W(z_3,z'_3) b_{\ell_1, \ell_2, \ell_3}(z'_1, z'_2,z'_3) \, ,
\end{equation}
where $W(z,z')$ denotes the window function with center at $z$, normalized to the unity in the redshift bin,  as a function of $z'$.

Broad window functions smear out the signal along the line of sight. This typically leads to a reduction of the amplitude of the bispectrum. This reduction affects strongly the density and even more the redshift space distortions, if integrated over  large redshift bins. The lensing, instead, which is a coherent signal integrated along the line of sight, is almost insensitive to the presence of the window functions. This implies that the signal coming from the lensing can become comparable to the density signal for wide window functions.

\begin{figure}[!htbp]
\begin{center}
\includegraphics[width=0.7\textwidth]{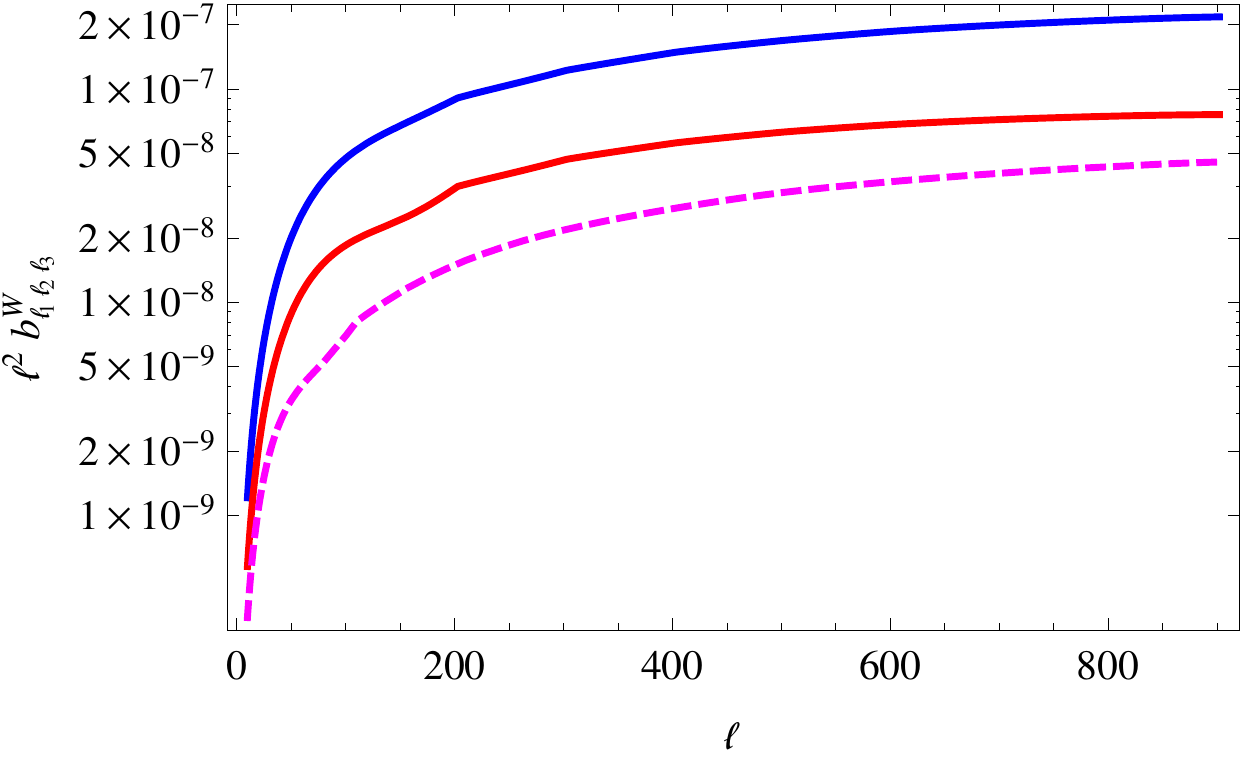}
\caption{We show the effect of a very large redshift bin, for which the bispectrum is integrated from $z_{\min}=0.2$ to $z_{\max}=3$, for a fix $\ell_3=3$ while varying $\ell=\ell_1=\ell_2$.
We plot the density (blue), redshift space distortions (red) and lensing (magenta) contributions. Dashed lines correspond to negative values.
}
\label{fig:b_W}
\end{center}
\end{figure}

In Fig.~\ref{fig:b_W} we show the different contributions to the bispectrum integrated along a very large redshift bin, $0.2<z<3$.
We fix one multipole value $\ell_3=3$ and vary $\ell=\ell_1=\ell_2$.
As expected, density and redshift space distortions signals decrease, and lensing is no longer a negligible term.
This is consistent with what was found, e.g., in~\cite{DiDio:2013sea}. The redshift space distortion contribution to the bispectrum is less reduced than the one to the power spectrum since in the signal considered here we integrate over the product $\de_\rho^{(1)}(z)\dd_rv_{\parallel}(z)$ which are taken at the same redshift.
Nevertheless, in Fig.~\ref{fig:b_W}  all three contributions  are of the same order of magnitude. This  suggests that the study of the bispectrum within a large redshift bin may lead to new independent constraints of lensing potential.

\section{Conclusions}
\label{Sec6}
\setcounter{equation}{0}
In this work we have computed the second order contribution to the perturbed galaxy number counts as a function of the observed redshift and the direction of observation. We have first expressed our result in the geodesic 
light-cone gauge~\cite{Gasperini:2011us} where it is simple, and exact for a given (exact) density fluctuation. Nevertheless, since so far the cosmological perturbation equations in this gauge are not known, we have translated our expression to the longitudinal or Poisson gauge. Our main final result is given in Eqs.~(\ref{e:Delta2}),~(\ref{e:SigmaIS}) and~(\ref{e:SigmaAS}). 
Furthermore, in Eq.~(\ref{SigmaLeading}) we also present a simplified formula containing only the dominant terms.
Since we have not used Einstein's equation in our derivation, the result is also valid for arbitrary models of non-interacting dark energy and for modified gravity theories as long as photons and dark matter particles move along geodesics.

We have then used our result to compute the probably dominant  contributions to the angular bispectrum of the number counts, $B_{\ell_1\ell_2\ell_3}(z_1,z_2,z_3)$. In particular, we have calculated the terms coming from second order density, $b^{(\de\de\de)}_{\ell_1\ell_2\ell_3}(z_1,z_2,z_3)$, from redshift space distortion combined with first order density terms, $b^{(v\de\de)}_{\ell_1\ell_2\ell_3}(z_1,z_2,z_3)$, and from lensing combined with first order density terms, $b^{(L\de\de)}_{\ell_1\ell_2\ell_3}(z_1,z_2,z_3)$.  As already with the power spectrum, $b^{(\de\de\de)}$ and $b^{(v\de\de)}$ are typically of the same order, namely $\ell^2b^{(\de\de\de)}_{\ell\ell\ell}(z,z,z)\simeq 10^{-4}-10^{-3}$ for small redshift $z\lesssim 0.5$ and $\ell\gtrsim 100$. The lensing term is about four orders of magnitude smaller. This results hold for a very narrow window function $\De z\ra 0$ which we approximate as a Dirac $\delta$-function. For broader window functions the density and redshift space distortion signals decrease while the lensing signal is virtually unchanged.
If we integrate over redshift from $z_{\min}=0.2$ to $z_{\max}=3$ the lensing signal contributes close to 20\% of the total and cannot be neglected.

This work mainly presents the derivation of the second order number counts expression and some 
preliminary evaluation of the bispectrum.
In the future~\cite{DDMM} we plan to analyze and compare the other terms in the bispectrum and to study whether the signal calculated here is measurable by planned surveys like Euclid. Especially, the new terms with four transverse derivatives are very promising and may well be at least as large as the lensing term discussed here. We shall also include the effects of biasing and magnification bias which have been neglected in this first discussion. The comparison of this bispectrum, which includes only the non-linear effects of gravity, with a possible primordial bispectrum will be necessary for any conclusions on primordial non-Gaussianities.


\section*{Acknowledgments}
FM has enjoyed the hospitality of the Institute for Theoretical Physics at the University of Heidelberg during the completion of this work. ED, RD and FM acknowledge support from the Swiss National Science Foundation.
GM is partially supported by the Marie Curie IEF, Project NeBRiC - ``Non-linear effects and backreaction in classical and quantum cosmology".

\appendix 
\section{Useful relations}
\label{AppA}
\setcounter{equation}{0}

We give here some additional relations that have been used in the evaluation of the observed redshift and of the galaxy number counts to second order.
 
In particular to compute the second order redshift perturbation we  need (see \cite{Marozzi:2014kua}):
\begin{eqnarray}
Q_s &=& -2 \int_{\eta_s}^{\eta_o} d \eta' \psi^I\left(\eta'\right) \label{Us1} 
\\
\partial_+ Q_s &=& - \psi^I_s -2 \int_{\eta_s}^{\eta_o} d \eta' \partial_{\eta'}\psi^I\left(\eta'\right) \label{Us2} 
\\
\partial_+\omega_s^{(2)} &=& -\frac{1}{4}\left(\phi_s^{(2)} + \psi_s^{(2)}\right)
- \psi^I_s \left[ - \psi^I_s -2 \int_{\eta_s}^{\eta_o} d \eta' \partial_{\eta'}\psi^I\left(\eta'\right) \right]
-2 \psi_s^I \psi_s^A  
\nonumber \\
&&
- {\gamma_{0s}^{ab}} \partial_a \left( \int_{\eta_s}^{\eta_o} d \eta' \psi^I\left(\eta'\right) \right) 
\partial_b \left( \int_{\eta_s}^{\eta_o} d \eta' \psi^I\left(\eta'\right) \right)  \nonumber \\
&& 
-\frac{1}{2} \int_{\eta_s}^{\eta_0} d\eta' \partial_{\eta'}  \left[ \phi^{(2)}\left( \eta'\right) + \psi^{(2)}\left( \eta'\right) 
+8 \psi^I\left(\eta'\right) \psi^A\left(\eta'\right)
- 4 \left(\psi^I\left(\eta'\right)\right)^2
\right.
\nonumber \\
&&
\left. 
-8\psi^I\left( \eta'\right) \int_{\eta'}^{\eta_o} d \eta'' \partial_{\eta''}\psi^I\left(\eta'' \right) 
+ 4 \gamma_0^{ab} \partial_a \left( \int_{\eta'}^{\eta_o} d \eta'' \psi^I\left(\eta''\right) \right) 
\partial_b \left( \int_{\eta'}^{\eta_o} d \eta'' \psi^I\left(\eta''\right) \right) \right] \,,\nonumber \\
\label{Us3}
\end{eqnarray}
as well as
\bea
& & \!\!\!\!\!\!\int_{\eta_s}^{\eta_0} d\eta' \partial_{\eta'}  \left[\psi^I\left(\eta'\right) \left(- \psi^I\left(\eta'\right) -2 \int_{\eta'}^{\eta_o} d \eta'' \partial_{\eta''}\psi^I\left(\eta''\right)\right) 
\right.
\nonumber \\ 
&&
\left. \hspace{2cm}
+\gamma_0^{ab} \partial_a \left( \int_{\eta'}^{\eta_o} d \eta'' \psi^I\left(\eta''\right) \right) 
\partial_b \left( \int_{\eta'}^{\eta_o} d \eta'' \psi^I\left(\eta''\right) \right) \right]
\nonumber 
\\
&=& 
-2\int_{\eta_s}^{\eta_0} d\eta' \left[
- \psi^I\left(\eta'\right) \partial_{\eta'}\psi^I\left(\eta'\right) 
- \partial_{\eta'}\psi^I\left(\eta'\right) \int_{\eta'}^{\eta_o} d \eta'' \partial_{\eta''}\psi^I\left(\eta''\right)
\right.
\nonumber \\ & & \left.
- \psi^I\left(\eta'\right) \int_{\eta'}^{\eta_o} d \eta'' \partial^2_{\eta''}\psi^I\left(\eta''\right)
+ \gamma_0^{ab} \partial_a \left( \!\int_{\eta'}^{\eta_o} \!\!d \eta'' \psi^I\!\left(\eta''\right) \right) 
\partial_b \left(\! \int_{\eta'}^{\eta_o} \!\!d \eta'' \partial_{\eta''}\psi^I \!\left(\eta''\right) \right) \right]\,. \quad
\label{Us4}
\eea
Useful relations for the evaluation of $\frac{1}{a(\eta_s^{(0)})} \frac{d \tau_s^{(2)} }{d\etasB}$ are
\bea
 \frac{d v_{||s}}{d\etasB}&=&\partial_r \left(\psi_s^I-\psi_s^A\right)-\partial_r v_{||s}-\HH_s v_{||s} \\
\frac{d v_{||s}^{(2)}}{d\etasB}&=& \frac{1}{2} \partial_r \phi_s^{(2)} -\partial_r v^{(2)}_{||s}-\HH_s v^{(2)}_{||s} 
+\left(\partial_\eta \psi_s^I+\partial_\eta \psi_s^A\right) v_{||s}+2 \psi_s^A\left(\partial_r \psi_s^I-\partial_r \psi_s^A\right)+v_{||s} \partial_r v_{||s}
\nonumber \\ & & 
-\frac{1}{\deta_s} v_{\perp a\,s} v_{\perp\,s}^a -a v_{\perp\,s}^a \partial_a v_{||s} 
\eea
and 
\bea
& & \!\!\!\!\!\!\!\frac{1}{2\HH_s} \frac{d }{d\etasB}\left[v_{\perp a\,s} v_{\perp\,s}^a+2 a v_{\perp\,s}^a \partial_a Q_s\right]
+\frac{1}{2\HH_s} \partial_\eta \left(\gamma_0^{ab}\partial_a Q_s\partial_b Q_s\right)  \nonumber \\
& =&
- \left( 1 - \frac{1}{\HH_s \deta_s} \right) v_{\perp a\,s} v_{\perp\,s}^a +\frac{1}{\HH_s} a v_{\perp\,s}^a \partial_a  v_{||s} +2 \left( 1 - \frac{2}{\HH_s \deta_s} \right)   a v_{\perp\,s}^a \partial_a \Qint 
\nonumber \\
&&
+\frac{1}{\HH_s}a v_{\perp\,s}^a \partial_a \left(\psi^I_s+ \psi_s^A\right)+\frac{2}{\HH_s} \gamma_{0s}^{ab}\partial_a (\psi_s^I-\psi_s^A-v_{||s})
\partial_b \Qint
\nonumber
\\
& & 
+\frac{4}{\HH_s} \gamma_{0s}^{ab}\partial_a\left(\Qint\right) \partial_b \left(\dQint\right)\,.
\eea

\end{document}